\documentclass[12pt]{elsarticle}

\usepackage{afterpage}
\usepackage[paper=A4,pagesize]{typearea}
\usepackage{tabularx}
\usepackage[T1]{fontenc}
\usepackage{amsmath,amssymb}
\usepackage{graphicx}
\usepackage{lscape}
\usepackage{longtable}
\usepackage{lineno,hyperref}
\usepackage{bm}
\usepackage{perpage}
\usepackage{footnote}
\usepackage{booktabs,caption}
\usepackage[flushleft]{threeparttable}
\usepackage{flexisym}
\usepackage{rotating}
\usepackage{float}
\usepackage{multirow}
\usepackage{makecell}
\usepackage{amssymb}
\usepackage{xcolor}
\usepackage[nameinlink]{cleveref}
\usepackage{mathtools}
\usepackage{cleveref}
\usepackage{caption}
\usepackage{csquotes}
\usepackage{geometry}
\hypersetup{
colorlinks   = true, 
urlcolor     = blue, 
linkcolor    = blue, 
citecolor   = red 
}
\MakePerPage{footnote}
\modulolinenumbers[5]
\biboptions{authoryear}

\begin{document}
\begin{frontmatter}
\title{An Overview of Meta-Analytic Methods for Economic Research}
\author[1]{Amin Haghnejad\texorpdfstring{\corref{cor1}}{}}
\address[1]{Assistant Professor of Economics, Faculty of Economics, University of Tehran, Tehran, Iran.\\
 E-mail: \texttt{amin.haghnejad@ut.ac.ir}}
\author[2]{Mahboobeh Farahati}
\address[2]{Assistant Professor of Economics, Faculty of Economics, Management and Administrative Sciences, Department of Economics, Semnan University, Semnan, Iran.\\
E-mail: \texttt{m.farahati@semnan.ac.ir}}
%
\journal{\textit{ar}\textsf{Xiv}}
\begin{abstract}
Meta-analysis employs statistical techniques to synthesize the results of individual studies, providing an estimate of the overall effect size for a specific outcome of interest. The direction and magnitude of this estimate, along with its confidence interval, offer valuable insights into the underlying phenomenon or relationship. As an extension of standard meta-analysis, meta-regression analysis incorporates multiple moderators---capturing key study characteristics---into the model to explain heterogeneity in true effect sizes across studies. This study provides a comprehensive overview of meta-analytic procedures tailored to economic research, addressing key challenges  such as between-study heterogeneity, publication bias, and effect size dependence. It equips researchers with essential tools and insights to conduct rigorous and informative meta-analyses in economics and related fields.
\end{abstract}

\end{frontmatter}

\section{Introduction} \label {Introduction}
Meta-analysis,  a powerful statistical technique, has become essential for synthesizing research findings across diverse disciplines. By integrating results from multiple studies, it enables researchers to draw comprehensive conclusions that overcome the limitations of individual investigations. The development of meta-analysis is deeply rooted in the history of scientific inquiry, with its origins in early statistical methods of the 17th century and its formalization in the late 20th century.

The foundation for quantitatively synthesizing study results dates back to the 17th century with the work of French mathematician Blaise Pascal, who devised mathematical methods to calculate probabilities in games of chance. These methods were later adopted by astronomers to combine observational data \citep{cheung2016guide}.

In the 18th and 19th centuries, mathematicians like Carl Friedrich Gauss and Pierre-Simon Laplace further developed these ideas. George Biddell Airy, the British Astronomer Royal, formalized these methods in his textbook, which influenced later statisticians such as Karl Pearson and Ronald Fisher \citep{o2007historical}. Karl Pearson In is credited with one of the earliest applications of meta-analytic techniques. In 1904, Pearson conducted a study to examine the relationship between mortality and inoculation with a vaccine for typhoid fever by combining correlation coefficients across multiple studies \citep{Pearson1904report}. Pearson's work laid the groundwork for the formalization of meta-analysis, even though his methods lacked the rigor characteristic of modern meta-analyses.

Ronald Fisher, working at the Agricultural Research Station in Rothamsted during the 1920s and 1930s, made significant contributions to the development of meta-analysis. In his 1935 textbook, Fisher illustrated how to effectively analyze multiple agricultural studies, addressing issues such as the variability of fertilizer effects across years and locations \citep{Fisher1935design}. His colleague, William Cochran, expanded these methods and applied them to both agricultural and medical research, further advancing the techniques of meta-analysis.

The term ``meta-analysis'' was coined by Gene Glass in 1976 to describe the statistical synthesis of results from multiple individual studies with the goal of integrating their findings \citep{glass1976primary}. Glass's work, along with that of his colleagues, drew significant attention to this approach, particularly in the field of psychology. For instance, Mary Smith and Glass published a meta-analysis on the effectiveness of psychotherapy, demonstrating that psychotherapy was effective and that there was little difference in effectiveness across different types of therapies \citep{smith1977meta}.

The use of meta-analysis in medical research began to take shape in the 1970s. A particularly influential study was the first randomized controlled trial conducted by Peter Elwood, Archie Cochrane, and their colleagues, which aimed to evaluate the effect of aspirin in reducing the recurrence of heart attacks. Although the initial results suggested a potential beneficial effect, they did not achieve statistical significance. As more trials became available, Elwood and Cochrane systematically aggregated and synthesized the findings through meta-analysis, thereby providing a more robust evaluation of aspirin's efficacy \citep{elwood1974randomized}.

Richard Peto and his colleagues further advanced the field by providing detailed examples of systematic reviews and meta-analyses, particularly in the context of randomized trials of beta-blockade following heart attacks \citep{peto1988randomised}. These contributions were instrumental in promoting the use of meta-analysis in clinical research.

In the decades following its formalization, meta-analysis has become a widely used method across various scientific disciplines, including social sciences, medicine, education, public health, environmental science, psychology, and economics. Notable publications and textbooks, such as Hedges and Olkin's \emph{Statistical Methods for Meta-Analysis} (1985), have provided rigorous statistical frameworks for conducting meta-analyses \citep{hedges1985statistical}.

Overall, meta-analysis has evolved from early statistical methods for combining observations to a sophisticated and widely accepted technique for synthesizing research findings across diverse fields. Its development has been marked by contributions from mathematicians, statisticians, and researchers who recognized the need for systematic approaches to integrating study results. Today, meta-analysis continues to be a vital tool for advancing scientific knowledge and informing evidence-based practice.

Meta-analysis has gained significant importance in economics, offering a structured approach to synthesizing a vast and diverse body of research. Economic studies often yield varying results due to differences in data, methodologies, and contexts. Meta-analysis helps in systematically combining these results, allowing economists to derive more robust and comprehensive conclusions. By integrating findings from multiple studies, meta-analysis enhances statistical power and provides more reliable estimates of effect sizes, which are crucial for understanding economic phenomena
\citep{stanley2012meta,borenstein2021introduction}.

Meta-analysis is a crucial tool in the evaluation of economic policies. Economic policies and interventions are often subject to rigorous analysis, and meta-analysis enables the aggregation of evidence from different studies to assess their overall effectiveness. This approach is particularly valuable in fields such as labour economics, health economics, public economics, environmental economics, resource economics, development economics, international economics, monetary economics, and behavioral economics, where policymakers rely on robust evidence to make informed decisions. For example, meta-analyses have been extensively used to evaluate the effectiveness of labour market policies \citep{card1995time, neumark2006minimum,card2010active,card2018works,vooren2019effectiveness,campolieti2020does,martinez2021effects}.

Meta-analysis also plays a crucial role in addressing publication bias, a common issue in economic research. Studies with statistically significant or theoretically well-supported findings are more likely to be published, while those with null or insignificant results, or inconsistent with theoretical expectations, are often overlooked. This bias can distort the overall understanding of an economic issue. Meta-analysis techniques, such as funnel plots and statistical tests for publication bias, help in detecting and correcting for these biases, leading to more accurate and credible conclusions \citep{egger1997bias,rothstein2005publication}.

Furthermore, meta-regression analysis, an extension of meta-analysis, allows economists to explore the heterogeneity in study results by incorporating multiple moderators representing identifiable study characteristics. This technique helps in understanding the factors that contribute to variations in effect sizes across different studies, providing deeper insights into economic relationships. For example, meta-regression can be used to analyze how different labor market conditions or policy environments influence the effectiveness of training programs \citep{stanley2012meta,borenstein2021introduction}.

In conclusion, meta-analysis has become an indispensable tool in economic research, offering a rigorous and systematic approach to synthesizing a wide range of study results. By addressing issues such as heterogeneity, publication bias, and effect size dependence, meta-analysis provides economists with the tools necessary to conduct comprehensive and informative analyses. This study aims to provide an overview of the meta-analytic procedures tailored for economic research, equipping researchers with the insights needed to conduct robust and impactful meta-analytic studies.

In the following sections, this paper will present a detailed exploration of various meta-analytic methods employed in economic research. \Cref{meta-analysis} delves into the methodologies for estimating effect sizes and confidence intervals in meta-analysis, alongside an examination of the different approaches for assessing and quantifying heterogeneity across studies. \Cref{meta-regression} expands on meta-analysis through meta-regression analysis, which incorporates study characteristics (moderators) to address heterogeneity, offering insights into fixed and random effects models, estimation tech-niques, tests for residual heterogeneity, and the influence of moderators on observed effects. \Cref{publication} addresses the challenges of publication bias in meta-analyses, including outcome reporting bias and language bias, and outlines methods such as funnel plots and statistical tests to detect and mitigate these biases. \Cref{uwls} introduces the Unrestricted Weighted Least Squares (UWLS) estimator for meta-analysis and meta-regression models, highlighting its advantages over traditional fixed effects models by accounting for residual heterogeneity. \Cref{dependence} examines the issue of effect size dependence in meta-analysis, providing a comprehensive analysis of its implications and presenting strategies to effectively manage this challenge. Finally, \Cref{conclusion} presents concluding remarks.

\section{Meta-Analysis} \label {meta-analysis}
Meta-analysis is a statistical method used to evaluate and combine findings from multiple studies addressing a common research question, with the goal of drawing overall conclusions. The standard meta-analysis process consists of four main steps: (i) selecting an appropriate measure of effect size, (ii) calculating the effect size for each study included in the meta-analysis, (iii) computing the mean effect size as a summary statistic by combining the individual effect sizes, and (iv) performing statistical inference about the true (population) mean effect, including confidence intervals and hypothesis testing, to draw overall conclusions.

The first step in conducting a meta-analysis is to standardize the results from different studies into a common metric, referred to as the ``effect size''. This process enables the comparison and aggregation of findings from studies that utilize varying measures of the same construct. For instance, the partial correlation coefficient is commonly used as an effect size measure. It quantifies the strength and direction of the relationship between two quantitative variables while controlling for other variables, and its values range from -1 to +1.\footnote{The partial correlation coefficient is widely employed in empirical economic research as a measure of effect size. It is calculated using the following formula:
	\begin{equation*}
		r_i= \frac{t_{i}}{\sqrt{t_{i}^{2} + df_{i}}}
	\end{equation*}
	with standard errors given by:
	\begin{equation*}
		S_i= \sqrt{\frac{1-r_i^2}{df_{i}}},
	\end{equation*}
	where \(t_{i}\) represents the $t$-statistic for the $i$th estimate ($i=1,2,...,k$) reported in the primary studies, and \(df_{i}\) denotes the corresponding degrees of freedom \citep{greene2012econometric}. For studies that report $z$-statistics (instead of $t$-statistics) for regression coefficients, the partial correlation coefficient can be derived as \(r = \sqrt{z^2/n} = z/\sqrt{n}\), where \(n\) is the sample size \citep{rosenthal1991meta}.
	Unfortunately, the sampling distribution of the correlation coefficient \( r \) becomes increasingly skewed as the absolute value of the population effect size (\( \rho \)) increases \citep{hedges1985statistical, rosenthal1991meta}. To render the variance of \( r \) independent of its population counterpart and to normalize its distribution, it is advisable to apply Fisher's \( z \)-transformation, as specified by:
	
	\[
	z_i = \frac{1}{2} \ln \left( \frac{1 + r_i}{1 - r_i} \right),
	\]
	where \( \ln \) denotes the natural logarithm. The corresponding transformation for the population correlation coefficient \( \rho_i \) is:
	
	\[
	\zeta_i = \frac{1}{2} \ln \left( \frac{1 + \rho_i}{1 - \rho_i} \right).
	\]
	
	In this context, \( z_i \) represents the Fisher-transformed correlation coefficient, which is approximately normally distributed with a mean of \( \zeta_i \) and a variance of \( \frac{1}{n_i - 3 - p_i} \), irrespective of the value of \( \zeta_i \). Here, \( n_i \) refers to the sample size associated with the \( i \)-th effect estimate, and \( p_i \) is the number of variables conditioned on (see \citealp{hedges1985statistical}). For ease of interpretation, the transformation can be reversed, with \( z \) converted back to \( r \) using the formula:
	
	\[
	r = \frac{e^{2z} - 1}{e^{2z} + 1},
	\]
	where \( e^x = \exp(x) \) denotes the exponential function.
	
	However, both \cite{hunter1996meta} and \cite{hunter2004methods} caution against the indiscriminate use of Fisher's \( z \)-transformation. They argue that while the transformation mitigates a negative bias in the untransformed correlation coefficients, it introduces a positive bias in the transformed values. Notably, this bias is generally larger (in absolute magnitude) than the original bias in the untransformed coefficients, particularly when population correlation coefficients exhibit substantial variation across studies.}
 Other examples include the (log) odds ratio, (log) risk ratio, raw mean difference, and standardized mean difference (see \citealp{hox2010multilevel}).
 
After obtaining estimates of the individual effect sizes, the subsequent step is to calculate an estimate of the overall (or combined) effect size. All meta-analytic procedures derive the overall effect size by computing a weighted average of the individual effect size estimates, with the primary distinction between methods lying in the specific weighting scheme employed. Meta-analysis typically involves two primary approaches: fixed-effect models and random-effects model \citep{hedges1998fixed, hunter2000fixed, schmidt2009fixed}. These approaches are outlined in the following discussion.

Consider a meta-analysis involving \( k \) independent studies, where study \( i \) (\( i = 1, 2, \dots, k \)) provides an estimate \( y_i \) of the corresponding population effect size \( \beta_i \), and this estimate has an associated within-study sampling variance \( \sigma_i^2 \). The model for this situation can be expressed as:

\begin{equation}\label{eq1}
y_i = \beta_i + \varepsilon_i, \quad i = 1, 2, \dots, k,
\end{equation}
where the sampling errors \( \varepsilon_i \) are assumed to be independently distributed as \( N(0, \sigma_i^2) \), with \( \sigma_i^2 \) typically assumed to be fixed and known. The fixed-effect and random-effects models differ in their treatment of \( \beta_i \) in this general model. The fixed-effect meta-analysis model is derived under the assumption that \( \beta_i = \mu_F \), a common (or fixed) effect size. Specifically,

\begin{equation}\label{eq2}
y_i = \mu_F + \varepsilon_i, \quad \varepsilon_i \sim N(0, \sigma_i^2),
\end{equation}
from which it follows that the \( y_i \)'s are independently distributed as \( N(\mu_F, \sigma_i^2) \). In the context of the fixed-effect model, all studies included in the meta-analysis are assumed to share a common true effect size \( \mu_F \). That is, all studies estimate the same true effect size, and variation in the observed effect sizes arises solely due to random sampling error (within-study variance \( \sigma_i^2 \)).

In this framework, the goal of the meta-analysis is to estimate the common effect size \( \mu_F \), accompanied by a confidence interval. A natural estimate of \( \mu_F \) is the pooled estimate, which can be obtained by computing a weighted average of the individual effect estimates:

\begin{equation}\label{eq3}
\hat{\mu}_F = \frac{\sum_{i=1}^k \hat{w}_i y_i}{\sum_{i=1}^k \hat{w}_i}, \quad \text{with} \quad \widehat{Var}(\hat{\mu}_F) = \frac{1}{\sum_{i=1}^k \hat{w}_i},
\end{equation}
where the ``true'' weights are the inverse of the within-study sampling variances, \( w_i = 1/\sigma_i^2 \), and \( \widehat{Var}(\hat{\mu}_F) \) is the estimated variance of \( \hat{\mu}_F \). Thus, the weight assigned to each effect size estimate is inversely proportional to its variance \( \sigma_i^2 \). This weighting scheme ensures that effect size estimates with smaller variances (and thus higher precision) receive larger weights in the calculation of the overall estimate, thereby having a greater influence on the resulting combined effect size.

If the within-study sample sizes are sufficiently large (e.g., \( n_i \geq 30 \) for each study, as suggested by \citealp{raudenbush2002hierarchical}), it is reasonable to assume that the effect size estimates are approximately normally distributed with within-study sampling variances, \( \sigma_i^2 \), which are estimated in practice by the sample variances \( S_i^2 \) from primary studies, though they are conventionally treated as fixed and known in the analysis. Thus, \( \hat{w}_i = 1/S_i^2 \) is used in \cref{eq3} as an estimate for \( w_i = 1/\sigma_i^2 \).
The estimator \( \hat{\mu}_F \) is equivalent to the weighted least squares (WLS) estimator with weights \( \hat{w}_i = 1/S_i^2 \) (which corresponds to applying ordinary least squares (OLS) to the transformed model \( y_i / S_i = \mu (1 / S_i) + \varepsilon_i / S_i \)).
Assuming that the effect size estimates (or \( \varepsilon_i \)'s) are normally distributed, a Wald-type \( 100(1-\alpha)\% \) confidence interval for the common effect size \( \mu_F \) can be constructed as \( \hat{\mu}_F \pm z_{\alpha/2} \sqrt{\widehat{Var}(\hat{\mu}_F)} \), where \( z_{\alpha/2} \) is the upper \( \alpha/2 \) quantile of the standard normal distribution \citep{van2002advanced, borenstein2021introduction}.

In contrast, under the random effects meta-analysis model, the assumption of a common effect size across all studies is relaxed. Instead, the effect size parameters (\(\beta_1, \beta_2, \dots, \beta_k\)) are assumed to be a random sample from a population following a normal distribution \(N(\mu_R, \tau^2)\). Specifically, the \(\beta_i\)'s are independent and identically distributed random variables, with a common mean \(\mu_R\) and variance \(\tau^2\), where \(\mu_R\) represents the mean of the true effect sizes, and \(\tau^2\) is the between-study variance, indicating the variability in true effects across studies. Thus, the true effects still share a common mean.

Assuming that $\beta_i = \mu_R + u_i$, the model can be reformulated as:  
\begin{equation}\label{eq4}  
	y_i = \mu_R + u_i + \varepsilon_i, \qquad u_i \sim N(0, \tau^2) \;\; \text{and} \;\; \varepsilon_i \sim N(0, \sigma_i^2),  
\end{equation}  
where the error terms $u_i$ and $\varepsilon_i$ are assumed to be uncorrelated. This formulation implies that the $y_i$'s follow independent distributions as $N(\mu_R, \sigma_i^2 + \tau^2)$. Consequently, the variation in the observed effect sizes is attributed not only to random sampling error (within-study variance, $\sigma_i^2$) but also to true variation in effect sizes (between-study variance, $\tau^2$). The objective of random-effects meta-analysis is, therefore, not to estimate a single common effect across all studies. Instead, it aims to estimate both the mean of the distribution of true effects ($E(\beta_i) = \mu_R$) and the degree of heterogeneity among these effects ($\tau^2$).  

The random-effects estimator of $\mu_R$, like the fixed-effect estimator, is a weighted average of the observed effect sizes (or WLS), expressed as:  
\begin{equation}\label{eq5}  
	\hat{\mu}_R = \frac{\sum_{i=1}^k \hat{w}_i^\ast y_i}{\sum_{i=1}^k \hat{w}_i^\ast}, \qquad \qquad \text{with} \qquad \widehat{Var}(\hat{\mu}_R) = \frac{1}{\sum_{i=1}^k \hat{w}_i^\ast},  
\end{equation}  
where the weights are the inverse of the variance of the observed effect sizes under the random-effects assumption. Specifically,  
$w_i^\ast = 1/\sigma_i^2 + \tau^2$.

The random-effects meta-analysis model incorporates between-study variance (also referred to as excess variance), $\tau^2$, into the weights. This approach reduces to the fixed-effect meta-analysis model when there is no heterogeneity among the true effects (i.e., $\tau^2 = 0$), such that any differences in the observed effect sizes are solely attributed to within-study sampling error. 
As in the fixed-effect model, the within-study variances $\sigma_i^2$ are estimated by $S_i^2$. Although these estimates are used in practice, $\sigma_i^2$ are typically treated as fixed and known for analytical purposes.  

The estimator $\hat{\mu}_R$ is equivalent to the WLS estimator, with the weights given by $\hat{w}_i^\ast = 1 / (S_i^2 + \hat{\tau}^2)$.  

Several procedures are available for estimating the between-study variance parameter, $\tau^2$, which serves as a measure of heterogeneity in the true effects (see \citealp{sidik2007comparison, sanchez2008confidence, veroniki2016methods, langan2017comparative, petropoulou2017comparison}). Among these, the most widely used is the method of moments (MM) estimator proposed by \cite{dersimonian1986meta}. This approach is both computationally simple and non-iterative, and it is calculated as:  
\begin{equation}\label{eq6}  
	\hat\tau^2 = \max \left\{0, \frac{Q - (k-1)}{\sum_{i=1}^k \hat w_i - \frac{\left(\sum_{i=1}^k \hat w_i\right)^2}{\sum_{i=1}^k \hat w_i}}\right\},  
\end{equation}  
where $\hat w_i = 1/S_i^2$ represents the weight for the $i$th effect estimate under a fixed-effect meta-analysis model.  
The term \( Q \), Cochran's heterogeneity statistic, is defined as:  
\begin{equation}\label{eq7}  
	Q = \sum_{i=1}^k \hat w_i (y_i - \hat{\mu}_F)^2 = \sum_{i=1}^k \left(\frac{y_i - \hat{\mu}_F}{S_i}\right)^2,  
\end{equation}  
and approximately follows a $\chi^2$ distribution with $k-1$ degrees of freedom under the null hypothesis that the true effects are homogeneous ($\tau^2 = 0$) (see \citealp{biggerstaff1997incorporating}).  
The resulting estimate $\hat\tau^2$ is incorporated into the weighting factor $\hat w_i^\ast = 1/S_i^2 + \hat\tau^2$, which serves as an estimate for the true weight, $w_i^\ast = 1/(\sigma_i^2 + \tau^2)$, in \cref{eq5}.  
	
In the context of the standard DerSimonian-Laird random-effects model, a \(100(1-\alpha)\%\) Wald-type confidence interval for the mean effect size (\(\mu_R\)) is constructed as 
$\hat{\mu}_R \pm z_{\alpha/2} \sqrt{\widehat{Var}(\hat{\mu}_R)}$,
where \(z_{\alpha/2}\) represents the upper \(\alpha/2\) quantile of the standard normal distribution. However, it has been observed that this confidence interval, based on the normal approximation, is often too narrow (with empirical coverage probabilities lower than the nominal value, particularly when the number of studies is small). This issue arises because the method does not account for the sampling variability (uncertainty) in estimating the heterogeneity parameter (\(\hat{\tau}^2\)). Furthermore, the approach assumes that within-study variances (\(\sigma_i^2\)) are fixed and known, an assumption that becomes untenable with small or moderate sample sizes.
	
To address these limitations, several alternatives to the standard DerSimonian-Laird method have been proposed to incorporate the uncertainty in estimating the between-study variance (\(\tau^2\)) (see \citealp{brockwell2001comparison, brockwell2007simple, sidik2002simple, sidik2006robust, ioannidis2007uncertainty, sanchez2008confidence, cornell2014random, rover2015hartung, bodnar2017bayesian, guolo2017random}).
	
One common approach is to construct confidence intervals for the true mean effect using the \(t\)-distribution rather than the normal distribution, as 
$\hat{\mu}_R \pm t_{df, \alpha/2} \sqrt{\widehat{Var}(\hat{\mu}_R)}$,
where \(t_{df, \alpha/2}\) represents the upper \(\alpha/2\) quantile of the \(t\)-distribution with \(df\) degrees of freedom. Determining an effective number of degrees of freedom (\(df\)) for the distribution of the test statistic 
$T =\left( \hat{\mu}_R - \mu_R \right)  /   \sqrt{\widehat{Var}(\hat{\mu}_R)}$
is challenging, as it depends on the number of studies included in the meta-analysis (\(k\)), the extent of between-study heterogeneity, and the magnitude of within-study variances \citep{higgins2009re}. Consequently, researchers have suggested several degrees of freedom for this test statistic, such as \(k-1\) \citep{follmann1999valid}, \(k-2\) \citep{raghunathan1993analysis, higgins2009re}, and \(k-4\) \citep{berkey1995random}.
	
Alternatively, \cite{hartung2001tests, hartung2001refined} introduced a modification to the DerSimonian-Laird method to construct an improved confidence interval for the true mean effect by (i) deriving a refined variance estimator adjusted for small-sample bias and (ii) replacing the normal approximation with a \(t\)-distribution with \(k-1\) degrees of freedom. The test statistic in this approach is given by  
\begin{equation}\label{eq8}  
	t_{k-1} = \frac{\hat{\mu}_R - \mu_R}{\sqrt{\widehat{Var}_{S}(\hat{\mu}_R)}},
\end{equation}
	where \(\widehat{Var}_{S}(\hat{\mu}_R)\) is the refined variance estimator defined as  
\begin{equation}\label{eq9}  
	\widehat{Var}_{S}(\hat{\mu}_R) = q \cdot \widehat{Var}(\hat{\mu}_R), \quad \text{with} \quad q = \left\{\frac{1}{k-1} \sum_{i=1}^k \hat{w}_i^\ast (y_i - \hat{\mu}_R)^2 \right\}.
\end{equation}

A \(100(1-\alpha)\%\) confidence interval for the true mean effect \(\mu_R\) would then be calculated as 
$\hat{\mu}_R \pm t_{k-1,\alpha/2} \sqrt{\widehat{Var}_{S}(\hat{\mu}_R)}$,
where \(t_{k-1,\alpha/2}\) is the upper \(\alpha/2\) quantile of the \(t\)-distribution with \(k-1\) degrees of freedom. This procedure was independently proposed by \cite{sidik2002simple}. 
The so-called Hartung-Knapp-Sidik-Jonkman method typically produces confidence intervals that are wider than those based on the standard normal distribution. However, if \(q\) is sufficiently small, the modified confidence interval may paradoxically become shorter than the standard one. To address this limitation, \cite{rover2015hartung} proposed a simple modification: replacing \(q\) in \cref{eq9} with \(q^\ast = \max\{1, q\}\). It is worth noting that the estimator of \(\tau^2\) used in this method is structurally identical to that employed in the DerSimonian-Laird approach. Nevertheless, the proposed confidence intervals are relatively robust to variations in the procedure used to estimate \(\tau^2\) \citep{sidik2003constructing, sidik2006robust}.

Additionally, \cite{sidik2006robust} proposed an alternative, approximately unbiased estimator for the variance of the overall effect estimate (\(\hat{\mu}_R\)) in random effects meta-analysis. This estimator is robust to errors in the estimated marginal variances (\(S_i^2\) and \(\hat{\tau}^2\)) that underlie the random effects weights. The authors advocate for using the robust variance estimator or the weighted sample variance estimator instead of the conventional variance estimator when making inferences about the overall effect size in the random effects meta-analysis framework.

Another approach is the method proposed by \cite{kenward1997small}, which adjusts the variance of the overall effect estimate using the expected information and modifies the degrees of freedom of the \(t\)-distribution accordingly. This method provides an alternative for constructing confidence intervals and testing hypotheses for the overall effect size (see \citealp{partlett2017random}).

Several alternative estimators for \(\tau^2\) have also been developed to incorporate the uncertainty associated with its estimation in the standard DerSimonian-Laird approach (see \citealp{veroniki2016methods}). For instance, the likelihood ratio (LR) method described by \cite{hardy1996likelihood} employs a profile likelihood function to simultaneously estimate \(\tau^2\) and \(\mu_R\), as well as to construct likelihood-based confidence intervals. 

\cite{partlett2017random} conducted a simulation study to compare the coverage probabilities of confidence intervals generated by these methods, specifically under restricted maximum likelihood (REML) estimation of \(\tau^2\) instead of the standard DerSimonian-Laird estimation. Their findings suggest that the Hartung-Knapp-Sidik-Jonkman method outperforms other approaches when between-study heterogeneity is large and/or within-study sample sizes are similar. However, ven this method is problematic  when heterogeneity is low, and within-study sample sizes vary significantly.

The primary objective of the meta-analysis is to estimate the mean effect size and its confidence interval. The confidence interval provides an indication of the precision of the estimate of the mean effect size (based on the standard error of the mean), as it shows the range within which the true mean effect is likely to lie \citep{lipsey2001practical}. Ideally, this interval should be as narrow as possible (the narrower the interval, the more precise the estimate). However, in a random effects meta-analysis, such intervals do not convey information about the dispersion of true effect sizes, which is also of significant interest. The prediction interval is used to address this, as it indicates the range within which the true effect size in a new (or future) study, similar to those included in the meta-analysis, is likely to lie.
An approximate $100(1-\alpha)\%$ prediction interval for the new effect can be formed using
$\hat{\mu}_R \pm t_{k-2,\alpha/2} \sqrt{\widehat{Var}(\hat{\mu}_R) + \hat{\tau}^2}$,
where $t_{k-2,\alpha/2}$ is the upper $\alpha/2$ quantile of the $t$-distribution with $k-2$ degrees of freedom \citep{higgins2009re}. \cite{partlett2017random} conducted a simulation study to compare the coverage of several versions of this so-called Higgins-Thompson-Spiegelhalter prediction interval, constructed using different estimators for the variance of the mean effect estimate (Hartung-Knapp-Sidik-Jonkman, Sidik-Jonkman, and Kenward-Roger variance estimators), following REML estimation of $\tau^2$. They conclude that all procedures perform reasonably well when $k \geq 5$ and the degree of heterogeneity is sufficiently large.
It is evident that the uncertainty in the estimate of $\tau^2$ affects both terms in the estimated variance of the new effect drawn from the true random effects distribution, $\left[\widehat{Var}(\hat{\mu}_R) + \hat{\tau}^2\right]$. \cite{nagashima2019prediction} propose a parametric bootstrap approach to construct prediction intervals that account for this uncertainty and are valid even when the number of studies included in the meta-analysis ($k$) is small. Using Monte Carlo simulations, they compare the coverage probabilities of the proposed prediction interval with the Higgins-Thompson-Spiegelhalter prediction intervals based on the standard, Hartung-Knapp-Sidik-Jonkman, and Sidik-Jonkman variance estimators for $\hat{\mu}_R$, following REML estimation of $\tau^2$. Their results indicate that the coverage performance of the Higgins-Thompson-Spiegelhalter prediction intervals may be poor, particularly when the relative degree of heterogeneity is low or moderate. In contrast, the coverage probability of the proposed prediction interval remains nearly constant and close to the nominal level, regardless of the amount of heterogeneity.

Assessment of statistical heterogeneity is another important aspect of meta-analysis, which may be conducted by testing for the presence of heterogeneity, measuring the extent of heterogeneity, quantifying its impact, or using a combination of these methods \citep{higgins2008commentary}. As discussed by \cite{thompson1994sources}, the impact of heterogeneity on the overall conclusions of the meta-analysis needs to be considered carefully.
Tests for heterogeneity are commonly used to decide which statistical model is appropriate for a particular meta-analysis. If there is evidence of heterogeneity in the true effect sizes, then a random effects model should be used, whereas a fixed effect model is more appropriate when there is no significant heterogeneity across studies.\footnote{\cite{hedges1998fixed} point out that the fixed and random effects procedures are designed for different inference goals, implying that the choice of a statistical procedure should be based on the nature of the inference the analyst wishes to make about the effect size parameters. They distinguish between conditional and unconditional inferences. In the former case, the analyst wishes to make statistical inferences about the effect size parameters in the set of studies being analyzed (or a set of studies identical to those included in the analysis except for uncertainty arising from the sampling of participants into the studies) and say nothing about other studies. In this case, the fixed effects model is chosen to treat the effect size parameters as fixed unknown constants (fixed effects) to be estimated. In contrast, the random effects model is appropriate when the analyst wishes to make an explicit generalization beyond the collection of studies being analyzed (referred to as unconditional inference). The random effects procedure treats each observed study's effect size as a random draw from an underlying population or distribution of effect size parameters (or a superpopulation), and makes inferences about the mean and variance of this distribution.}
A simple way to assess heterogeneity in the true effect sizes is through the ``forest plot.'' The forest plot is a graphical tool used to represent the point estimates of the effect sizes (both individual and overall) and the corresponding confidence intervals in a simple visual display. Less overlap between confidence intervals is associated with greater heterogeneity between studies \citep{lewis2001forest,panesar2010systematic,hanji2017meta}.
In addition, Cochran's $Q$ statistic ($\sim \chi_{k-1}^2$), as described in \cref{eq7}, is commonly used in meta-analyses to test the null hypothesis of homogeneity — $\text{H}_0: \tau^2 = 0$ or $\beta_i = \mu_F$ for all $i$ — against the alternative of heterogeneity. If $Q$ is greater than the upper $\alpha$ quantile of the $\chi_{k-1}^2$ distribution, then the null hypothesis would be rejected at the significance level $\alpha$. If $\text{H}_0$ is rejected, the meta-analyst may conclude that the true effects are heterogeneous and a random effects model is more appropriate for the given meta-analysis.
Furthermore, the confidence interval provides a useful tool for testing a hypothesis about $\tau^2$. If the confidence interval for $\tau^2$ does not contain 0, the null hypothesis that $\tau^2 = 0$ is rejected. The confidence interval provides a measure of the degree of uncertainty associated with the point estimate of the amount of heterogeneity. In other words, it is a measure of precision (based on the standard error), which gives an indication of how precisely the heterogeneity parameter, $\tau^2$, is estimated. There are a number of different approaches to constructing confidence intervals for $\tau^2$. These include Wald-type approaches, likelihood-based approaches, approaches using forms of generalized Cochran's $Q$ statistics, bootstrap approaches, Bayesian approaches, and others (see \citealp{knapp2006assessing,viechtbauer2007confidence,jackson2013confidence,jackson2015approximate,jackson2016confidence,veroniki2016methods}).

Having tested for and found evidence of heterogeneity across studies, the logical next step in the meta-analysis is to quantify this heterogeneity. It is important to quantify between-study heterogeneity because the degree of heterogeneity determines the extent to which the findings of the meta-analysis can be generalized, thus influencing the overall conclusions of the meta-analysis \citep{higgins2008commentary}. 
The $p$-value of the $Q$-test is often quoted as an indication of the degree of heterogeneity. A low $p$-value provides evidence of significant heterogeneity in the effect estimates (beyond that expected due to sampling error alone). However, it has been shown that the statistical power of the $Q$-test for detecting heterogeneity depends on the number and size of studies included in the meta-analysis, as well as the between-study variance parameter, $\tau^2$. In particular, this test has poor power when the number of studies or within-study sample sizes is small, and excessive power when there are too many studies, especially with large sample sizes \citep{sackett1986seeking,spector1987meta,alexander1989statistical,sanchez1997homogeneity,hardy1998detecting,viechtbauer2007hypothesis}. 
Another approach for quantifying heterogeneity is estimating the between-study variance, $\tau^2$, as done in a random effects meta-analysis. The problem with this approach is that it depends mainly on the particular effect size metric used in the meta-analysis. Therefore, it is not possible to make comparisons of the degree of heterogeneity across meta-analyses that have used different measures of effect size \citep{higgins2002quantifying,huedo2006assessing}.

\cite{higgins2002quantifying} propose a simple measure to quantify between-study heterogeneity in a meta-analysis as follows:\footnote{In addition to $I^2$, \cite{higgins2002quantifying} develop two further measures of the impact of heterogeneity: $H = \sqrt{Q/(k-1)}$ and $R = \sqrt{\left\{\frac{\sum_{i=1}^k \hat w_i}{\sum_{i=1}^k \hat w_i^\ast} \right\}} = \sqrt{\left\{\frac{\sum_{i=1}^k \hat w_i}{\sum_{i=1}^k (\hat w_i^{-1} + \hat\tau^2)^{-1}} \right\}}$. They conclude that $H$ and $I^2$ are particularly useful measures of the impact of heterogeneity. However, the most popular measure is $I^2$, which is a transformation of $H$, defined as $I^2 = \frac{H^2 - 1}{H^2}$.}
\begin{equation}\label{eq10}
	I^2 = \frac{\hat\tau^2}{S^2 + \hat\tau^2},
\end{equation}
(as an estimate of $\tau^2/(\sigma^2 + \tau^2)$), where $\hat\tau^2$ is an estimate of the between-study variance, $\tau^2$, which can be obtained by the method of moments as shown in \cref{eq6}, and $S^2$ provides an estimate of the so-called ``typical'' within-study sampling variance, $\sigma^2$, which can be derived from the individual within-study variances as follows:
\begin{equation}\label{eq11}
	S^2 = \frac{(k-1) \sum_{i=1}^k \hat w_i}{\left(\sum_{i=1}^k \hat w_i\right)^2 - \sum_{i=1}^k \hat w_i^2},
\end{equation}
where $\hat w_i = 1/S_i^2$ and $k$ is the number of effect estimates. By substituting \cref{eq6} and \cref{eq11} into \cref{eq10}, the $I^2$ statistic can be rewritten as
\begin{equation}\label{eq12}
	I^2 = \max\left\{0, \frac{Q - (k - 1)}{Q}\right\}.
\end{equation}

As pointed out by \cite{higgins2008commentary}, $I^2$ is not a point estimate of ``between-study variance'' or a measure of the degree of ``between-study heterogeneity''. It represents the proportion of the total variation in observed effect sizes that is due to heterogeneity in the true effect sizes rather than sampling error. That is, the ratio of the between-study variance to the variance of the observed effects (i.e., the sum of the between-study variance and the within-study sampling variance). \cite{borenstein2017basics} provide many of the practical details. 

$I^2$ can also be viewed as a measure of ``inconsistency'' in the studies' results in a meta-analysis, since it reflects the degree of overlap of confidence intervals across effect size estimates \citep{higgins2003measuring,higgins2008commentary,borenstein2021introduction}. Thus, this measure quantifies the impact rather than the extent of between-study heterogeneity in a meta-analysis, ranging from 0\% to 100\%, with higher values indicating greater impact of true heterogeneity on the conclusions of the meta-analysis.
A useful guide for the interpretation of $I^2$ is as follows: 0-40\% may not be important, 30-60\% may represent moderate heterogeneity, 50-90\% may represent substantial heterogeneity, and 75-100\% represents considerable heterogeneity \citep{deeks2008analysing}.
In addition, \cite{higgins2003measuring} suggest that $I^2$ values on the order of 25\%, 50\%, and 75\% may be regarded as low, moderate, and high degree of heterogeneity, respectively.
A confidence interval for $\tau^2/(\sigma^2+\tau^2)$ can be obtained by substituting the endpoints of a confidence interval for $\tau^2$ into \cref{eq10}, whereby statistical homogeneity is rejected if the confidence interval does not contain zero. For practical application, \cite{higgins2002quantifying} also suggest a simple computation of an interval only based on $Q$ and $k$.

In general, the main advantages of $I^2$ are that it (i) is easy to calculate and has an intuitive interpretation, (ii) can be readily calculated for the majority of published meta-analyses, (iii) is characterized by an uncertainty interval, (iv) is inherently independent of the number of studies included in the meta-analysis, (v) can be directly compared across meta-analyses with different types of outcome data and different measures of effect size, and (vi) has a wide range of applications, such as the investigation of the nature and causes of heterogeneity \citep{higgins2002quantifying,higgins2003measuring}.

\section{Meta-Regression Analysis} \label {meta-regression}
In general, as discussed above, the total variation in the observed effect size can be attributed to within-study variance (sampling error variance) and between-study variance (heterogeneity in the true effects). An important aspect of a meta-analysis is identifying potential sources of heterogeneity between studies. Meta-regression has become a widely used tool for addressing this issue. In particular, the fixed and random effects meta-regression models have been developed to investigate whether identifiable study characteristics (moderators) may explain at least part of the heterogeneity among the true effects (see \citealp{thompson1999explaining,higgins2004controlling,harbord2008meta,viechtbauer2015comparison}).

The choice between these two models depends on whether or not all the heterogeneity beyond chance (sampling error) can be systematically accounted for by the moderator variables. 

The fixed effects meta-regression model extends the fixed effect meta-analysis model by replacing \( \mu_F \) with a linear predictor, \( \boldsymbol{x}_i' \boldsymbol{\beta}_F \), in \cref{eq2}:
\begin{equation}\label{eq13}
	y_i = \boldsymbol{x}_i' \boldsymbol{\beta}_F + \varepsilon_i, \qquad \varepsilon_i \sim N(0, \sigma_i^2),
\end{equation}
where \( \boldsymbol{x}_i \) is a \( (p+1) \times 1 \) vector of \( p \) explanatory variables (covariates) plus a \( 1 \) in the first position corresponding to the model intercept, and \( \boldsymbol{\beta}_F \) is a \( (p+1) \times 1 \) vector of unknown constants to be estimated. This formulation implies that the \( y_i \)'s are distributed as \( N(\boldsymbol{x}_i' \boldsymbol{\beta}_F, \sigma_i^2) \). To correct for heteroscedasticity, the parameters \( \boldsymbol{\beta}_F = (\beta_0, \beta_1, ..., \beta_p)' \) can be estimated easily by weighted least squares (WLS) as follows:
\begin{equation}\label{eq14}
	\boldsymbol{\hat{\beta}}_F = \left[\sum_{i=1}^k \hat{w}_i \boldsymbol{x}_i \boldsymbol{x}_i' \right]^{-1} \left[\sum_{i=1}^k \hat{w}_i \boldsymbol{x}_i y_i\right] = (\boldsymbol{X}' \boldsymbol{\hat{W}} \boldsymbol{X})^{-1} (\boldsymbol{X}' \boldsymbol{\hat{W}} \boldsymbol{y}),
\end{equation}
where \( \boldsymbol{X} \) is a \( k \times (p+1) \) matrix of observations with the \( i \)-th row equal to \( \boldsymbol{x}_i' = (1, x_{1i}, ..., x_{pi}) \), \( \boldsymbol{y} = (y_1, ..., y_k)' \) is a \( k \times 1 \) vector of the observed effect sizes, and \( \boldsymbol{\hat{W}} \) is a \( k \times k \) diagonal weighting matrix whose \( i \)-th diagonal element is \( \hat{w}_i = \frac{1}{S_i^2} \). 

In general, the variance-covariance matrix of this WLS estimator is given by
\begin{equation}\label{eq15}
	\widehat{Var}(\boldsymbol{\hat{\beta}}_F) = \left(\boldsymbol{X}' \hat{\boldsymbol{W}} \boldsymbol{X}\right)^{-1} \left(\boldsymbol{X}' \hat{\boldsymbol{W}} \boldsymbol{\Sigma} \hat{\boldsymbol{W}} \boldsymbol{X}\right) \left(\boldsymbol{X}' \hat{\boldsymbol{W}} \boldsymbol{X}\right)^{-1},
\end{equation}
where \( \boldsymbol{\Sigma} \) is the variance-covariance matrix of the effect size estimates. The diagonal elements of \( \boldsymbol{\Sigma} \) are the variances associated with the effect sizes, and the off-diagonal elements are the covariances. Assuming that the effect size estimates are statistically independent (i.e., the off-diagonal elements are all equal to zero), and that inverse variance weights are used, then
\begin{equation}\label{eq16}
	\widehat{Var}(\boldsymbol{\hat{\beta}}_F) = (\boldsymbol{X}' \boldsymbol{\hat{W}} \boldsymbol{X})^{-1},
\end{equation}
where the \( j \)-th diagonal element represents the variance of the estimator \( \hat{\beta}_{j-1} \).

The fixed effects meta-regression model assumes that between-study heterogeneity can be fully explained by the covariates \( X_1, ..., X_p \). This model is generally not recommended because it leads to unacceptably inflated Type I error rates when there is unexplained, or residual, heterogeneity among the true effects, which is often the case in practical applications \citep{higgins2004controlling}.

The random effects meta-regression model is obtained by replacing \( \mu_R \) with \( \boldsymbol{x}_i' \boldsymbol{\beta}_R \) in \cref{eq4}:
\begin{equation}\label{eq17}
	y_i = \boldsymbol{x}_i' \boldsymbol{\beta}_R + u_i + \varepsilon_i, \qquad u_i \sim N(0, \tau_{res}^2) \quad \text{and} \quad \varepsilon_i \sim N(0, \sigma_i^2),
\end{equation}
which implies that the \( y_i \)'s are distributed as \( N(\boldsymbol{x}_i' \boldsymbol{\beta}_R, \sigma_i^2 + \tau_{res}^2) \). This model is referred to as the mixed effects meta-analysis model (the meta-regression coefficients \( \beta_0, \beta_1, ..., \beta_p \) are fixed effects, and the \( u_i \)'s are mutually independent random effects). It assumes that the variation in the observed effect size beyond that expected by sampling error (i.e., the variation in the true effect size) has both systematic and random components. The systematic variation is explained by the moderator variables included in the meta-regression model, while the random variation, known as the residual between-study variance, remains unexplained. In other words, the mixed effects meta-analysis model assumes that between-study heterogeneity (\( \tau^2 \)) is partitioned into two components: unexplained or residual heterogeneity (\( \tau_{res}^2 \)) and explained heterogeneity (\( \tau^2 - \tau_{res}^2 \)).\footnote{Note that the variance \( \tau_{res}^2 \) is the conditional variance of the true effect sizes given the covariates.} Therefore, all models discussed earlier are restricted versions of the mixed effects meta-analysis model.

For the mixed effects meta-analysis model, it is necessary to estimate both the regression coefficients and the amount of residual heterogeneity in the true effect sizes. Similar to the fixed effects meta-regression model, the estimates of $\boldsymbol{\beta}_R$ and the corresponding variance-covariance matrix can be obtained using the weighted least squares estimator. The only difference is that, for the random meta-regression model, the weights are $\hat w_i^\ast$ instead of $\hat w_i$ as in \cref{eq14,eq16}. Thus,

\begin{equation}\label{eq18}
	\boldsymbol{\hat\beta}_R = \left[\sum_{i=1}^k \hat w_i^\ast \boldsymbol{x}_i \boldsymbol{x}_i'\right]^{-1} \left[\sum_{i=1}^k \hat w_i^\ast \boldsymbol{x}_i y_i\right] = (\boldsymbol{X}' \boldsymbol{\hat{W}}^\ast \boldsymbol{X})^{-1} (\boldsymbol{X}' \boldsymbol{\hat{W}}^\ast \boldsymbol{y}),
\end{equation}

and

\begin{equation}\label{eq19}
	\widehat{Var}(\boldsymbol{\hat\beta}_R) = (\boldsymbol{X}' \boldsymbol{\hat{W}}^\ast \boldsymbol{X})^{-1},
\end{equation}

where the weighting matrix is now $\boldsymbol{\hat{W}}^\ast = \text{diag}[\hat w_1^\ast, \dots, \hat w_k^\ast]$ and $\hat w_i^\ast = 1/(S_i^2 + \hat\tau_{res}^2)$, with $\hat\tau_{res}^2$ being an estimate of the amount of between-study variance $\tau_{res}^2$, that is, the degree of heterogeneity in the true effects not accounted for by the moderators included in the model. Many algorithms have been developed for the estimation of this parameter \citep{thompson1999explaining, raudenbush2009analyzing, panityakul2013estimating, chung2013avoiding, lopez2014estimation, viechtbauer2015comparison, van2018multistep}. The most common ones include the method of moments (MM), maximum likelihood (ML), restricted maximum likelihood (REML), and empirical Bayes (EB) estimators.

In the context of a meta-regression, meta-analysts can use a generalized version of Cochran's $Q$-test, as described in \cref{eq7}, to choose between the fixed and random/mixed models. The test is based on the residual heterogeneity statistic

\begin{equation}\label{eq20}
	Q_{res} = \sum_{i=1}^k \hat w_i (y_i - \boldsymbol{x}_i \boldsymbol{\hat\beta}_F)^2 = \sum_{i=1}^k \left(\frac{y_i - \boldsymbol{x}_i \boldsymbol{\hat\beta}_F}{S_i}\right)^2,
\end{equation}

which is distributed as $\chi^2$ with $k - p - 1$ degrees of freedom under the null hypothesis of no residual/unexplained heterogeneity ($\tau_{res}^2 = 0$). \footnote{
	The method of moments (MM) estimator for $\tau^2$ in the context of a meta-regression is
	
	\begin{equation*}
		\hat\tau_{res}^2 = \frac{Q_{res} - (k - p - 1)}{tr(\boldsymbol{M})},
	\end{equation*}
	
	where $tr(\boldsymbol{M})$ is the trace of the matrix
	
	\begin{equation*}
		\boldsymbol{M} = \boldsymbol{\hat{W}} - \boldsymbol{\hat{W}} \boldsymbol{X} (\boldsymbol{X}' \boldsymbol{\hat{W}} \boldsymbol{X})^{-1} \boldsymbol{X}' \boldsymbol{\hat{W}},
	\end{equation*}
	(see \citealp{raudenbush2009analyzing, lopez2017assessing}).
}

If $Q_{res}$ is statistically significant, then some of the between-study heterogeneity remains even after accounting for study characteristics as moderators. In other words, even after accounting for moderators, the true effect sizes remain heterogeneous \citep{lipsey2001practical, borenstein2021introduction}. In this case, the mixed effects meta-analysis model would be preferred. However, as previously mentioned, it may be more useful to quantify the heterogeneity in the true effects rather than rely on an overall test to detect its presence \citep{viechtbauer2007confidence, harbord2008meta}. 

In addition to the amount of residual heterogeneity, $\hat\tau_{res}^2$, a measure quantifying the impact of residual heterogeneity on the results can be defined as follows:

\begin{equation}\label{eq21}
	I_{res}^2 = \frac{\hat\tau_{res}^2}{S^2 + \hat\tau_{res}^2} = \max \left\{0, \frac{Q_{res} - (k - p - 1)}{Q_{res}}\right\}.
\end{equation}

This statistic represents the proportion of variation in the observed effect sizes, beyond what is explained by the covariates, that is attributable to residual (unexplained) heterogeneity rather than sampling error. It ranges from 0\% to 100\%, with higher values indicating a greater influence of unexplained heterogeneity on the conclusions of the meta-regression analysis.

As in random effects meta-analysis, it is essential to construct confidence intervals for the between-study variance within the context of a meta-regression model (i.e., the residual heterogeneity parameter, $\tau_{res}^2$) as an alternative method of inference for the heterogeneity parameter. \cite{jackson2014methods} present two exact methods for constructing confidence intervals for $\tau_{res}^2$: confidence intervals based on a form of generalized Cochran $Q$ statistics and the $Q$ profile method. They also advocate for the use of informative prior distributions for this parameter in Bayesian meta-regression. All three methods extend the approaches proposed for random effects meta-analysis. Furthermore, \cite{jackson2015approximate} propose a straightforward method for constructing approximate confidence intervals for the between-study variance component (heterogeneity) that can be applied to moment-based estimators under both random effects meta-analysis and meta-regression models.

An additional index for assessing the impact of moderators, pseudo $R^2$, is defined as follows:
\begin{equation}\label{eq22}
	R^2= \frac{\hat{\tau}^2-\hat{\tau}_{res}^2}{\hat{\tau}^2}=1-\frac{\hat{\tau}_{res}^2}{\hat{\tau}^2},
\end{equation}
where $\hat{\tau}^2$ represents the estimate of total between-study heterogeneity derived from the random effects meta-analysis model (refer to \cref{eq4}), and $\hat{\tau}_{res}^2$ represents the estimate of residual (unexplained) heterogeneity derived from the mixed effects meta-analysis model (refer to \cref{eq17}). This index quantifies the proportion of the total between-study heterogeneity explained by the moderators included in the mixed effects meta-analysis model. It ranges from 0 to 1 (or 0\% to 100\%), with higher values indicating that the moderators explain a larger share of the heterogeneity in the true effects \citep{pastor2018multilevel,borenstein2021introduction}.

Several procedures have been proposed in the literature for testing the statistical significance of individual moderator effects and constructing corresponding confidence intervals. These include standard (Wald-type) tests based on the $t$ and $z$ distributions, the likelihood ratio ($LR$) test, permutation tests, resampling tests, the Knapp-Hartung method (an extension of the Hartung-Knapp-Sidik-Jonkman method), and the robust (Huber-White-type) method (see \citealp{knapp2003improved,higgins2004controlling,raudenbush2009analyzing,huizenga2011testing,viechtbauer2015comparison}).

Additionally, determining whether the regression model as a whole is statistically significant is an important question. This can be addressed by testing the joint hypothesis that all coefficients, excluding the constant term, are zero ($\beta_1 = \dots = \beta_p = 0$). A common procedure for this is the omnibus (Wald-type) test. The test statistic is computed as
$Q_m = \boldsymbol{\hat\beta}'_p \hat{V}^{-1} \boldsymbol{\hat\beta}_p$,
where $\boldsymbol{\hat\beta}_p = (\hat\beta_1, \dots, \hat\beta_p)'$ is a $p \times 1$ vector of estimated coefficients, and $\hat{V}$ is the estimated variance-covariance matrix of $\boldsymbol{\hat\beta}_p$. Under the null hypothesis that all $p$ slope coefficients in the meta-regression model are zero, the $Q_m$ statistic follows a $\chi^2$ distribution with $p$ degrees of freedom, which is equal to the difference between the degrees of freedom of $Q$ and $Q_{res}$,  expressed as $p = (k - 1) - (k - p - 1)$.
Rejecting the null hypothesis indicates that at least one of the individual regression coefficients is significantly different from zero, rendering the fitted model significant \citep{lipsey2001practical,borenstein2021introduction}.

An alternative approach to test the joint hypothesis that all coefficients for the moderator variables are zero is the $LR$ or chi-square difference test. The $LR$ test statistic is $LR = -2[\ln L_1 - \ln L_0]$,
where $\ln L_1$ and $\ln L_0$ are the log-likelihood values for the models with and without the imposed restrictions, respectively. The test statistic follows a $\chi^2$ distribution with $p$ degrees of freedom, corresponding to the number of restrictions imposed \citep{greene2012econometric}. However, it is important to note that the $LR$ test is invalid when the restricted maximum likelihood (REML) procedure is used for model estimation. This limitation arises because REML only permits comparisons of models with identical fixed parts/regression coefficients \citep{hox2010multilevel}.

The omnibus and $LR$ tests are analogous to the $F$ test, which examines whether all slope coefficients in a regression model are zero. These tests can also be applied to assess whether one or a subset of coefficients for the moderator variables in a multiple meta-regression model are zero.

\section{Publication Bias} \label {publication}
Publication selection bias occurs when the selection of studies for publication is influenced by the statistical significance and/or direction of the results, such that studies reporting statistically significant results and/or those consistent with theoretical expectations are more likely to be published than studies with insignificant or unexpected results \citep{begg1989publication,rothstein2005publication,doucouliagos2005publication2,scherer2007full,sutton2009publication,jennions2013publication,stanley2017finding,borenstein2021introduction,page2022investigating}. Unfortunately, publication bias is prevalent across most research areas \citep{stanley2013better}.
This phenomenon is commonly referred to as the ``file drawer'' problem \citep{rosenthal1979file}, with file drawers metaphorically representing the location of the missing/unpublished studies. Because the missing studies are a non-random subset of all relevant studies (i.e., they systematically differ from the published ones), the studies included in a meta-analysis form a biased, unrepresentative sample of the entire body of research (both published and unpublished). In other words, publication bias leads to the selective and biased inclusion of studies in meta-analyses. This issue presents a significant threat to the validity of meta-analytic reviews and other synthesis methods \citep{rothstein2008publication,sutton2009publication,borenstein2021introduction}. 
Moreover, many authors may refrain from submitting studies with statistically insignificant and/or inconsistent (i.e., wrong-signed) effects for publication, anticipating rejection by journals \citep{egger1998bias}. This issue, known as ``outcome reporting bias'', has garnered considerable attention in recent years \citep{hutton2000bias,hahn2002investigation,chan2004empirical,chan2004outcome,williamson2005identification,dwan2008systematic,smyth2011frequency,pigott2013outcome,copas2013model,jennions2013publication,stanley2017finding,andrews2019identification,bom2019kinked,ioannidis2020historical,page2022investigating,vaert2024correcting}. Selective reporting can take several forms, including the selective reporting of certain outcomes from a given study based on the nature and direction of the results, as well as the analytical manipulation of unfavorable results to make them statistically significant or theoretically interesting. This type of bias also contributes to publication bias.

In addition to publication bias, other factors may contribute to bias in the inclusion of individual studies in a meta-analysis. Among published studies, those with significant effects in the expected direction are more likely to be published in English and therefore more likely to be searched (English language bias; \citealp{egger1997language,egger1998bias,juni2002direction}), more likely to be published more than once (duplication bias; \citealp{easterbrook1991publication,tramer1997impact}), more likely to be cited by other authors and therefore easier to find (citation bias; \citealp{gotzsche1987reference,kjaergard2002citation,ravnskov1992frequency}), more likely to be published quickly (time-lag bias; \citealp{stern1997publication,ioannidis1998effect,jennions2002relationships}), and more likely to be published in journals indexed in major electronic databases, particularly in the case of developing countries (database bias; \citealp{egger1998bias,kjaergard2002citation}). 
In addition, published studies that are easily accessible to researchers, available free or at low cost, or limited to one's own discipline are more likely to be included in a meta-analysis, leading to availability bias, cost bias, and familiarity bias, respectively \citep{rothstein2005publicationch1,wade2006information}. 
There are two strategies to address these and other related biases: conducting comprehensive searches for relevant studies and the prospective registration of research \citep{song2002asymmetric}. These biases fall under the broader umbrella of publication selection bias \citep{borenstein2021introduction}.\footnote{The combination of publication bias and related selection biases is often referred to as ``dissemination bias'' \citep{song2002asymmetric}.}

Several graphical and statistical methods have been proposed for detecting and correcting for publication bias in meta-analysis. The funnel plot is the most commonly used approach for identifying the presence of publication bias and for providing a visual summary of a meta-dataset \citep{song2000publication,sutton2000empirical,sterne2001investigating,stanley2005beyond,sterne2005funnel,sutton2009publication}. It is a scatter plot of the effect estimates from individual studies on the horizontal axis against a measure of sample size on the vertical axis \citep{light1984summing,begg1988publication,light1994visual}. Standard error, variance, and their inverses are valid choices for the vertical axis in a funnel plot \citep{sterne2001funnel}, among which the inverse of the standard error (precision) is the most commonly used.
Accordingly, the precision of the effect estimate increases as the sample size grows larger (i.e., larger samples provide more precise estimates of effect sizes). In this context, studies with large samples and, therefore, greater precision (smaller standard errors) appear at the top of the plot and tend to cluster near the mean effect size under the fixed-effect meta-analysis model (represented as a vertical line). In contrast, smaller studies with lower precision appear at the bottom of the plot and are more widely distributed because they are subject to greater sampling variation in effect estimates (i.e., the dispersion of effect estimates is inversely correlated with their precision).
When there is no publication bias, the resulting plot should be approximately funnel-shaped, hence its name, with its major axis aligned with the mean effect size and the studies symmetrically distributed around this value. However, in the presence of publication bias, the studies are expected to exhibit symmetry at the top, a few missing data points in the middle, and more missing data near the bottom \citep{doucouliagos2005publication1,borenstein2021introduction}.
Under publication selection, researchers conducting smaller-sample studies may need to search more extensively across model specifications, proxies/indices, econometric procedures, and datasets to find estimates large enough to offset their larger standard errors, ensuring that the test statistic reaches a statistically significant level. By contrast, researchers with larger samples and smaller standard errors need not search as rigorously to achieve the desired significance level, as smaller estimates suffice. Consequently, in the presence of publication bias, the reported effect estimates are positively related to their standard errors \citep{stanley2004does,doucouliagos2009publication,doucouliagos2012estimates}.
Furthermore, if the expected sign of the effect is positive, small studies providing small (insignificant) and negative estimates with large standard errors may be missing, resulting in a plot skewed to the right. Conversely, if the expected sign of the effect is negative, small studies providing small (insignificant) and positive estimates with large standard errors may be missing, causing the plot to be skewed to the left.

\cite{doucouliagos2005publication2}, \cite{doucouliagos2013all}, and \cite{doucouliagos2018credible} argue that publication bias is less likely to occur in a contested research field. Specifically, in cases where there is theoretical support for both significant positive and negative effects, researchers are likely to submit statistically significant results in either direction for publication without fear of rejection by journals. Consequently, the resulting published literature is expected to be less prone to publication bias. In such cases, the majority of studies in the meta-analysis yield statistically significant effect estimates in either direction, positioned at the margins of the funnel plot, while small effect estimates from smaller studies are often missing from the insignificant central part of the plot. This phenomenon is referred to as the ``tunnel'' effect \citep{peters2008contour,martyn2010effects,onishi2014publication}. The tunnel effect is unlikely to produce a highly biased estimate of the mean effect size, as the funnel plot remains symmetrical, albeit hollow \citep{doucouliagos2005publication1,peters2008contour}.

It is important to note that funnel plot asymmetry may arise from factors other than publication bias. For example, confounding variables, heterogeneity in true effects, data irregularities, variations in methodological quality, and chance can all contribute to asymmetry \citep{egger1997bias,sterne2001funnel,sterne2005funnel,sterne2008addressing,moreno2009assessment,sterne2011recommendations,callot2011problem}. Thus, funnel plot asymmetry cannot be regarded as definitive evidence of publication bias in meta-analysis \citep{sterne2001funnel}. For this reason, many researchers attribute funnel plot asymmetry to ``small-study effects'' rather than publication bias, where smaller studies (i.e., those with lower precision) tend to produce larger estimates of a treatment effect compared to larger studies. In other words, smaller studies often report larger effects \citep{sterne2000publication,sterne2001investigating,sterne2001funnel,brockwell2001comparison}.
To address this issue, \cite{peters2008contour} propose an enhancement to the traditional funnel plot by incorporating contours that represent conventional ``milestone'' levels of statistical significance. This ``contour-enhanced'' funnel plot helps distinguish between asymmetry caused by publication bias and asymmetry attributable to other factors. Briefly, the statistical significance of each point (i.e., effect estimate) on the funnel plot is calculated, and regions of statistical significance are delineated on the plot. If studies appear to be missing from regions of statistical non-significance (or significance), the observed asymmetry is attributed to publication bias (or factors other than publication bias).

The primary limitation of funnel plots is that their visual interpretation, as graphical tools, is inherently subjective. To address this, several statistical methods have been developed to formally test for funnel plot asymmetry. These tests evaluate whether the observed association between the estimated effect size and a measure of sample size (such as the standard error of the effect estimate or its inverse) is greater than what would be expected by chance \citep{sterne2008addressing}.
\cite{egger1997bias} introduced a widely used regression-based approach to test for funnel plot asymmetry. This method relies on the following specification:
\begin{equation}\label{eq23}
	y_i = \beta_1 + \beta_0 S_i + \varepsilon_i,
\end{equation}
where $y_i$ represents the estimated empirical effect, and $S_i$ denotes its standard error. In the presence of publication selection bias, the reported effect estimates are positively correlated with their standard errors, resulting in an asymmetrical funnel plot. Consequently, a test for funnel plot asymmetry involves evaluating the null hypothesis that the asymmetry coefficient $\beta_0$ equals zero. If the hypothesis $H_0: \beta_0 = 0$ cannot be rejected, the funnel plot is deemed symmetric \citep{ashenfelter1999review,gorg2001multinational,sterne2001funnel}.
However, the OLS estimator for \cref{eq23} is often inefficient due to heteroscedasticity, as the standard error of the effect estimates varies widely \citep{stanley2005beyond}. A more suitable approach is weighted least squares (WLS), using the inverse of the variance of the effect estimates ($1/S_i^2$) as weights. Applying this weighting scheme to \cref{eq23} is equivalent to performing OLS on the transformed model obtained by dividing both sides of \cref{eq23} by $S_i$:
\begin{equation}\label{eq24}
	\frac{y_i}{S_i} \equiv t_i = \beta_0 + \beta_1 \left(\frac{1}{S_i}\right) + v_i,
\end{equation}
where $t_i$ represents the $t$-value for each estimate from the primary studies, $1/S_i$ is the precision of the estimate, and $v_i = \varepsilon_i / S_i$ is the transformed error term \citep{egger1997bias,sterne2000publication,sterne2005regression,stanley2005beyond,stanley2008meta1}.
In this formulation, testing for funnel plot asymmetry is equivalent to testing whether the intercept $\beta_0$ equals zero ($H_0: \beta_0 = 0$) using a conventional $t$-test.\footnote{Although the expected sign of $\beta_0$ is typically positive, the possibility of a negative sign necessitates the use of two-sided $p$-values \citep{sterne2005regression}.} Failure to reject the null hypothesis indicates funnel plot symmetry, suggesting the absence of publication bias. Conversely, rejecting the null implies asymmetry in the funnel plot, indicating the presence of publication bias or, more generally, ``small-study'' bias. The magnitude and direction of asymmetry are captured by the estimated intercept $\hat\beta_0$: the greater the deviation of $\hat\beta_0$ from zero, the more pronounced the asymmetry. If $\hat\beta_0 > 0$ ($\hat\beta_0 < 0$), the funnel plot is skewed to the right (left). This method is commonly referred to as the ``funnel-asymmetry test (FAT)'' or the ``Egger test'' \citep{egger1997bias,sterne2000publication,sterne2005regression,stanley2005beyond,stanley2008meta1}.
Simulations confirm that the FAT is valid, though it exhibits limited power in detecting publication selection bias \citep{stanley2008meta2}. A practical guideline for interpreting FAT results is as follows: publication selectivity is considered ``little to modest'' if FAT is not statistically significant or $\lvert\hat\beta_0\rvert < 1$, ``substantial'' if FAT is significant and $1 \leq \lvert\hat\beta_0\rvert \leq 2$, and ``severe'' if FAT is significant and $2 < \lvert\hat\beta_0\rvert$ \citep{stanley2010picture,doucouliagos2013all}.

As discussed earlier, there are two sources of publication selection bias: statistical significance and direction of results. The funnel plot is a useful tool for detecting directional selection (type I selection). In cases where there is theoretical support for both significant positive and negative effects, publication selectivity is based on statistical significance. That is, studies with statistically significant effect estimates are more likely to be published, irrespective of their sign (type II selection). In these cases, the funnel plot remains symmetrical but appears hollow and excessively wide. This type of publication selection is generally neglected in meta-analyses because it does not induce a bias in the mean effect size and is thus more benign \citep{stanley2005beyond}.
The Galbraith plot \citep{galbraith1988note} is a visual tool for identifying the presence of type II publication bias. It is a scatter plot of standardized effect estimates (often a $t$-value) on the vertical axis against the precision of the estimates ($1/S_i$) on the horizontal axis. In the absence of type II publication bias, the proportion of $t$-values, $(y_i-\mu)/S_i$, that fall outside the range of $\pm2$ should not exceed 5\% of all cases. Here, $\pm2$ are the approximate critical values at the 5\% significance level, and $\mu$ represents the true mean effect, which can be estimated, for instance, using the funnel plot and its top 10\% of points that are less prone to selection bias or using PET and PEESE (see below). When there is no true effect beyond publication selection, the $t$-values should be randomly distributed around zero, with no systematic relationship to the precision of the estimates \citep{stanley2005beyond}.
In addition to the graphical approach, a statistical test for type II publication bias can be performed by testing the null hypothesis that the slope coefficient in the following regression equals zero ($\beta_0=0$):
\begin{equation}\label{eq25}
	\lvert y_i\rvert=\beta_1 + \beta_0 S_i + \varepsilon_i,
\end{equation}
where $\lvert y_i\rvert$ indicates the absolute value of the $i$th effect estimate and $S_i$ is its standard error. As before, this model suffers from heteroscedasticity, rendering the OLS estimator inefficient. This issue can be addressed using WLS with the inverse of the variance of the effect estimates (i.e., $1/S_i^2$) as the weight, or equivalently, by applying OLS to the transformed model:
\begin{equation}\label{eq26}
	\left\lvert\frac{y_i}{S_i}\right\rvert \equiv\lvert t_i \rvert=\beta_0 + \beta_1 \left(\frac{1}{S_i}\right) + v_i.
\end{equation}

In the same manner as the FAT, if the null hypothesis $H_0: \beta_0=0$ is rejected, there is evidence of type II publication selection bias \citep{stanley2005beyond}.

Furthermore, a test for the existence of a genuine pooled effect beyond publication bias can be conducted by testing the null hypothesis that the slope coefficient in \cref{eq24} equals zero ($H_0: \beta_1=0$). This test is designed to determine whether the true mean effect is statistically different from zero after accounting for publication selection bias. Rejection of the null hypothesis implies the presence of an empirical effect beyond publication bias, with the estimated slope coefficient ($\hat\beta_1$) indicating the direction and magnitude of the effect. Conversely, failure to reject the null hypothesis suggests no significant evidence of a genuine empirical effect beyond publication bias. This approach is referred to as the ``precision-effect test (PET)'' \citep{stanley2005beyond,stanley2008meta2,sterne2005regression}. Simulation studies demonstrate that PET is highly effective in detecting publication bias and remains robust regardless of the intensity of the bias \citep{stanley2008meta2}.\footnote{Another meta-regression procedure for testing a genuine empirical effect corrected for publication selection relies on a well-established property of statistical power, which states that the absolute value of the standardized test statistic (the $t$-ratio) is positively associated with its degrees of freedom only if a genuine underlying empirical effect exists \citep{stanley2001wheat,stanley2005beyond}. Accordingly, testing $H_0: 0\leq \delta_1$ in the following regression model provides a formal statistical test, known as the ``meta-significance test (MST)'', to detect any genuine empirical effect beyond publication bias:
	\begin{equation*}
		\text{ln}\lvert t_i\rvert=\delta_0 + \delta_1 \text{ln}(df_i) + \varepsilon_i,
	\end{equation*}
where $\text{ln}$ denotes the natural logarithm, $t_i$ is the $t$-value corresponding to the $i$th estimate, and $df_i$ represents its degrees of freedom. Rejection of this null hypothesis indicates a statistically significant positive relationship between the absolute value of the test statistic and its degrees of freedom, providing evidence of a genuine nonzero effect \citep{stanley2005beyond}.}

Although the conventional $t$-test of the null hypothesis $\beta_1=0$ in \cref{eq24} often serves as a valid and powerful test for the existence of a genuine effect beyond publication bias, simulations suggest that $\hat\beta_1$ tends to underestimate the true mean effect when it is nonzero \citep{stanley2008meta2,stanley2014meta}. 
To address this issue, the following quadratic approximation is recommended, where the variance of the effect estimate ($S_i^2$) replaces the standard error ($S_i$) in \cref{eq23}, for estimating the genuine pooled effect corrected for publication selection:
\begin{equation}\label{eq27}
	y_i=\lambda_1 + \lambda_0 S_i^2 + \varepsilon_i,
\end{equation}
so that the intercept estimate, $\hat\lambda_1$, in this formulation serves as a more accurate (less biased) estimate of the true bias-corrected effect. This method is referred to as the ``precision-effect estimate with standard error (PEESE)'' \citep{stanley2007identifying,moreno2009assessment,stanley2012meta,stanley2014meta}. 
Given the presence of heteroscedasticity, the WLS estimator can be applied to efficiently estimate the parameters of \cref{eq27}, with weights set to $1/S_i^2$. Alternatively, this is equivalent to dividing both sides of \cref{eq27} by $S_i$, resulting in:
\begin{equation}\label{eq28}
	\frac{y_i}{S_i} \equiv t_i=\lambda_0 S_i+ \lambda_1 \left(\frac{1}{S_i}\right) + v_i,
\end{equation}
and subsequently applying OLS to the transformed regression equation. In this case, the coefficient of $1/S_i$ in \cref{eq28} corresponds to the intercept term in \cref{eq27}, whose estimate, $\hat\lambda_1$, represents the genuine effect corrected for publication bias (PEESE). 
In summary, if PET results indicate the presence of a true mean effect beyond publication selection (rejecting $H_0: \beta_1=0$), PEESE, represented by $\hat\lambda_1$ from \cref{eq28}, should be used. Otherwise, $\hat\beta_1$ from \cref{eq24} serves as a less biased estimator of the true bias-corrected effect \citep{stanley2017finding}. This conditional estimator is known as the ``precision-effect test and precision-effect estimate with standard error (PET-PEESE)'' \citep{stanley2014meta}.

To avoid omitted variable bias, the meta-regression model given by \cref{eq23} can be extended to include a set of moderator variables explaining the heterogeneity in effect sizes ($Z_k$'s) and other factors related to the publication selection process ($K_j$'s), as follows:
\begin{equation}\label{eq29}
	y_i = \beta_1 + \sum_k\alpha_kZ_{ki} + \beta_0 S_i + \sum_j\gamma_j K_{ji} S_i + \varepsilon_i,
\end{equation}
where $\beta_1$ and $\beta_0 S_i$ in \cref{eq23} are replaced by $\beta_1 + \sum_k\alpha_kZ_{ki}$ and $\beta_0 S_i + \sum_j\gamma_j K_{ji} S_i$, respectively \citep{stanley2008meta1,doucouliagos2009publication,stanley2012meta}. 
Due to the presence of heteroscedasticity, this model can be estimated using WLS with $1/S_i^2$ as weights or, equivalently, by applying OLS to the transformed model. 
\footnote{Similarly, the extended versions of \cref{eq27,eq28} can be expressed as:
	\begin{equation*}
		y_i = \lambda_1 + \sum_k\alpha_kZ_{ki} + \lambda_0 S_i^2 + \sum_j\gamma_j K_{ji} S_i^2 + \varepsilon_i,
	\end{equation*}
	and
	\begin{equation*}
		\frac{y_i}{S_i} \equiv t_i = \lambda_0 S_i + \sum_j\gamma_j K_{ji} S_i + \lambda_1 \left(\frac{1}{S_i}\right) + \sum_k\alpha_k \left(\frac{Z_{ki}}{S_i}\right) + v_i,
	\end{equation*}
	respectively \citep{doucouliagos2012estimates,stanley2014meta}. These models can be used to identify factors explaining the heterogeneity in effect sizes. While \cref{eq27,eq28} are preferable to \cref{eq23,eq24} for estimating the true mean effect beyond publication selection, it remains unclear whether the multiple meta-regression models described above are superior to \cref{eq29,eq30} for explaining variation in true effects \citep{doucouliagos2012estimates}.}
Similarly, the transformed version of \cref{eq29} can be written as:
\begin{equation}\label{eq30}
	\frac{y_i}{S_i} \equiv t_i = \beta_0 + \sum_j\gamma_j K_{ji} + \beta_1 \left(\frac{1}{S_i}\right) + \sum_k\alpha_k \left(\frac{Z_{ki}}{S_i}\right) + v_i.
\end{equation}

The coefficients $\gamma_j$ represent the effects of the corresponding variables ($K_j$'s) on publication selection, while the coefficients $\alpha_k$ capture the effects of the corresponding moderator variables ($Z_k$'s) on the true effect size, after accounting for publication selection. 
A Wald test can be conducted to evaluate the joint hypothesis that $\beta_0$ and $\gamma_j$'s are equal to zero. Rejection of this null hypothesis indicates the presence of publication selection bias. Similarly, rejecting the joint hypothesis that the $\alpha_k$ coefficients are zero suggests that the $Z$-variables account for at least part of the heterogeneity in the true effects.

A variety of methods methods and techniques have been developed for the identification of and correction for publication bias in meta-analyses.\footnote{For a detailed review of methods for detecting, assessing, and correcting for publication bias, see \cite{vevea2019publication}.} Some of the most common approaches are introduced in the following. \cite{stanley2010discard} proposed the ``Top 10'' approach, where only the most precise 10\% of reported estimates are used in meta-analyses. This approach mitigates publication bias by emphasizing high-precision studies, which are less prone to selection bias. By focusing on precision, the ``Top 10'' approach reduces the influence of less reliable studies, thereby enhancing the accuracy and reliability of meta-analytic conclusions. 
 \cite{stanley2017finding} proposed a meta-analytic method designed to mitigate selective reporting bias by focusing on estimates with adequate statistical power, typically defined as having at least 80\% power to detect a nonzero effect. The Weighted Average of Adequately Powered (WAAP) method computes a weighted average of effect sizes using optimal weights based on the inverse of variance but applies these weights exclusively to estimates that meet the power threshold. By focusing on adequately powered studies, WAAP diminishes the influence of smaller, less reliable studies that may exaggerate effect sizes due to selective reporting. Overall, WAAP generally outperforms traditional meta-analytic models by significantly reducing selective reporting bias while maintaining strong statistical properties when no bias is present. \cite{andrews2019identification} proposed two approaches to identify the probability of publication conditional on a study’s results. The first uses data from systematic replications of original studies, while the second uses data from meta-studies. The authors initially propose methods for calculating bias-corrected estimators and confidence sets for the parameters of interest when the conditional probability of publication is known. They then demonstrate how this conditional publication probability can be nonparametrically identified using data from replication studies and meta-analyses. \cite{bom2019kinked} introduced the endogenous kink (EK) meta-regression model as a novel approach for correcting publication bias in meta-analysis. The EK model fits a piecewise linear regression of effect size estimates on their standard errors, featuring a kink at an endogenous cutoff value of the standard error. This cutoff represents the point below which publication selection is unlikely, as studies are deemed sufficiently significant, and increases in standard errors do not influence publication likelihood. Above this cutoff, publication selection becomes more likely, resulting in a positive linear relationship. The model endogenously determines this cutoff based on a first-stage estimate of the true effect size and a predefined threshold of statistical significance. \cite{stanley2021detecting} proposed three tests for detecting publication selection bias based on excess statistical significance (ESS): the Proportion of Statistical Significance Test (PSST), the Test of Excess Statistical Significance (TESS), and their combination (TESSPSST). These tests explicitly account for between-study heterogeneity by incorporating it into the formulas for expected and excess statistical significance calculations. The PSST assesses whether the observed proportion of statistically significant findings is notably larger than what could theoretically be expected given the mean statistical power. If it is, this is taken as an indication of publication bias. The TESS evaluates whether the proportion of ESS exceeds 5\%, a threshold commonly accepted as the rate of false positives (i.e., Type I errors). The combined TESSPSST test identifies publication selection bias if either TESS or PSST is significant, thereby enhancing detection power beyond conventional methods such as the Egger test.

A prominent class of techniques for addressing publication bias is selection models \citep{hedges1984estimation,iyengar1988selection,hedges1992modeling,vevea1995general,mcshane2016adjusting,maier2022using}. These models estimate the relative probability that studies with p-values within predefined intervals have been published, as well as correct the meta-analytic effect size estimates. In other words, by modeling the relationship between statistical significance and the likelihood of publication, selection models explicitly address missing data. They primarily differ in their specified weight functions or are applicable only to the statistically significant results \citep{maier2023robust,bartos2023robust}. 
One widely used selection method for addressing selective reporting in scientific research is the p-curve, proposed by \cite{simonsohn2014pcurve}. The p-curve represents the distribution of statistically significant p-values (i.e., p-values < 0.05) across a set of independent findings. A right-skewed p-curve indicates the findings contain evidential value, suggesting that the observed effects are genuine rather than artifacts of selective reporting or p-hacking. In contrast, p-curves that lack a right-skewed distribution suggest the findings lack evidential value, while a left-skewed p-curve indicates intense p-hacking. The p-curve method employs binomial and continuous tests to assess whether the distribution is right-skewed. Despite its limitations, the p-curve remains a powerful tool for evaluating the impact of selective reporting on hypothesis testing and distinguishing between true effects and artifacts. The `$\textit{p-curve}$' package in R facilitates p-curve analysis, enabling users to perform tests and generate visualizations. A notable limitation of the p-curve method was its inability to estimate the average effect size. To address this, \cite{simonsohn2014effect} extended the p-curve to address this limitation. However, the p-curve does not provide a confidence interval (CI) around the estimate nor a test for publication bias. To overcome these limitations, \cite{vanassen2015meta} developed p-uniform, a method that simultaneously (a) tests for publication bias, (b) tests the null hypothesis of no effect, and (c) estimates the average effect size along with a CI for this estimate, using only statistically significant p-values from primary studies. Extending the p-uniform method, \cite{van2021correcting} introduced  $\text{p-uniform}^*$, which, unlike its predecessor, incorporates both significant and nonsignificant effect sizes. This extension introduces three major improvements: (i) $\text{p-uniform}^*$ provides a more efficient estimation than p-uniform, (ii) it reduces the overestimation of average effect size caused by between-study heterogeneity (variance) in true effect sizes, and (iii) it allows for the estimation and testing of between-study heterogeneity. Both p-uniform and  $\text{p-uniform}^*$ methods can be executed using the `$\textit{p-uniform}$' package in R.

\cite{mcshane2016adjusting} reviewed model selection methods and conducted an extensive simulation study to assess different aspects of their performance, such as estimation accuracy and confidence interval coverage. The results indicate that while the p-curve and p-uniform methods perform reasonably well in restrictive settings (characterized by rigid publication/selection rules and homogeneous effect sizes), they do not perform as well as the original \cite{hedges1984estimation} approach, due to the alternative estimation strategies they employ. Moreover, these approaches perform poorly in more realistic settings (involving more flexible publication rules and heterogeneous effect sizes), whereas variants of the Hedges method exhibit better performance. The authors conclude that, given the idealized assumptions underlying selection methods and the sensitivity of effect size estimates to these assumptions, such methods should not be used solely to obtain a single adjusted effect size estimate, but rather as part of a broader sensitivity analysis to explore the impact of different forms and degrees of publication bias.

As has been discussed so far, there is a wide range of methods available for detecting and correcting publication bias, each with its own strengths and limitations. However, in practice, researchers often lack a clear understanding of the data-generating process and do not have sufficient information to confidently select from these methods. Furthermore, this broad array of methods can often lead to conflicting conclusions. The combination of uncertainty regarding the data-generating process and these contradictions can create a ``breeding ground'' for confirmation bias, where researchers may inadvertently choose methods that align with their desired outcomes. This flexibility in method selection can significantly inflate the rate of false positives, posing a serious challenge to conventional meta-analytic methods \citep{maier2023robust,bartos2023robust}.

Given these challenges, one promising alternative is to explicitly combine different models and allow the data to determine the weight of each model based on its predictive accuracy for the observed data. In this context, \cite{maier2023robust} introduced a robust Bayesian meta-analysis (RoBMA) that model-averages selection models, fixed effects, and random effects models to address publication bias. By averaging over 12 models, RoBMA accounts for the uncertainty inherent in selecting meta-analytical models and is more robust to model misspecification. It provides quantifiable evidence for the absence of publication bias, performs well even under high between-study heterogeneity, and reliably avoids false positives. RoBMA also allows for the sequential updating of evidence as new studies become available, effectively addressing concerns about accumulation bias. The method outperforms other approaches in simulations and real data applications. Although RoBMA tends to underestimate effect sizes when the alternative hypothesis is true and overestimate them in the presence of publication bias, these biases are slight and disappear with larger sample sizes, particularly when the majority of the weight is assigned to the correct models. 

\cite{bartos2023robust} extended the RoBMA framework by integrating PET-PEESE regression models (as well as one-sided weight functions) with p-value-based selection models, leading to the development of RoBMA-PSMA (publication selection model-averaging). This extension optimally combines both approaches to correct for publication bias, resolving the tension between them and enabling inference through a model-averaged weighting scheme that reflects how well each model fits the data. \cite{bartos2023robust} demonstrated that RoBMA-PSMA outperforms previous methods in meta-analyses for which a gold standard is available and performs well in simulations. However, the method exhibits limitations in settings with prominent p-hacking, where it tends to overestimate effect sizes. Overall, Bayesian model averaging, in the form of RoBMA-PSMA, significantly improves both PET-PEESE and selection models by reducing the MSE and bias of PET-PEESE, as well as the MSE of the selection models. Importantly, RoBMA-PSMA accounts for uncertainty regarding the type of publication bias and seamlessly integrates the strengths of both approaches. Moreover, the framework’s flexibility allows researchers to specify alternative prior distributions or incorporate additional models based on prior knowledge. Both RoBMA and RoBMA-PSMA can be implemented using the latest version of the `$\textit{RoBMA}$' R package.

\section{Unrestricted Weighted Least Squares (UWLS) Estimator for Meta-Analysis and Meta-Regression Models}\label {uwls}
As discussed so far, in the context of meta-analysis and meta-regression analysis, random effects models account for between-study heterogeneity in true effect sizes. A random effects meta-analysis model incorporates an additive between-study variance component to capture the heterogeneity in true effect sizes, beyond the sampling variability. Also, in random effects meta-regression model, moderators are introduced to explain some of the variability in true effect sizes, but the model still incorporates an additive variance component to account for residual heterogeneity that remains unexplained by the moderators. In contrast, a fixed effect meta-analysis model does not allow for any between-study heterogeneity, and a fixed effects meta-regression model does not account for residual heterogeneity. Some authors  (e.g., \citealp{thompson1999explaining,higgins2004controlling}) caution against using a fixed effects meta-regression. They argue that, in practice, the included moderators often do not account for all the between-study heterogeneity. This can lead to residual heterogeneity that is not captured by the model, resulting in coefficient standard errors that are generally too small. Therefore, relying on conventional fixed effects meta-analysis and meta-regression models can lead to misleading conclusions, as they may ignore all or part of the variability in true effect sizes across studies. To address this issue, one approach is to incorporate total heterogeneity into the fixed effect meta-analysis model or residual heterogeneity into fixed effects meta-regression model by applying a common multiplicative factor, $\sigma^2$ (which is denoted as the overdispersion parameter $\phi$ in some studies), to each of the variances $\sigma_i^2$, resulting in $\sigma^2 \sigma_i^2$ or $\phi \sigma_i^2$. This is equivalent to estimating fixed effects models using unrestricted weighted least squares (UWLS) instead of the restricted weighted least squares (WLS) estimator ($\sigma^2 = 1$) that was used in the previous sections \citep{thompson1999explaining,stanley2015neither,stanley2016neither}. In the following discussion, the UWLS estimator for the fixed effect meta-analysis and fixed effects meta-regression models will be explained.

The fixed effect meta-analysis model can be rewritten as follows:
\begin{equation}\label{eq31}
\boldsymbol{y} = \mu_F \boldsymbol{1} + \boldsymbol{\varepsilon}, \quad \boldsymbol{\varepsilon} \sim N(\boldsymbol{0}, \boldsymbol{\Sigma}),
\end{equation}
where $\boldsymbol{1}$ is a $ k$-dimensional vector of ones, $\boldsymbol{y} = (y_1, y_2, \dots, y_k)^\top$ represents the column vector of observed effect sizes, and $\boldsymbol{\varepsilon} = (\varepsilon_1, \varepsilon_2, \dots, \varepsilon_k)^\top$ is the vector of errors. Additionally, $ E(\boldsymbol{\varepsilon}) = \boldsymbol{0}$, and $Var(\boldsymbol{\varepsilon}) = E(\boldsymbol{\varepsilon} \boldsymbol{\varepsilon}') = \sigma^2 \boldsymbol{\Omega} = \boldsymbol{\Sigma}$.
For heteroscedastic but uncorrelated error terms, the error variance-covariance matrix ($\boldsymbol{\Omega}$) is diagonal:
\begin{equation*}
 \boldsymbol \Omega =
	\begin{pmatrix}
		\sigma_1^2  & 0 & \cdots & 0 \\
		0 & \sigma_2^2 & \cdots &  0 \\
		\vdots  & \vdots  & \ddots & \vdots  \\
		0 &  0 & \cdots & \sigma_k^2 
	\end{pmatrix},
\end{equation*}
with the inverse matrix given by:
\begin{equation}\label{eq32}
	\boldsymbol \Omega^{-1} = 
	\begin{pmatrix}
		\frac{1}{\sigma_1^2}  & 0 & \cdots & 0 \\
		0 & \frac{1}{\sigma_2^2} & \cdots &  0 \\
		\vdots  & \vdots  & \ddots & \vdots  \\
		0 &  0 & \cdots & \frac{1}{\sigma_k^2} 
	\end{pmatrix}=
\underbrace{
	\begin{pmatrix}
		w_1  & 0 & \cdots & 0 \\
		0 & w_2 & \cdots &  0 \\
		\vdots  & \vdots  & \ddots & \vdots  \\
		0 &  0 & \cdots & w_k
	\end{pmatrix}
}_{\boldsymbol{W}}
\end{equation}
and $w_i = \frac{1}{\sigma_i^2}$ for $i = 1, 2, \dots, k$. Then, the inverse of $\boldsymbol{\Sigma}$ is:
\begin{equation}\label{eq33}
	\boldsymbol \Sigma^{-1} = 
	\begin{pmatrix}
		\frac{1}{\sigma^2\sigma_1^2}  & 0 & \cdots & 0 \\
		0 & \frac{1}{\sigma^2\sigma_2^2} & \cdots &  0 \\
		\vdots  & \vdots  & \ddots & \vdots  \\
		0 &  0 & \cdots & \frac{1}{\sigma^2\sigma_k^2} 
	\end{pmatrix}=
	\underbrace{
		\begin{pmatrix}
			w_1^u  & 0 & \cdots & 0 \\
			0 & w_2^u & \cdots &  0 \\
			\vdots  & \vdots  & \ddots & \vdots  \\
			0 &  0 & \cdots & w_k^u
		\end{pmatrix}
	}_{\boldsymbol{W^u}}
\end{equation}
with $w_i^u=\frac{1}{\sigma^2\sigma_i^2}= \frac{1}{\sigma^2} w_i$ for $i = 1, 2, \dots, k$.

In practice, $\sigma^2_i$'s are estimated by the sample variances obtained from primary studies ($S^2_i$'s), although they are conventionally treated as fixed and known in the analysis. In addition, $\sigma^2 > 0$ as a multiplicative variance parameter can be routinely estimated using the mean squared error (MSE) of the estimated intercept-only regression model by the WLS estimator as follows:
\begin{equation}\label{eq34}
s^2 = \frac{\hat{\boldsymbol{\varepsilon}}^{*'} \hat{\boldsymbol{\varepsilon}}^*}{k - 1} = \frac{(\boldsymbol{y} - \hat{\mu}_F \boldsymbol{1})' \hat{\boldsymbol{W}} (\boldsymbol{y} - \hat{\mu}_F \boldsymbol{1})}{k - 1}
\end{equation}
where $\hat{\boldsymbol{\varepsilon}}^* = \boldsymbol{y}^* - \hat{\mu}_F \boldsymbol{1}$, $\boldsymbol{y}^* = \hat{\boldsymbol{W}}^{1/2} \boldsymbol{y}$, and $\hat{\boldsymbol{W}} = \text{diag}[\hat{w}_1, \dots, \hat{w}_k] = \text{diag}\left[\frac{1}{S_1^2}, \dots, \frac{1}{S_k^2}\right]$. Therefore, $\hat{w}_{u_i} = \frac{1}{S^2 S_i^2}$ serves as an estimate of the ``true'' weight $w_i^u = \frac{1}{\sigma^2 \sigma_i^2}$.

Then, the UWLS estimator for the meta-analysis model is given by:
\begin{equation}\label{eq35}
\hat{\mu}_U = \frac{\sum_{i=1}^{k} \hat{w}_i^u y_i}{\sum_{i=1}^{k} \hat{w}_i^u} 
= \frac{\sum_{i=1}^{k} \frac{1}{S^2 S_i^2} y_i}{\sum_{i=1}^{k} \frac{1}{S^2 S_i^2}} 
= \frac{\sum_{i=1}^{k} \frac{1}{S_i^2} y_i}{\sum_{i=1}^{k} \frac{1}{S_i^2}} 
= \frac{\sum_{i=1}^{k} \hat{w}_i y_i}{\sum_{i=1}^{k} \hat{w}_i} 
= \hat{\mu}_F,
\end{equation}
whose variance is estimated as:
\begin{equation}\label{eq36}
\widehat{Var}(\hat{\mu}_U) = \frac{1}{\sum_{i=1}^{k} \hat{w}_i^u} 
= \frac{S^2}{\sum_{i=1}^{k} \hat{w}_i} 
= S^2 \widehat{Var}(\hat{\mu}_F).
\end{equation}
Consequently, the WLS and UWLS estimators provide identical point estimates for the fixed effect meta-analysis model (i.e., $\hat{\mu}_U = \hat{\mu}_F$), but their estimated variances differ only by a multiplicative factor of $S^2$. In fact, WLS is a special case of UWLS, with the constraint $\sigma^2 = 1$ imposed on it.

Similarly, the fixed effects meta-regression model can be expressed as:
\begin{equation}\label{eq37}
\boldsymbol{y} = \boldsymbol{X}\boldsymbol{\beta}_F + \boldsymbol{\varepsilon}, \quad \boldsymbol{\varepsilon} \sim N(\boldsymbol{0}, \boldsymbol{\Sigma}),
\end{equation}
where $\boldsymbol{X}$ is the $k \times (p+1)$ design matrix containing the moderators (independent variables) as columns, with the first column consisting of ones to account for the intercept term. $\boldsymbol{\beta}_F$ is a $(p+1) \times 1$ vector of fixed effect coefficients to be estimated.

The UWLS estimator for this model is:
\begin{equation*}
\hat{\boldsymbol{\beta}}_U = \left[\boldsymbol{X}'\boldsymbol{\Sigma}^{-1}\boldsymbol{X}\right]^{-1} \left[\boldsymbol{X}'\boldsymbol{\Sigma}^{-1}\boldsymbol{y}\right] = \left[\boldsymbol{X}'(\sigma^2 \boldsymbol{\Omega})^{-1}\boldsymbol{X}\right]^{-1} \left[\boldsymbol{X}'(\sigma^2 \boldsymbol{\Omega})^{-1}\boldsymbol{y}\right].
\end{equation*}

After canceling out the term $\sigma^2$, this becomes:
\begin{equation*}
\hat{\boldsymbol{\beta}}_U = \left(\boldsymbol{X}'\boldsymbol{\Omega}^{-1}\boldsymbol{X}\right)^{-1} \left(\boldsymbol{X}'\boldsymbol{\Omega}^{-1}\boldsymbol{y}\right) = \left(\boldsymbol{X}'\boldsymbol{W} \boldsymbol{X}\right)^{-1} \left(\boldsymbol{X}'\boldsymbol{W} \boldsymbol{y}\right).
\end{equation*}

Then, replacing $\boldsymbol{W} = \text{diag}[w_1, \dots, w_k]$ with $\hat{\boldsymbol{W}} = \text{diag}[\hat{w}_1, \dots, \hat{w}_k]$ results in the following:
\begin{equation}\label{eq38}
\hat{\boldsymbol{\beta}}_U = 
\left(
\boldsymbol{X}' \hat{\boldsymbol{W}} \boldsymbol{X}
\right)^{-1} 
\left(
\boldsymbol{X}' \hat{\boldsymbol{W}} \boldsymbol{y}
\right)
= \hat{\boldsymbol{\beta}}_F,
\end{equation}
with the estimated variance-covariance matrix:
\begin{equation}\label{eq39}
\widehat{Var}(\hat{\boldsymbol{\beta}}_U) = 
\left(\boldsymbol{X}' \hat{\boldsymbol{W}}^U \boldsymbol{X}\right)^{-1} 
= S^2 \left(\boldsymbol{X}' \hat{\boldsymbol{W}} \boldsymbol{X}\right)^{-1} 
= S^2 \widehat{Var}(\hat{\boldsymbol{\beta}}_F),
\end{equation}
where $S^2$ is an estimator of $\sigma^2$, computed as the mean squared error (MSE) of the estimated meta-regression model by the WLS estimator, as follows:
\begin{equation}\label{eq40}
S^2 = \frac{\hat{\boldsymbol{\varepsilon}}^{*'} \hat{\boldsymbol{\varepsilon}}^*}{k - p - 1} = 
\frac{
	\left(
	\boldsymbol{y} - \boldsymbol{X} \hat{\boldsymbol{\beta}}_F
	\right)' 
	\hat{\boldsymbol{W}} 
	\left(
	\boldsymbol{y} - \boldsymbol{X} \hat{\boldsymbol{\beta}}_F
	\right)
}{k - p - 1}, 
\end{equation}
where $\hat{\boldsymbol{\varepsilon}}^* = \boldsymbol{y}^* - \boldsymbol{X}^* \hat{\boldsymbol{\beta}}_F$, $\boldsymbol{y}^* = \hat{\boldsymbol{W}}^{1/2} \boldsymbol{y}$, and $\boldsymbol{X}^* = \hat{\boldsymbol{W}}^{1/2} \boldsymbol{X}$.

Consequently, the WLS and UWLS estimators provide identical point estimates for the fixed effects meta-regression model (i.e., $\hat{\boldsymbol{\beta}}_U = \hat{\boldsymbol{\beta}}_F$), but their estimated variances differ only by a multiplicative factor of $S^2$.

Overall, both conventional random effects models and multiplicative fixed effects models account for between-study heterogeneity. However, \cite{stanley2015neither,stanley2016neither} demonstrate that in both meta-analysis and meta-regression analysis, a multiplicative fixed effects (UWLS) estimator outperforms a conventional random effects estimator (i.e., has smaller bias and MSE) when there is publication selection bias or selective reporting bias. Even when the random effects model is known to be true and there is no publication selection bias, simulations show that the random effects model has no significant advantage over UWLS \citep{stanley2017finding}. 
Further supporting this, \cite{stanley2023unrestricted} conducted a comprehensive analysis of 67,308 meta-analyses from the Cochrane Database of Systematic Reviews (CDSR) published between 1997 and 2020. They compared UWLS with the conventional random effects model to evaluate which better represents medical research. Their findings reveal that UWLS frequently outperforms the random effects model, regardless of the level of heterogeneity, the number of studies, and the outcome measures. The authors conclude that UWLS is generally a superior model for medical research and should be routinely reported in all meta-analyses.

\section{Effect Size Dependence} \label {dependence}
So far, it has been assumed that all effect size estimates are statistically independent of one another. This assumption holds when all studies are independent and each study contributes only one effect estimate to the meta-analysis \citep{hedges2010robust}. 
In practice, however, it is quite common for effect size estimates to be dependent for various reasons. According to \cite{hedges2010robust}, dependence among effect size estimates across different studies may occur either through the estimation errors ($\varepsilon_i$'s in \cref{eq1}), the effect size parameters ($\beta_i$'s in \cref{eq1}), or both. Dependence among the estimation errors arises when a primary study reports more than one effect size estimate---i.e., multiple effect sizes are clustered (or nested) within one or more studies, which is referred to as ``correlated effects.'' This situation arises, for example, when effect size estimates share a common control group, when multiple outcomes are measured for the same individuals, or when the same outcome is measured at multiple time points for the same individuals \citep{becker2000multivariate,hedges2010robust,cheung2015meta,tipton2015small,lopez2017assessing}. This type of dependence is always present in the data, even when a fixed effect model, where the true effect sizes are fixed, is employed \citep{cheung2015meta}. 
Dependence among the (random) effect parameters occurs when there are groups or clusters of studies that share common characteristics, such as those conducted by the same investigator/research group or in the same laboratory, country, or region (i.e., multiple studies are nested within larger clusters). This type of dependence is referred to as ``hierarchical effects'' or ``hierarchical dependence'' \citep{stevens2009hierarchical,hedges2010robust}.

Ignoring or misspecifying dependence in meta-analytic data can lead to invalid estimates of the mean effect size (or the meta-regression coefficients) and their standard errors, resulting in inflated Type I error rates \citep{becker2000multivariate,olkin2009stochastically}. 
Several strategies have been employed to address statistically dependent effect sizes. The simplest of these involve ignoring the dependence among effect size estimates, selecting only one effect size estimate per study, computing a single summary effect size per study by averaging the effect size estimates clustered within studies, and shifting the unit of analysis. Although these strategies can effectively remove the dependence among effect size estimates, they result in a loss of information (see \citealp{rosenthal1986meta,cooper1998synthesizing,marin1999averaging,becker2000multivariate,jackson2011multivariate,scammacca2014meta,cheung2014modeling,cheung2015meta,borenstein2021introduction}). A more principled approach is to explicitly model the correlations among effect size estimates using multivariate meta-analytic methods. The challenge with this approach is that it generally requires knowledge of the covariance structure of within-study effect size estimates, which is rarely reported in primary studies (see \citealp{raudenbush1988modeling,kalaian1996multivariate,becker2000multivariate,olkin2009stochastically,jackson2011multivariate,mavridis2013practical}).
Fortunately, three common approaches---the robust variance estimation (RVE) approach, the multilevel meta-analysis model, and the generalized weights (GW) approach---address this issue. These recently developed methods are discussed in the following subsections.

\subsection{The Robust Variance Estimation (RVE)} \label{RVE}  
\cite{hedges2010robust} propose an appealing meta-analytic method, known as robust variance estimation (RVE), which can be used to handle both types of clustering or dependence structures in meta-analysis (i.e., correlated effects and hierarchical effects) without requiring knowledge of the covariance structure of the effect estimates and without making strong assumptions about the error distribution. Instead of modeling dependence as done in multivariate meta-analytic methods, the RVE method adjusts the variance-covariance matrix of meta-regression coefficients to account for the dependence of effect sizes, thereby improving statistical inference, including confidence intervals and hypothesis testing. This method is briefly described in the following discussion. For further details, the interested reader is referred to \cite{hedges2010robust,tipton2013robust,tanner2014robust,tipton2015small,tipton2015smallsample,tanner2016handling}.

Consider the general case of a meta-analysis of \(m\) studies, with \(k_j\) effect size estimates nested within the \(j\)-th study (\(j = 1, \dots, m\)), such that the total number of effect size estimates is \(\sum_{j=1}^m k_j = k\). Then, a general meta-regression model can be written as
\begin{equation}\label{eq41}
	\boldsymbol y = \boldsymbol X \boldsymbol \beta + \boldsymbol \epsilon,
\end{equation}
where \(\boldsymbol y = (\boldsymbol y_1^{'}, \dots, \boldsymbol y_m^{'})'\) is a \(k \times 1\) vector of \(m\) vectors of effect size estimates, each with \(k_j\) elements; \(\boldsymbol \epsilon = (\boldsymbol \epsilon_1^{'}, \dots, \boldsymbol \epsilon_m^{'})'\) is a stack of \(m\) error vectors, each with mean \(\boldsymbol 0\) and variance-covariance matrix \(\boldsymbol \Sigma_j\); \(\boldsymbol X = (\boldsymbol X_1^{'}, \dots, \boldsymbol X_m^{'})'\) is a stack of \(k_j \times (p+1)\) matrices of 1s in the first column and observations on \(p\) covariates in the other columns (note that the covariates may vary across effect sizes or studies); and \(\boldsymbol \beta = (\beta_0, \beta_1, \dots, \beta_p)'\) is a \((p+1) \times 1\) vector of parameters, which can be estimated by weighted least squares as 
\begin{equation}\label{eq42}
	\boldsymbol{\hat{\beta}} = \left(\sum_{j=1}^m \boldsymbol X_j' \boldsymbol W_j \boldsymbol X_j\right)^{-1} \left(\sum_{j=1}^m \boldsymbol X_j' \boldsymbol W_j \boldsymbol y_j\right),
\end{equation}
where \(\boldsymbol W_j\) is the weight matrix associated with the \(j\)-th study. The exact variance-covariance matrix of \(\boldsymbol{\hat{\beta}}\) can be written as
\begin{equation}\label{eq43}
	Var(\boldsymbol{\hat{\beta}}) = \left(\sum_{j=1}^m \boldsymbol X_j' \boldsymbol W_j \boldsymbol X_j\right)^{-1} \left(\sum_{j=1}^m \boldsymbol X_j' \boldsymbol W_j \boldsymbol \Sigma_j \boldsymbol W_j \boldsymbol X_j\right) \left(\sum_{j=1}^m \boldsymbol X_j' \boldsymbol W_j \boldsymbol X_j\right)^{-1}.
\end{equation}

The problem with this estimator is that the within-study variance-covariance matrices, $ \boldsymbol \Sigma_1, \dots, \boldsymbol \Sigma_m $, are usually unknown and need to be estimated. In the standard fixed and random effects models, as discussed earlier, the variance elements are assumed to be fully known (in the fixed effect case) or partially known and partially estimated (in the random effects case). However, the covariance elements are assumed to be equal to zero. In practice, this assumption usually does not hold, especially when some studies contribute more than one effect size estimate to the meta-analysis. The complications arise when the covariance structure of the effect size estimates is unknown, which is the case in most applications.
In the RVE method, the cross-products of observed residuals within study $j$ are used to estimate the elements of $ \boldsymbol \Sigma_j $. Thus, the RVE estimator can be written as
\begin{equation}\label{eq44}
	\widehat{Var}(\boldsymbol{\hat{\beta}}) = \left( \sum_{j=1}^m \boldsymbol X_j' \boldsymbol W_j \boldsymbol X_j \right)^{-1} \left( \sum_{j=1}^m \boldsymbol X_j' \boldsymbol W_j \boldsymbol{\hat{\epsilon}}_j \boldsymbol{\hat{\epsilon}}_j' \boldsymbol W_j \boldsymbol X_j \right) \left( \sum_{j=1}^m \boldsymbol X_j' \boldsymbol W_j \boldsymbol X_j \right)^{-1},
\end{equation}
where $ \boldsymbol{\hat{\epsilon}}_j = \boldsymbol y_j - \boldsymbol X_j \boldsymbol{\hat{\beta}} $ is a $k_j \times 1$ vector containing all $k_j$ residuals for the $k_j$ effect size estimates nested within study $j$. According to \cite{hedges2010robust}, although the $ \boldsymbol{\hat{\Sigma}}_j = \boldsymbol{\hat{\epsilon}}_j \boldsymbol{\hat{\epsilon}}_j' $ are rather poor estimates of the $ \boldsymbol \Sigma_j $'s, $ \widehat{Var}(\boldsymbol{\hat{\beta}}) $ will converge in probability to the true covariance matrix, $ Var(\boldsymbol{\hat{\beta}}) $, if $ m \to \infty $.

An important property of the RVE estimator is that it provides (asymptotically) accurate estimates of the standard errors and a useful framework for statistical inference, regardless of the weights used. That is, RVE is valid for any given choice of $ \boldsymbol W_j $'s. As noted by \cite{hedges2010robust}, this means that the only criterion for choosing the weighting scheme is efficiency. In this case, the ``optimal'' weighting matrices, that is, the $ \boldsymbol W_j $'s that produce the most efficient estimator of $ \boldsymbol{\beta} $, would be $ \boldsymbol W_j = \boldsymbol \Sigma_j^{-1} $. Unfortunately, as discussed above, the true covariance structure of the effect size estimates is usually unknown (thus necessitating a robust method). For this reason, \cite{hedges2010robust} suggest approximating the optimal weights based on a ``working'' model of the unknown covariance structure, which produces an approximately efficient estimator of $ \boldsymbol{\beta} $. The resulting approximately optimal weighting matrix for the $j$-th study is defined as $ \boldsymbol W_j \approx \boldsymbol \Sigma_{aj}^{-1} $, where $ \boldsymbol \Sigma_{aj} $ is the working covariance matrix. They propose two working models for handling and coping with two types of dependence structures (i.e., correlated and hierarchical effects) in meta-analytic research separately (see \citealp{hedges2010robust,tipton2015small,tipton2015smallsample}).

In the correlated effects case, the working model can be written as:
\begin{equation}\label{eq45}
	\boldsymbol\Sigma_{aj}=\tau^2 \boldsymbol J_j+\rho \sigma_j^2(\boldsymbol J_j-\boldsymbol I_j)+\sigma_j^2 \boldsymbol I_j,
\end{equation}
where $\boldsymbol J_j$ is a $k_j\times k_j$ matrix of ones, $\boldsymbol I_j$ is a $k_j\times k_j$ identity matrix, and $\tau^2$ represents the between-study variance as a measure of heterogeneity in the study-average true effect sizes. In this model, the sampling variances associated with multiple effect sizes nested within each study are assumed to be equal ($\sigma_{ij}^2=\sigma_j^2$), which is the case when the sample sizes are equal. Moreover, the simplified working model assumes a common correlation ($\rho$) for all pairs of effect size estimates. Based on this working model of the known covariance structure, \cite{hedges2010robust} propose the approximately inverse variance (optimal) weights for study $j$ as follows:
\begin{equation}\label{eq46}
	\boldsymbol W_j= w_j \boldsymbol I_j=\left\{1/[k_j(\sigma_j^2+\tau^2)]\right\}\boldsymbol I_j.
\end{equation}

In other words, $\boldsymbol W_j=diag[w_{1j},...,w_{k_jj}]$, where $w_{ij}=w_j=1/k_j(\sigma_j^2+\tau^2)$. \cite{hedges2010robust} provide a method of moments (MM) estimator of $\tau^2$, depending on the unknown value of $\rho \in(0,1)$. They show that in many practical situations, the choice of $\rho$ plays a very minor role in the estimation of $\tau^2$. However, they recommend performing a sensitivity analysis to ensure that $\hat\tau^2$ is not highly sensitive to the choice of $\rho$. In addition, the parameter $\sigma_j^2$ can be estimated by the average of within-study sampling (error) variances, $S_j^2= \sum_{i=1}^{k_j} \frac{1}{k_j} S_{ij}^2$. Note that the fixed effects weights can be obtained by setting $\tau^2$ equal to zero.

In the hierarchical effects case, the working model can be written as
\begin{equation}\label{eq47}
	\boldsymbol\Sigma_{aj}=\tau^2 \boldsymbol J_j+\omega^2 \boldsymbol I_j+\boldsymbol\sigma_j^2,
\end{equation}
where $\boldsymbol J_j$ is a $k_j\times k_j$ matrix of ones, $\boldsymbol I_j$ is a $k_j\times k_j$ identity matrix, $\tau^2$ is the between-study variance as a measure of heterogeneity in the study-average true effects (as before), $\omega^2$ is the within-study variance as a measure of heterogeneity in the within-study true effects, and $\boldsymbol\sigma_j^2=diag[\sigma_{1j}^2,...,\sigma_{k_jj}^2]$ is a $k_j\times k_j$ diagonal matrix whose diagonal elements are the sampling (error) variances in study $j$. Based on this working model, \cite{hedges2010robust} propose the approximately inverse variance weights for study $j$ as follows:
\begin{equation}\label{eq48}
	\boldsymbol W_j= diag[w_{1j},...,w_{k_jj}], \qquad \text{where} \qquad w_{ij}= 1/(\sigma_{ij}^2+\omega^2+\tau^2).
\end{equation}

\cite{hedges2010robust} develop the method of moments (MM) estimators for the parameters $\omega^2$ and $\tau^2$. The $\sigma_{ij}^2$ are estimated by the sample variances $S_{ij}^2$ in practice, while they are commonly considered to be fixed and known in the analysis.

In practice, it is common for both types of dependence to occur in the same meta-analysis. In this case, the choice between the two working models can be based on the most common type of dependence in the data to be meta-analyzed, as recommended by \cite{tanner2014robust} and \cite{tipton2015smallsample}.

The RVE estimator can be used to construct Wald-type test statistics for testing hypotheses or forming confidence intervals about one or more elements of $\boldsymbol\beta$ (see \citealp{hedges2010robust,tipton2015smallsample}). For example, a $100(1-\alpha)\%$ confidence interval for the parameter $\beta_k$ ($k = 0, 1, \dots, p$) can be calculated as $\hat\beta_k \pm t_{m - p - 1, \alpha/2} \sqrt{\widehat{Var}(\hat\beta_k)}$, where $\widehat{Var}(\hat\beta_k)$ is the $k+1$th diagonal element of $\widehat{Var}(\boldsymbol{\hat\beta})$, and $t_{m - p - 1, \alpha/2}$ is the upper $\alpha/2$ quantile of the $t$-distribution with $m - p - 1$ degrees of freedom.
However, when the sample size is small or moderate, simulation studies suggest that tests based on the RVE estimator may produce inflated Type I error rates \citep{hedges2010robust,tipton2013robust,tipton2015small,tipton2015smallsample}. Therefore, several small-sample adjustments have been proposed by \cite{hedges2010robust,tipton2015small,tipton2015smallsample}. These adjustments fall into two categories: one for the residuals used in the RVE estimator, given by
\begin{equation}\label{eq49}
	Adj.\widehat{Var}(\boldsymbol{\hat\beta}) = \left( \sum_{j=1}^{m} \boldsymbol X_j' \boldsymbol W_j \boldsymbol X_j \right)^{-1} \left( \sum_{j=1}^{m} \boldsymbol X_j' \boldsymbol W_j \boldsymbol A_j \boldsymbol{\hat\epsilon}_j \boldsymbol{\hat\epsilon}_j' \boldsymbol A_j \boldsymbol W_j \boldsymbol X_j \right) \left( \sum_{j=1}^{m} \boldsymbol X_j' \boldsymbol W_j \boldsymbol X_j \right)^{-1},
\end{equation}
where $\boldsymbol A_j$ is a $k_j \times k_j$ adjustment matrix. The other adjustment concerns the degrees of freedom associated with the distribution of Wald-type test statistics ($t$, $F$, and $\chi^2$) used for constructing confidence intervals or testing hypotheses about single and multiparameter meta-regression coefficients (and the overall mean effect). 

RVE can be implemented in R (via the `$robumeta$' package, \citealp{fisher2015robumeta,tanner2016handling,fisher2017package}), Stata (via the `$ROBUMETA$' macro, \citealp{hedberg2011robumeta,tanner2014robust}), and, to some extent, SPSS (see \citealp{tanner2014robust}).

\subsection{The Multilevel Model} \label{Multilevel }
An alternative way to handle the hierarchical dependence of the effect sizes is to use a multilevel (three-level) meta-analysis model (see \citealp{hox2010multilevel, konstantopoulos2011fixed, van2013three, van2015meta, cheung2014modeling, cheung2015meta}).
It is evident that the random effects meta-analysis model in \cref{eq4} has a two-level structure as follows:
\begin{align}\label{eq50}
	\begin{split}
		\text{Level\;1}: \qquad y_i &=  \beta_i + \varepsilon_i, \qquad\qquad\;\;\;\;\;     \varepsilon_i \sim N(0, \sigma_i^2), \\
		\text{Level\;2}: \qquad \beta_i &=  \mu_R + u_i, \qquad\qquad\;\;\;     u_i \sim N(0, \tau^2), \\
		\text{Combined}: \;\;\; y_i &=  \mu_R + u_i + \varepsilon_i. \qquad\qquad\qquad\qquad\quad
	\end{split}
\end{align}

Similarly, the random/mixed effects meta-regression model in \cref{eq17} can be expressed as a two-level model (simply by adding the moderator variables to the second level of the previous model). The two-level random effects meta-analysis model can be extended to a three-level model by adding another level to account for the clustering of effect sizes within studies, as follows:
\begin{align}\label{eq51}
	\begin{split}
		\text{Level\;1}: \qquad y_{ij}=  \beta_{ij} + \varepsilon_{ij}, \qquad\qquad     \varepsilon_{ij}\sim N(0,\sigma_{ij}^2),
		\\
		\text{Level\;2}: \qquad \beta_{ij}=  \theta_j + \upsilon_{ij}, \qquad\qquad\;     \upsilon_{ij}\sim N(0,\omega^2),
		\\
		\text{Level\;3}: \qquad \theta_j=  \mu_R + u_j, \qquad\qquad\;\;     u_{ij}\sim N(0,\tau^2),
	\end{split}
\end{align}
where $y_{ij}$ is an estimate of the $i$th true (population) effect size in the $j$th study ($\beta_{ij}$), $\theta_j$ is the average true effect in the $j$th study, and $\mu_R$ is the overall average true effect (i.e., the average of all true effect sizes). Substituting the level 3 equation into the level 2 equation, and then substituting the resulting equation into the level 1 equation yields the following combined equation:
\begin{equation}\label{eq52}
	y_{ij} =  \mu_R + u_j + \upsilon_{ij} + \varepsilon_{ij}, \qquad\qquad\qquad\;
\end{equation}
where the random effects at different hierarchical levels ($\upsilon_{ij}$ and $u_j$) and the sampling error ($\varepsilon_{ij}$) are assumed to be independent of each other, i.e., 
$Cov(\upsilon_{ij}, u_j) = Cov(\upsilon_{ij}, \varepsilon_{ij}) = Cov(u_j, \varepsilon_{ij}) = 0$.
In this formulation, there are three potential sources of variation in effect size estimates: $\varepsilon_{ij}$ and $u_j$ as in the traditional two-level meta-analytic model, and $\upsilon_{ij}$ as an additional source of random variation. Thus, the variance of the $ij$th effect size estimate is given by
$Var(y_{ij}) = \sigma_{ij}^2 + \omega^2 + \tau^2$,
where $Var(y_{ij})$ is the variance of the $ij$th effect size estimate, $\sigma_{ij}^2$ is the sampling (error) variance associated with the $ij$th effect size estimate (which is considered to be known), $\omega^2$ is the within-study variance, representing the heterogeneity in the within-study true effect sizes, and $\tau^2$ is the between-study variance, representing the heterogeneity in the study-average true effect sizes. That is, the total variation in effect size estimates is the sum of the level 2 and level 3 heterogeneity, along with the sampling (error) variance. 
Although the multilevel (three-level) model proposed for dealing with multiple outcomes within the same study is based on the assumption that there is no sampling covariation (i.e., independent samples at the first level), a simulation study by \cite{van2013three} suggests that adding an intermediate level of outcomes within studies plays an important role in accurately accounting for the sampling covariance, providing appropriate standard errors and offering a useful framework for statistical inference about the effects.

In a similar fashion to the traditional two-level random effects meta-analysis, the first step in this extended meta-analysis is to estimate the model parameters (i.e., the true mean effect, $\mu_R$, and the variance components of the random effects, $\omega^2$ and $\tau^2$). The usual estimators of the parameters of multilevel regression models are maximum likelihood (ML) and restricted maximum likelihood (REML), which estimate all parameters simultaneously \citep{hox2010multilevel, konstantopoulos2011fixed, van2013three, van2015meta, cheung2014modeling, cheung2015meta}. 
In general, the ML and REML procedures produce very similar estimates of variance components. However, if they do not, the REML procedure is preferred because it accounts for the loss of degrees of freedom due to the estimation of regression coefficients as fixed effects, and thus provides less biased estimates of variance components \citep{hox2010multilevel}.
As in the case of traditional two-level random effects meta-analysis, the Hartung-Knapp-Sidik-Jonkman method can be used to test hypotheses or form confidence intervals about the mean effect size, $\mu_R$, instead of the Wald test based on the $z$ distribution.

The next step involves testing for the heterogeneity of the true effects based on Cochran's $Q$ statistic. When the true effect sizes are homogeneous, cluster effects do not exist, and thus the conventional $Q$ statistic given by \cref{eq7} can be applied directly. Under the null hypothesis of homogeneity, this statistic is asymptotically distributed as $\chi^2$ with $k-1$ degrees of freedom, where $k=\sum_{j=1}^m k_j$ is the total number of effect sizes. Large values of $Q$ provide evidence against the validity of the null hypothesis. 
The rejection of the null hypothesis, however, is not very informative, as the total heterogeneity in the true effects is the sum of two components: within-study heterogeneity at level 2 and between-study heterogeneity at level 3. Therefore, one might be interested in testing the statistical significance of the two heterogeneity parameters separately. This can be achieved by using the likelihood ratio (LR) or chi-square difference test, which is based on the likelihood function \citep{van2015meta, cheung2014modeling, cheung2015meta}. 
The null hypotheses to be tested are: $\text{H}_0: \omega^2=0$ and $\text{H}_0: \tau^2=0$. If the null hypothesis of $\omega^2=0$ is rejected, there is within-study heterogeneity in the true effects. Similarly, the rejection of the null hypothesis of $\tau^2=0$ indicates the presence of between-study heterogeneity. However, there is a problem with this testing procedure. The hypothesis that $\omega^2=0$ or $\tau^2=0$ is not well-defined for the LR test because the true parameter under the null hypothesis lies on the boundary of the parameter space ($\omega^2 \geq 0$ and $\tau^2 \geq 0$). In this situation, the test statistic does not follow a $\chi^2$ distribution with one degree of freedom, even when the null hypothesis is true \citep{stoel2006likelihood, hox2010multilevel, greene2012econometric, cheung2014modeling, cheung2015meta}. 
One simple way that has been suggested to resolve this issue is to use $2\alpha$ instead of $\alpha$ as the significance level, or equivalently, to divide the (two-sided) p-value obtained from the LR test by two \citep{pinheiro2000mixed, berkhof2001variance, cheung2014modeling, cheung2015meta}. That is, the null hypothesis is rejected at $\alpha=0.05$ if the calculated p-value is smaller than $2\alpha=0.10$, or equivalently, if the p-value divided by two is smaller than $\alpha=0.05$. 
In addition to testing the hypotheses, constructing confidence intervals for the variance components is of great interest. It has been suggested to use likelihood-based confidence intervals instead of Wald confidence intervals. 
In general, the likelihood-based approach for statistical inference about the variance parameters is preferred over Wald-type procedures, as it can account for the asymmetry or non-normality of the sampling distribution of the parameter estimates \citep{hox2010multilevel, cheung2014modeling, cheung2015meta}.

Having tested for and found evidence of heterogeneity among the true effect sizes, the logical next step in the random effects meta-analysis is to quantify such heterogeneity. In doing so, the $I^2$ index of traditional two-level meta-analysis, as given by \cref{eq10}, can reasonably be extended to the three-level case. Thus, the $I^2$ measures at levels 2 and 3 can be calculated as
\begin{equation}\label{eq53}
	I_2^2 = \frac{\hat\omega^2}{S^2 + \hat\omega^2 + \hat\tau^2},
\end{equation}
and
\begin{equation}\label{eq54}
	I_3^2 = \frac{\hat\tau^2}{S^2 + \hat\omega^2 + \hat\tau^2},
\end{equation}
where $\hat\omega^2$ is an estimate of the amount of level 2 heterogeneity (i.e., within-study variance), $\hat\tau^2$ is an estimate of the amount of level 3 heterogeneity (i.e., between-study variance), and $S^2$ is an estimate of the typical within-study sampling variance ($\sigma^2$), as defined in \cref{eq11}. Thus, $I_2^2$ and $I_3^2$ can be interpreted as the proportions of the total variation in observed effect sizes that are due to within-study heterogeneity (level 2) and between-study heterogeneity (level 3), respectively, rather than sampling error.

According to \cite{cheung2014modeling,cheung2015meta}, the sampling variance $S^2$ depends on the sample sizes on which the effect size estimates are based and can therefore take different values for each set of studies selected. As $S^2$ is included in the calculation of $I_2^2$ and $I_3^2$, these indices depend on the sample sizes and thus do not estimate any population parameters. Consequently, \cite{cheung2014modeling,cheung2015meta} also define two intraclass correlations (ICCs) that include only the level-2 and level-3 variances. The ICC statistics are given as follows:
\begin{equation}\label{eq55}
	{ICC}_2^2 = \frac{\hat\omega^2}{\hat\omega^2 + \hat\tau^2} = \frac{I_2^2}{I_2^2 + I_3^2},
\end{equation}
and
\begin{equation}\label{eq56}
	{ICC}_3^2 = \frac{\hat\tau^2}{\hat\omega^2 + \hat\tau^2} = \frac{I_3^2}{I_2^2 + I_3^2},
\end{equation}
where ${ICC}_2^2$ and ${ICC}_3^2$ can be interpreted as the proportions of the total heterogeneity in the true effects attributed to level 2 (within-study heterogeneity) and level 3 (between-study heterogeneity), respectively. It is clear that these indices are independent of sample size.

The three-level random effects meta-analysis model can be extended to a mixed effects model by including effect size- and study-level characteristics as moderator variables ($x_{ij}$'s and $x_j$'s) in the Level 2 and Level 3 equations, respectively, to model the heterogeneity, if detected early. The resulting three-level equation can be expressed as follows:
\begin{equation}\label{eq57}
	y_{ij} = \boldsymbol{x}_{ij}' \boldsymbol{\beta}_R + u_j + \upsilon_{ij} + \varepsilon_{ij}.
\end{equation}
where $\boldsymbol{x}_{ij}$ is a $(p+1) \times 1$ vector of $p$ explanatory variables (which may include the study-level variables, $x_j$'s), plus a $1$ in the first position corresponding to the model intercept, and $\boldsymbol{\beta}_R$ is a $(p+1) \times 1$ vector of unknown constants to be estimated.

The model parameters (i.e., the variance components of the random effects and the fixed effects/regression coefficients) can be estimated simultaneously using maximum likelihood procedures (ML or REML). The variance components, denoted $\omega_{res}^2$ and $\tau_{res}^2$, represent the level 2 and level 3 residual (unexplained) heterogeneity, respectively. As in the two-level mixed effects model, the generalized Cochran's $Q$ statistic given by \cref{eq20} (denoted $Q_{res}$) can be used to test the null hypothesis that the moderator variables included in the model fully explain the heterogeneity in the true effects, against the alternative that the moderators only partially explain the heterogeneity. It is also of interest to separately test whether the residual heterogeneity at each of the second and third levels is statistically different from zero. This can be done, as before, by using the likelihood ratio (LR) test for each hypothesis: $\text{H}_0: \omega_{res}^2=0$ and $\text{H}_0: \tau_{res}^2=0$ (as discussed earlier, this test is preferred over Wald-type tests). Rejection of the null hypothesis $\omega_{res}^2=0$ implies that the effect size-level covariates cannot fully explain the within-study heterogeneity in the true effects. Similarly, rejection of the null hypothesis $\tau_{res}^2=0$ implies that the study-level covariates cannot fully explain the between-study heterogeneity. 
Additionally, the $R^2$ discussed for two-level mixed effects meta-analysis can be extended to three-level meta-analysis, as follows:
\begin{equation}\label{eq58}
	R_2^2= \frac{\hat\omega^2-\hat\omega_{res}^2}{\hat\omega^2}=1-\frac{\hat\omega_{res}^2}{\hat\omega^2},
\end{equation}
and
\begin{equation}\label{eq59}
	R_3^2= \frac{\hat\tau^2-\hat\tau_{res}^2}{\hat\tau^2}=1-\frac{\hat\tau_{res}^2}{\hat\tau^2},
\end{equation}
where $\hat\omega^2$ and $\hat\tau^2$ denote the estimates of total within-study and between-study heterogeneity based on the meta-analysis model given by \cref{eq52}, and $\hat\omega_{res}^2$ and $\hat\tau_{res}^2$ denote the estimates of within-study and between-study residual heterogeneity based on the meta-analysis model given by \cref{eq57}. The $R_2^2$ index represents the proportion of the total within-study (level 2) heterogeneity that is explained by the effect size-level covariates included in the three-level mixed effects meta-analysis model, and the $R_3^2$ index represents the proportion of the total between-study (level 3) heterogeneity that is explained by the study-level covariates included in the model (see \citealp{cheung2014modeling,cheung2015meta}).

It would also be of interest to test whether the coefficients on the moderator variables are statistically significant. A common procedure used in multilevel analyses is the LR test, which is conducted by comparing the log-likelihood values of the constrained (reduced) and unconstrained (full) models. This test can be performed with either ML or REML. As mentioned above, REML is generally preferred to ML as it provides more accurate estimates of variance components. However, this procedure only allows for comparison of models with the same fixed part (regression coefficients) and can thus be used to compare models that differ only in the random part (variance components). This means that the LR test for the significance of one or more of the slope coefficients can be performed with ML but not with REML. Similarly, the omnibus procedure is commonly used to test the null hypothesis that one or more of the slope coefficients are zero. Additionally, the Knapp-Hartung method can be used to test the statistical significance of a single slope coefficient, which is based on the t-distribution with an adjusted standard error.

The three-level meta-analysis can be conducted in R using the `$metafor$' package \citep{viechtbauer2010package,viechtbauer2023metafor}. Another option is the `$metaSEM$' package \citep{cheung2015metasem}, which offers a suite of functions for performing univariate, multivariate, and three-level meta-analyses through a structural equation modeling (SEM) approach \citep{cheung2013multivariate,cheung2014modeling,cheung2015meta}.\footnote{\cite{polanin2017review} provide a review of 63 meta-analysis packages in R.}

\subsection{The Generalized--Weights (GW) Approach} \label{GW}
Meta-studies are often conducted on empirical outcomes derived from studies that use overlapping samples. This is particularly common in fields such as economics and finance, which rely heavily on aggregated observational data. In these fields, the same or similar datasets are reused across multiple studies. This characteristic of meta-analysis data is referred to as sample overlap \citep{bom2020generalized,bom2024accounting}. 
\cite{bom2020generalized} demonstrate that sample overlap creates a dependence structure among primary estimates, which can result in asymptotic biases, significant efficiency losses, and increased rates of false positives at the meta-analysis level. 
To address this issue, they proposed the generalized-weights (GW) meta-estimator, which follows a two-step procedure. The first step involves constructing the complete variance-covariance matrix to capture the correlation structure between primary outcomes. This matrix can be feasibly estimated using readily available information from primary studies, such as sample sizes, standard errors, and the number of overlapping observations. In the second step, the estimated matrix is used to optimally weight each primary outcome based on its contribution of independent (nonoverlapping) sampling information.
\cite{bom2020generalized} employ Monte Carlo simulations to quantify the efficiency gains of the GW meta-estimator relative to conventional meta-estimators and to demonstrate how the GW meta-estimator brings the false positive rate closer to its nominal level.
The GW meta-estimator is a simple and versatile tool designed to handle all types of sample overlap, whether within a study or across studies. It can accommodate complex cases without the need for additional estimation, provided that the necessary information to approximate the variance-covariance matrix of the estimates can be extracted from the primary studies, which is typically available in most applications.

The GW estimator provides several significant advantages over cluster-robust methods and multilevel models in meta-analysis, particularly when dealing with sample overlap. Unlike cluster-robust methods, which require defining disjoint clusters of correlated (overlapping) estimates and assuming a constant within-cluster correlation, the GW estimator addresses these challenges by utilizing exogenously determined overlap information. This method provides a more flexible way to correct for overlap-induced dependencies, especially when clusters are small, unbalanced, or irregular in size--situations in which cluster-robust methods tend to perform poorly. While multilevel models enhance efficiency by capturing hierarchical dependencies, they are limited by the need to define a nested structure and estimate numerous parameters, which can be technically demanding and impractical in complex overlap scenarios. In contrast, the GW estimator simplifies the process by eliminating the need for joint parameter estimation. Furthermore, the GW method addresses both inference and efficiency concerns by effectively utilizing overlap information, making it especially useful for complex and unbalanced overlap structures. Although it does not directly address all types of dependency, it can complement other methods, including multilevel and cluster-robust models, in meta-analytic applications \citep{bom2020generalized,bom2024accounting}.

Sample overlap is common in economics meta-analysis and can give rise to different structures of estimate dependence, ranging from balanced to unbalanced, and simple to complex. If not properly addressed, sample overlap can result in efficiency losses and inflated false positive rates in meta-analyses. To tackle these issues, \cite{bom2024accounting} demonstrate the application of the GW method in meta-analytic research within economics, addressing common practical challenges such as variations in data aggregation levels, estimation techniques, and effect size metrics. They derive explicit covariance formulas for various overlapping scenarios, assess the accuracy of the covariance approximations, and employ Monte Carlo simulations to demonstrate how the method improves efficiency and reduces false positive rates. The GW estimator is versatile and applicable to both simple meta-analysis models and meta-regression models.
\cite{bom2024accounting} have developed Stata and R codes that fully automate the implementation of the GW estimator in meta-analyses, particularly in cases involving sample overlap. These software codes, along with detailed instructions, are freely available for download at \href{https://osf.io/g5t2j}{https://osf.io/g5t2j}.

\section{Concluding Remarks}\label{conclusion}
Meta-analysis, a specific type of systematic review, is a robust statistical method used to combine quantitative findings from multiple studies, providing an estimate of the overall or average effect size for a specific outcome of interest. The direction and magnitude of the estimated effect, along with its confidence interval and hypothesis test results, provide valuable insights into the phenomenon or relationship under investigation.

As an essential tool for synthesizing empirical evidence across various scientific disciplines, including economics, meta-analysis plays a pivotal role in addressing the challenges posed by the rapid growth of research output. Individual studies often yield conflicting results due to variations in data sources, methodologies, and contextual factors. By systematically combining these results, meta-analysis bridges such gaps and offers a more precise and generalized understanding of economic phenomena. Its structured approach enhances statistical power, refines effect size estimates, and supports evidence-based decision-making.

A key strength of meta-analysis lies in its ability to address between-study heterogeneity---the differences in true effect sizes across studies. By employing meta-regression analysis, researchers can incorporate moderators representing study characteristics, thereby identifying the underlying sources of variability. This nuanced approach allows economists to draw more accurate conclusions about the factors influencing economic relationships. For example, it facilitates the investigation of how varying labor market conditions shape the effectiveness of training programs, or how policy interventions produce differential impacts across regions and contexts.

Publication bias, a pervasive challenge in empirical research, undermines the validity of meta-analytic findings by favoring studies with statistically significant results and strong theoretical justification. This bias skews the available evidence, inflating effect size estimates and potentially misleading conclusions. The selective publication of favorable outcomes creates a misleading picture of the research landscape, impacting the reliability of synthesized results. Techniques such as funnel plots, statistical tests, the EK meta-regression model, the Top 10 approach, the p-curve method, the p-uniform and $\text{p-uniform}^*$ methods, the WAAP method, conditional publication probability, and Bayesian model averaging approaches have been developed to detect and correct for publication selection effects, enhancing the credibility of meta-analytic findings.

Effect size dependence, another critical issue in meta-analysis, refers to the statistical interdependence of effect sizes reported across or within studies, which violates the assumption of independence in traditional meta-analytic methods. Failing to account for this dependence can lead to distorted conclusions. In economic research, effect size dependence often arises due to factors such as multiple effect sizes within a single study or between-study sample overlap. Advanced techniques, such as robust variance estimation (RVE), the multilevel model, and the generalized weights (GW) meta-estimator, account for the dependence among primary estimates, ensuring that the synthesized results more accurately reflect the true relationships under investigation.

This study provides a comprehensive review of key tools and techniques for applying meta-analysis in economics, with a particular emphasis on addressing critical challenges such as publication bias, between-study heterogeneity, and effect size dependence. It also discusses the strengths and weaknesses of various methods. In conclusion, the study equips researchers with the necessary knowledge to apply these methods effectively, thereby enhancing the reliability of empirical evidence and supporting evidence-based policy development.

\section*{References}
\bibliography{references}

\begin{thebibliography}{216}
\expandafter\ifx\csname natexlab\endcsname\relax\def\natexlab#1{#1}\fi
\providecommand{\url}[1]{\texttt{#1}}
\providecommand{\href}[2]{#2}
\providecommand{\path}[1]{#1}
\providecommand{\DOIprefix}{doi:}
\providecommand{\ArXivprefix}{arXiv:}
\providecommand{\URLprefix}{URL: }
\providecommand{\Pubmedprefix}{pmid:}
\providecommand{\doi}[1]{\href{http://dx.doi.org/#1}{\path{#1}}}
\providecommand{\Pubmed}[1]{\href{pmid:#1}{\path{#1}}}
\providecommand{\bibinfo}[2]{#2}
\ifx\xfnm\relax \def\xfnm[#1]{\unskip,\space#1}\fi
\bibitem[{van Aert \& van Assen(2021)}]{van2021correcting}
\bibinfo{author}{van Aert, R.~C.}, \& \bibinfo{author}{van Assen, M.~A.}
  (\bibinfo{year}{2021}).
\newblock \bibinfo{title}{Correcting for publication bias in a meta-analysis
  with the p-uniform* method}.
\newblock \bibinfo{note}{Working paper}.
\bibitem[{van Aert \& Jackson(2018)}]{van2018multistep}
\bibinfo{author}{van Aert, R.~C.}, \& \bibinfo{author}{Jackson, D.}
  (\bibinfo{year}{2018}).
\newblock \bibinfo{title}{Multistep estimators of the between-study variance:
  {The} relationship with the {Paule-Mandel }estimator}.
\newblock {\it \bibinfo{journal}{Statistics in Medicine}\/},  {\it
  \bibinfo{volume}{37}\/}, \bibinfo{pages}{2616--2629}.
\bibitem[{van Aert \& Wicherts(2024)}]{vaert2024correcting}
\bibinfo{author}{van Aert, R. C.~M.}, \& \bibinfo{author}{Wicherts, J.~M.}
  (\bibinfo{year}{2024}).
\newblock \bibinfo{title}{Correcting for outcome reporting bias in a
  meta-analysis: {A} meta-regression approach}.
\newblock {\it \bibinfo{journal}{Behavior Research Methods}\/},  {\it
  \bibinfo{volume}{56}\/}, \bibinfo{pages}{1994--2012}.
\bibitem[{Alexander et~al.(1989)Alexander, Scozzaro \&
  Borodkin}]{alexander1989statistical}
\bibinfo{author}{Alexander, R.~A.}, \bibinfo{author}{Scozzaro, M.~J.}, \&
  \bibinfo{author}{Borodkin, L.~J.} (\bibinfo{year}{1989}).
\newblock \bibinfo{title}{Statistical and empirical examination of the
  chi-square test for homogeneity of correlations in meta-analysis}.
\newblock {\it \bibinfo{journal}{Psychological Bulletin}\/},  {\it
  \bibinfo{volume}{106}\/}, \bibinfo{pages}{329--331}.
\bibitem[{Andrews \& Kasy(2019)}]{andrews2019identification}
\bibinfo{author}{Andrews, I.}, \& \bibinfo{author}{Kasy, M.}
  (\bibinfo{year}{2019}).
\newblock \bibinfo{title}{Identification of and correction for publication
  bias}.
\newblock {\it \bibinfo{journal}{American Economic Review}\/},  {\it
  \bibinfo{volume}{109}\/}, \bibinfo{pages}{2766--2794}.
\bibitem[{Ashenfelter et~al.(1999)Ashenfelter, Harmon \&
  Oosterbeek}]{ashenfelter1999review}
\bibinfo{author}{Ashenfelter, O.}, \bibinfo{author}{Harmon, C.}, \&
  \bibinfo{author}{Oosterbeek, H.} (\bibinfo{year}{1999}).
\newblock \bibinfo{title}{A review of estimates of the schooling/earnings
  relationship, with tests for publication bias}.
\newblock {\it \bibinfo{journal}{Labour Economics}\/},  {\it
  \bibinfo{volume}{6}\/}, \bibinfo{pages}{453--470}.
\bibitem[{van Assen et~al.(2015)van Assen, van Aert \&
  Wicherts}]{vanassen2015meta}
\bibinfo{author}{van Assen, M. A. L.~M.}, \bibinfo{author}{van Aert, R. C.~M.},
  \& \bibinfo{author}{Wicherts, J.~M.} (\bibinfo{year}{2015}).
\newblock \bibinfo{title}{Meta-analysis using effect size distributions of only
  statistically significant studies}.
\newblock {\it \bibinfo{journal}{Psychological Methods}\/},  {\it
  \bibinfo{volume}{20}\/}, \bibinfo{pages}{293--309}.
\bibitem[{Bartoš et~al.(2023)Bartoš, Maier, Wagenmakers, Doucouliagos \&
  Stanley}]{bartos2023robust}
\bibinfo{author}{Bartoš, F.}, \bibinfo{author}{Maier, M.},
  \bibinfo{author}{Wagenmakers, E.-J.}, \bibinfo{author}{Doucouliagos, H.}, \&
  \bibinfo{author}{Stanley, T.~D.} (\bibinfo{year}{2023}).
\newblock \bibinfo{title}{Robust bayesian meta-analysis: {Model-averaging}
  across complementary publication bias adjustment methods}.
\newblock {\it \bibinfo{journal}{Research Synthesis Methods}\/},  {\it
  \bibinfo{volume}{14}\/}, \bibinfo{pages}{99--116}.
\bibitem[{Becker(2000)}]{becker2000multivariate}
\bibinfo{author}{Becker, B.~J.} (\bibinfo{year}{2000}).
\newblock \bibinfo{title}{Multivariate meta-analysis}.
\newblock In \bibinfo{editor}{H.~E. Tinsley}, \& \bibinfo{editor}{S.~D. Brown}
  (Eds.), {\it \bibinfo{booktitle}{Handbook of applied multivariate statistics
  and mathematical modeling}\/} chapter~\bibinfo{chapter}{17}. (pp.
  \bibinfo{pages}{499--525}).
\newblock \bibinfo{publisher}{Elsevier}. (\bibinfo{edition}{1st} ed.).
\bibitem[{Begg \& Berlin(1988)}]{begg1988publication}
\bibinfo{author}{Begg, C.~B.}, \& \bibinfo{author}{Berlin, J.~A.}
  (\bibinfo{year}{1988}).
\newblock \bibinfo{title}{Publication bias: {A} problem in interpreting medical
  data}.
\newblock {\it \bibinfo{journal}{Journal of the Royal Statistical Society.
  Series A (Statistics in Society)}\/},  {\it \bibinfo{volume}{151}\/},
  \bibinfo{pages}{419--463}.
\bibitem[{Begg \& Berlin(1989)}]{begg1989publication}
\bibinfo{author}{Begg, C.~B.}, \& \bibinfo{author}{Berlin, J.~A.}
  (\bibinfo{year}{1989}).
\newblock \bibinfo{title}{Publication bias and dissemination of clinical
  research}.
\newblock {\it \bibinfo{journal}{JNCI: Journal of the National Cancer
  Institute}\/},  {\it \bibinfo{volume}{81}\/}, \bibinfo{pages}{107--115}.
\bibitem[{Berkey et~al.(1995)Berkey, Hoaglin, Mosteller \&
  Colditz}]{berkey1995random}
\bibinfo{author}{Berkey, C.~S.}, \bibinfo{author}{Hoaglin, D.~C.},
  \bibinfo{author}{Mosteller, F.}, \& \bibinfo{author}{Colditz, G.~A.}
  (\bibinfo{year}{1995}).
\newblock \bibinfo{title}{A random-effects regression model for meta-analysis}.
\newblock {\it \bibinfo{journal}{Statistics in Medicine}\/},  {\it
  \bibinfo{volume}{14}\/}, \bibinfo{pages}{395--411}.
\bibitem[{Berkhof \& Snijders(2001)}]{berkhof2001variance}
\bibinfo{author}{Berkhof, J.}, \& \bibinfo{author}{Snijders, T.~A.}
  (\bibinfo{year}{2001}).
\newblock \bibinfo{title}{Variance component testing in multilevel models}.
\newblock {\it \bibinfo{journal}{Journal of Educational and Behavioral
  Statistics}\/},  {\it \bibinfo{volume}{26}\/}, \bibinfo{pages}{133--152}.
\bibitem[{Biggerstaff \& Tweedie(1997)}]{biggerstaff1997incorporating}
\bibinfo{author}{Biggerstaff, B.}, \& \bibinfo{author}{Tweedie, R.}
  (\bibinfo{year}{1997}).
\newblock \bibinfo{title}{Incorporating variability in estimates of
  heterogeneity in the random effects model in meta-analysis}.
\newblock {\it \bibinfo{journal}{Statistics in Medicine}\/},  {\it
  \bibinfo{volume}{16}\/}, \bibinfo{pages}{753--768}.
\bibitem[{Bodnar et~al.(2017)Bodnar, Link, Arendack{\'a}, Possolo \&
  Elster}]{bodnar2017bayesian}
\bibinfo{author}{Bodnar, O.}, \bibinfo{author}{Link, A.},
  \bibinfo{author}{Arendack{\'a}, B.}, \bibinfo{author}{Possolo, A.}, \&
  \bibinfo{author}{Elster, C.} (\bibinfo{year}{2017}).
\newblock \bibinfo{title}{Bayesian estimation in random effects meta-analysis
  using a non-informative prior}.
\newblock {\it \bibinfo{journal}{Statistics in Medicine}\/},  {\it
  \bibinfo{volume}{36}\/}, \bibinfo{pages}{378--399}.
\bibitem[{Bom \& Rachinger(2019)}]{bom2019kinked}
\bibinfo{author}{Bom, P. R.~D.}, \& \bibinfo{author}{Rachinger, H.}
  (\bibinfo{year}{2019}).
\newblock \bibinfo{title}{A kinked meta-regression model for publication bias
  correction}.
\newblock {\it \bibinfo{journal}{Research Synthesis Methods}\/},  {\it
  \bibinfo{volume}{10}\/}, \bibinfo{pages}{497--514}.
\bibitem[{Bom \& Rachinger(2020)}]{bom2020generalized}
\bibinfo{author}{Bom, P. R.~D.}, \& \bibinfo{author}{Rachinger, H.}
  (\bibinfo{year}{2020}).
\newblock \bibinfo{title}{A generalized‐weights solution to sample overlap in
  meta‐analysis}.
\newblock {\it \bibinfo{journal}{Research Synthesis Methods}\/},  {\it
  \bibinfo{volume}{11}\/}, \bibinfo{pages}{812--832}.
  \DOIprefix\doi{10.1002/jrsm.1411}.
\bibitem[{Bom \& Rachinger(2024)}]{bom2024accounting}
\bibinfo{author}{Bom, P. R.~D.}, \& \bibinfo{author}{Rachinger, H.}
  (\bibinfo{year}{2024}).
\newblock \bibinfo{title}{Accounting for sample overlap in economics
  meta‐analyses: {The} generalized‐weights method in practice}.
\newblock {\it \bibinfo{journal}{Journal of Economic Surveys}\/}, .
  \DOIprefix\doi{10.1111/joes.12633}.
\newblock \bibinfo{note}{Advance online publication}.
\bibitem[{Borenstein et~al.(2021)Borenstein, Hedges, Higgins \&
  Rothstein}]{borenstein2021introduction}
\bibinfo{author}{Borenstein, M.}, \bibinfo{author}{Hedges, L.~V.},
  \bibinfo{author}{Higgins, J.~P.}, \& \bibinfo{author}{Rothstein, H.~R.}
  (\bibinfo{year}{2021}).
\newblock {\it \bibinfo{title}{Introduction to Meta-Analysis}\/}.
\newblock (\bibinfo{edition}{2nd} ed.).
\newblock \bibinfo{publisher}{John Wiley \& Sons}.
\bibitem[{Borenstein et~al.(2017)Borenstein, Higgins, Hedges \&
  Rothstein}]{borenstein2017basics}
\bibinfo{author}{Borenstein, M.}, \bibinfo{author}{Higgins, J.~P.},
  \bibinfo{author}{Hedges, L.~V.}, \& \bibinfo{author}{Rothstein, H.~R.}
  (\bibinfo{year}{2017}).
\newblock \bibinfo{title}{Basics of meta-analysis: {I2} is not an absolute
  measure of heterogeneity}.
\newblock {\it \bibinfo{journal}{Research Synthesis Methods}\/},  {\it
  \bibinfo{volume}{8}\/}, \bibinfo{pages}{5--18}.
\bibitem[{Brockwell \& Gordon(2001)}]{brockwell2001comparison}
\bibinfo{author}{Brockwell, S.~E.}, \& \bibinfo{author}{Gordon, I.~R.}
  (\bibinfo{year}{2001}).
\newblock \bibinfo{title}{A comparison of statistical methods for
  meta-analysis}.
\newblock {\it \bibinfo{journal}{Statistics in Medicine}\/},  {\it
  \bibinfo{volume}{20}\/}, \bibinfo{pages}{825--840}.
\bibitem[{Brockwell \& Gordon(2007)}]{brockwell2007simple}
\bibinfo{author}{Brockwell, S.~E.}, \& \bibinfo{author}{Gordon, I.~R.}
  (\bibinfo{year}{2007}).
\newblock \bibinfo{title}{A simple method for inference on an overall effect in
  meta-analysis}.
\newblock {\it \bibinfo{journal}{Statistics in Medicine}\/},  {\it
  \bibinfo{volume}{26}\/}, \bibinfo{pages}{4531--4543}.
\bibitem[{Callot \& Paldam(2011)}]{callot2011problem}
\bibinfo{author}{Callot, L.}, \& \bibinfo{author}{Paldam, M.}
  (\bibinfo{year}{2011}).
\newblock \bibinfo{title}{The problem of natural funnel asymmetries: {A}
  simulation analysis of meta-analysis in macroeconomics}.
\newblock {\it \bibinfo{journal}{Research Synthesis Methods}\/},  {\it
  \bibinfo{volume}{2}\/}, \bibinfo{pages}{84--102}.
\bibitem[{Campolieti(2020)}]{campolieti2020does}
\bibinfo{author}{Campolieti, M.} (\bibinfo{year}{2020}).
\newblock \bibinfo{title}{Does an increase in the minimum wage decrease
  employment? {A} meta-analysis of {Canadian} studies}.
\newblock {\it \bibinfo{journal}{Canadian Public Policy}\/},  {\it
  \bibinfo{volume}{46}\/}, \bibinfo{pages}{531--564}.
\bibitem[{Card et~al.(2010)Card, Kluve \& Weber}]{card2010active}
\bibinfo{author}{Card, D.}, \bibinfo{author}{Kluve, J.}, \&
  \bibinfo{author}{Weber, A.} (\bibinfo{year}{2010}).
\newblock \bibinfo{title}{Active labour market policy evaluations: {A}
  meta-analysis}.
\newblock {\it \bibinfo{journal}{The Economic Journal}\/},  {\it
  \bibinfo{volume}{120}\/}, \bibinfo{pages}{F452--F477}.
\bibitem[{Card et~al.(2018)Card, Kluve \& Weber}]{card2018works}
\bibinfo{author}{Card, D.}, \bibinfo{author}{Kluve, J.}, \&
  \bibinfo{author}{Weber, A.} (\bibinfo{year}{2018}).
\newblock \bibinfo{title}{What works? {A} meta analysis of recent active labor
  market program evaluations}.
\newblock {\it \bibinfo{journal}{Journal of the European Economic
  Association}\/},  {\it \bibinfo{volume}{16}\/}, \bibinfo{pages}{894--931}.
\bibitem[{Card \& Krueger(1995)}]{card1995time}
\bibinfo{author}{Card, D.}, \& \bibinfo{author}{Krueger, A.~B.}
  (\bibinfo{year}{1995}).
\newblock \bibinfo{title}{Time-series minimum-wage studies: {A} meta-analysis}.
\newblock {\it \bibinfo{journal}{The American Economic Review}\/},  {\it
  \bibinfo{volume}{85}\/}, \bibinfo{pages}{238--243}.
\bibitem[{Chan et~al.(2004{\natexlab{a}})Chan, Hr{\'o}bjartsson, Haahr,
  G{\o}tzsche \& Altman}]{chan2004empirical}
\bibinfo{author}{Chan, A.-W.}, \bibinfo{author}{Hr{\'o}bjartsson, A.},
  \bibinfo{author}{Haahr, M.~T.}, \bibinfo{author}{G{\o}tzsche, P.~C.}, \&
  \bibinfo{author}{Altman, D.~G.} (\bibinfo{year}{2004}{\natexlab{a}}).
\newblock \bibinfo{title}{Empirical evidence for selective reporting of
  outcomes in randomized trials: {Comparison} of protocols to published
  articles}.
\newblock {\it \bibinfo{journal}{Jama}\/},  {\it \bibinfo{volume}{291}\/},
  \bibinfo{pages}{2457--2465}.
\bibitem[{Chan et~al.(2004{\natexlab{b}})Chan, Krle{\v{z}}a-Jeri{\'c}, Schmid
  \& Altman}]{chan2004outcome}
\bibinfo{author}{Chan, A.-W.}, \bibinfo{author}{Krle{\v{z}}a-Jeri{\'c}, K.},
  \bibinfo{author}{Schmid, I.}, \& \bibinfo{author}{Altman, D.~G.}
  (\bibinfo{year}{2004}{\natexlab{b}}).
\newblock \bibinfo{title}{Outcome reporting bias in randomized trials funded by
  the {Canadian Institutes of Health Research}}.
\newblock {\it \bibinfo{journal}{Canadian Medical Association Journal}\/},
  {\it \bibinfo{volume}{171}\/}, \bibinfo{pages}{735--740}.
\bibitem[{Cheung(2013)}]{cheung2013multivariate}
\bibinfo{author}{Cheung, M. W.-L.} (\bibinfo{year}{2013}).
\newblock \bibinfo{title}{Multivariate meta-analysis as structural equation
  models}.
\newblock {\it \bibinfo{journal}{Structural Equation Modeling: A
  Multidisciplinary Journal}\/},  {\it \bibinfo{volume}{20}\/},
  \bibinfo{pages}{429--454}.
\bibitem[{Cheung(2014)}]{cheung2014modeling}
\bibinfo{author}{Cheung, M. W.-L.} (\bibinfo{year}{2014}).
\newblock \bibinfo{title}{Modeling dependent effect sizes with three-level
  meta-analyses: {A} structural equation modeling approach}.
\newblock {\it \bibinfo{journal}{Psychological Methods}\/},  {\it
  \bibinfo{volume}{19}\/}, \bibinfo{pages}{211}.
\bibitem[{Cheung(2015{\natexlab{a}})}]{cheung2015meta}
\bibinfo{author}{Cheung, M. W.-L.} (\bibinfo{year}{2015}{\natexlab{a}}).
\newblock {\it \bibinfo{title}{Meta-analysis: {A} structural equation modeling
  approach}\/}.
\newblock \bibinfo{publisher}{John Wiley \& Sons}.
\bibitem[{Cheung(2015{\natexlab{b}})}]{cheung2015metasem}
\bibinfo{author}{Cheung, M. W.-L.} (\bibinfo{year}{2015}{\natexlab{b}}).
\newblock \bibinfo{title}{{metaSEM}: {An R} package for meta-analysis using
  structural equation modeling}.
\newblock {\it \bibinfo{journal}{Frontiers in Psychology}\/},  {\it
  \bibinfo{volume}{5}\/}. \URLprefix
  \url{http://journal.frontiersin.org/article/10.3389/fpsyg.2014.01521/full}.
  \DOIprefix\doi{10.3389/fpsyg.2014.01521}.
\bibitem[{Cheung \& Vijayakumar(2016)}]{cheung2016guide}
\bibinfo{author}{Cheung, M. W.-L.}, \& \bibinfo{author}{Vijayakumar, R.}
  (\bibinfo{year}{2016}).
\newblock \bibinfo{title}{A guide to conducting a meta-analysis}.
\newblock {\it \bibinfo{journal}{Neuropsychology Review}\/},  {\it
  \bibinfo{volume}{26}\/}, \bibinfo{pages}{121--128}.
\bibitem[{Chung et~al.(2013)Chung, Rabe-Hesketh \& Choi}]{chung2013avoiding}
\bibinfo{author}{Chung, Y.}, \bibinfo{author}{Rabe-Hesketh, S.}, \&
  \bibinfo{author}{Choi, I.-H.} (\bibinfo{year}{2013}).
\newblock \bibinfo{title}{Avoiding zero between-study variance estimates in
  random-effects meta-analysis}.
\newblock {\it \bibinfo{journal}{Statistics in Medicine}\/},  {\it
  \bibinfo{volume}{32}\/}, \bibinfo{pages}{4071--4089}.
\bibitem[{Cooper(1998)}]{cooper1998synthesizing}
\bibinfo{author}{Cooper, H.~M.} (\bibinfo{year}{1998}).
\newblock {\it \bibinfo{title}{Synthesizing research: {A} guide for literature
  reviews}\/} volume~\bibinfo{volume}{2}.
\newblock (\bibinfo{edition}{3rd} ed.).
\newblock \bibinfo{publisher}{SAGE Publications, Inc}.
\bibitem[{Copas et~al.(2013)Copas, Dwan, Kirkham \&
  Williamson}]{copas2013model}
\bibinfo{author}{Copas, J.}, \bibinfo{author}{Dwan, K.},
  \bibinfo{author}{Kirkham, J.}, \& \bibinfo{author}{Williamson, P.}
  (\bibinfo{year}{2013}).
\newblock \bibinfo{title}{A model-based correction for outcome reporting bias
  in meta-analysis}.
\newblock {\it \bibinfo{journal}{Biostatistics}\/},  {\it
  \bibinfo{volume}{15}\/}, \bibinfo{pages}{370--383}.
\bibitem[{Cornell et~al.(2014)Cornell, Mulrow, Localio, Stack, Meibohm, Guallar
  \& Goodman}]{cornell2014random}
\bibinfo{author}{Cornell, J.~E.}, \bibinfo{author}{Mulrow, C.~D.},
  \bibinfo{author}{Localio, R.}, \bibinfo{author}{Stack, C.~B.},
  \bibinfo{author}{Meibohm, A.~R.}, \bibinfo{author}{Guallar, E.}, \&
  \bibinfo{author}{Goodman, S.~N.} (\bibinfo{year}{2014}).
\newblock \bibinfo{title}{Random-effects meta-analysis of inconsistent effects:
  {A} time for change}.
\newblock {\it \bibinfo{journal}{Annals of Internal Medicine}\/},  {\it
  \bibinfo{volume}{160}\/}, \bibinfo{pages}{267--270}.
\bibitem[{Deeks et~al.(2008)Deeks, Higgins \& Altman}]{deeks2008analysing}
\bibinfo{author}{Deeks, J.~J.}, \bibinfo{author}{Higgins, J.~P.}, \&
  \bibinfo{author}{Altman, D.~G.} (\bibinfo{year}{2008}).
\newblock \bibinfo{title}{Analysing data and undertaking meta-analyses}.
\newblock In \bibinfo{editor}{J.~P. Higgins}, \& \bibinfo{editor}{S.~Green}
  (Eds.), {\it \bibinfo{booktitle}{Cochrane handbook for systematic reviews of
  interventions: {Cochrane} book series}\/} chapter~\bibinfo{chapter}{9}. (pp.
  \bibinfo{pages}{243--296}).
\newblock \bibinfo{publisher}{Wiley Online Library}.
\bibitem[{DerSimonian \& Laird(1986)}]{dersimonian1986meta}
\bibinfo{author}{DerSimonian, R.}, \& \bibinfo{author}{Laird, N.}
  (\bibinfo{year}{1986}).
\newblock \bibinfo{title}{Meta-analysis in clinical trials}.
\newblock {\it \bibinfo{journal}{Controlled Clinical Trials}\/},  {\it
  \bibinfo{volume}{7}\/}, \bibinfo{pages}{177--188}.
\bibitem[{Doucouliagos(2005)}]{doucouliagos2005publication2}
\bibinfo{author}{Doucouliagos, C.} (\bibinfo{year}{2005}).
\newblock \bibinfo{title}{Publication bias in the economic freedom and economic
  growth literature}.
\newblock {\it \bibinfo{journal}{Journal of Economic Surveys}\/},  {\it
  \bibinfo{volume}{19}\/}, \bibinfo{pages}{367--387}.
\bibitem[{Doucouliagos \& Stanley(2013)}]{doucouliagos2013all}
\bibinfo{author}{Doucouliagos, C.}, \& \bibinfo{author}{Stanley, T.~D.}
  (\bibinfo{year}{2013}).
\newblock \bibinfo{title}{Are all economic facts greatly exaggerated? {Theory}
  competition and selectivity}.
\newblock {\it \bibinfo{journal}{Journal of Economic Surveys}\/},  {\it
  \bibinfo{volume}{27}\/}, \bibinfo{pages}{316--339}.
\bibitem[{Doucouliagos et~al.(2012)Doucouliagos, Stanley \&
  Giles}]{doucouliagos2012estimates}
\bibinfo{author}{Doucouliagos, C.}, \bibinfo{author}{Stanley, T.~D.}, \&
  \bibinfo{author}{Giles, M.} (\bibinfo{year}{2012}).
\newblock \bibinfo{title}{Are estimates of the value of a statistical life
  exaggerated?}
\newblock {\it \bibinfo{journal}{Journal of Health Economics}\/},  {\it
  \bibinfo{volume}{31}\/}, \bibinfo{pages}{197--206}.
\bibitem[{Doucouliagos et~al.(2018)Doucouliagos, Freeman, Laroche \&
  Stanley}]{doucouliagos2018credible}
\bibinfo{author}{Doucouliagos, H.}, \bibinfo{author}{Freeman, R.~B.},
  \bibinfo{author}{Laroche, P.}, \& \bibinfo{author}{Stanley, T.}
  (\bibinfo{year}{2018}).
\newblock \bibinfo{title}{How credible is trade union research? {Forty} years
  of evidence on the monopoly--voice trade-off}.
\newblock {\it \bibinfo{journal}{ILR Review}\/},  {\it \bibinfo{volume}{71}\/},
  \bibinfo{pages}{287--305}.
\bibitem[{Doucouliagos et~al.(2005)Doucouliagos, Laroche \&
  Stanley}]{doucouliagos2005publication1}
\bibinfo{author}{Doucouliagos, H.}, \bibinfo{author}{Laroche, P.}, \&
  \bibinfo{author}{Stanley, T.~D.} (\bibinfo{year}{2005}).
\newblock \bibinfo{title}{Publication bias in union-productivity research?}
\newblock {\it \bibinfo{journal}{Relations Industrielles/Industrial
  Relations}\/},  {\it \bibinfo{volume}{60}\/}, \bibinfo{pages}{320--347}.
\bibitem[{Doucouliagos \& Stanley(2009)}]{doucouliagos2009publication}
\bibinfo{author}{Doucouliagos, H.}, \& \bibinfo{author}{Stanley, T.~D.}
  (\bibinfo{year}{2009}).
\newblock \bibinfo{title}{Publication selection bias in minimum-wage research?
  {A} meta-regression analysis}.
\newblock {\it \bibinfo{journal}{British Journal of Industrial Relations}\/},
  {\it \bibinfo{volume}{47}\/}, \bibinfo{pages}{406--428}.
\bibitem[{Dwan et~al.(2008)Dwan, Altman, Arnaiz, Bloom, Chan, Cronin,
  Decullier, Easterbrook, Von~Elm, Gamble, Ghersi, Ioannidis, Simes \&
  R}]{dwan2008systematic}
\bibinfo{author}{Dwan, K.}, \bibinfo{author}{Altman, D.~G.},
  \bibinfo{author}{Arnaiz, J.~A.}, \bibinfo{author}{Bloom, J.},
  \bibinfo{author}{Chan, A.-W.}, \bibinfo{author}{Cronin, E.},
  \bibinfo{author}{Decullier, E.}, \bibinfo{author}{Easterbrook, P.~J.},
  \bibinfo{author}{Von~Elm, E.}, \bibinfo{author}{Gamble, C.},
  \bibinfo{author}{Ghersi, D.}, \bibinfo{author}{Ioannidis, J. P.~A.},
  \bibinfo{author}{Simes, J.}, \& \bibinfo{author}{R, W.~P.}
  (\bibinfo{year}{2008}).
\newblock \bibinfo{title}{Systematic review of the empirical evidence of study
  publication bias and outcome reporting bias}.
\newblock {\it \bibinfo{journal}{PLoS One}\/},  {\it \bibinfo{volume}{3}\/},
  \bibinfo{pages}{e3081}.
\bibitem[{Easterbrook et~al.(1991)Easterbrook, Gopalan, Berlin \&
  Matthews}]{easterbrook1991publication}
\bibinfo{author}{Easterbrook, P.~J.}, \bibinfo{author}{Gopalan, R.},
  \bibinfo{author}{Berlin, J.}, \& \bibinfo{author}{Matthews, D.~R.}
  (\bibinfo{year}{1991}).
\newblock \bibinfo{title}{Publication bias in clinical research}.
\newblock {\it \bibinfo{journal}{The Lancet}\/},  {\it
  \bibinfo{volume}{337}\/}, \bibinfo{pages}{867--872}.
\bibitem[{Egger \& Smith(1998)}]{egger1998bias}
\bibinfo{author}{Egger, M.}, \& \bibinfo{author}{Smith, G.~D.}
  (\bibinfo{year}{1998}).
\newblock \bibinfo{title}{Bias in location and selection of studies}.
\newblock {\it \bibinfo{journal}{BMJ: British Medical Journal}\/},  {\it
  \bibinfo{volume}{316}\/}, \bibinfo{pages}{61--66}.
\bibitem[{Egger et~al.(1997{\natexlab{a}})Egger, Smith, Schneider \&
  Minder}]{egger1997bias}
\bibinfo{author}{Egger, M.}, \bibinfo{author}{Smith, G.~D.},
  \bibinfo{author}{Schneider, M.}, \& \bibinfo{author}{Minder, C.}
  (\bibinfo{year}{1997}{\natexlab{a}}).
\newblock \bibinfo{title}{Bias in meta-analysis detected by a simple, graphical
  test}.
\newblock {\it \bibinfo{journal}{Bmj: British Medical Journal}\/},  {\it
  \bibinfo{volume}{315}\/}, \bibinfo{pages}{629--634}.
\bibitem[{Egger et~al.(1997{\natexlab{b}})Egger, Zellweger-Z{\"a}hner,
  Schneider, Junker, Lengeler \& Antes}]{egger1997language}
\bibinfo{author}{Egger, M.}, \bibinfo{author}{Zellweger-Z{\"a}hner, T.},
  \bibinfo{author}{Schneider, M.}, \bibinfo{author}{Junker, C.},
  \bibinfo{author}{Lengeler, C.}, \& \bibinfo{author}{Antes, G.}
  (\bibinfo{year}{1997}{\natexlab{b}}).
\newblock \bibinfo{title}{Language bias in randomised controlled trials
  published in {English} and {German}}.
\newblock {\it \bibinfo{journal}{The Lancet}\/},  {\it
  \bibinfo{volume}{350}\/}, \bibinfo{pages}{326--329}.
\bibitem[{Elwood et~al.(1974)Elwood, Cochrane, Burr, Sweetnam, Williams,
  Welsby, Hughes \& Renton}]{elwood1974randomized}
\bibinfo{author}{Elwood, P.~C.}, \bibinfo{author}{Cochrane, A.~L.},
  \bibinfo{author}{Burr, M.~L.}, \bibinfo{author}{Sweetnam, P.},
  \bibinfo{author}{Williams, G.}, \bibinfo{author}{Welsby, E.},
  \bibinfo{author}{Hughes, S.}, \& \bibinfo{author}{Renton, R.}
  (\bibinfo{year}{1974}).
\newblock \bibinfo{title}{A randomized controlled trial of acetyl salicyclic
  acid in the secondary prevention of mortality from myocardial infarction}.
\newblock {\it \bibinfo{journal}{British Medical Journal}\/},  {\it
  \bibinfo{volume}{1}\/}, \bibinfo{pages}{436--440}.
\bibitem[{Fisher(1935)}]{Fisher1935design}
\bibinfo{author}{Fisher, R.~A.} (\bibinfo{year}{1935}).
\newblock {\it \bibinfo{title}{The Design of Experiments}\/}.
\newblock \bibinfo{publisher}{Oliver and Boyd}.
\bibitem[{Fisher \& Tipton(2015)}]{fisher2015robumeta}
\bibinfo{author}{Fisher, Z.}, \& \bibinfo{author}{Tipton, E.}
  (\bibinfo{year}{2015}).
\newblock \bibinfo{title}{robumeta: {An R-package} for robust variance
  estimation in meta-analysis}.
\newblock \bibinfo{note}{\url{https://arxiv.org/abs/1503.02220v1}}.
\bibitem[{Fisher et~al.(2017)Fisher, Tipton \& Zhipeng}]{fisher2017package}
\bibinfo{author}{Fisher, Z.}, \bibinfo{author}{Tipton, E.}, \&
  \bibinfo{author}{Zhipeng, H.} (\bibinfo{year}{2017}).
\newblock \bibinfo{title}{Package \enquote{robumeta}: {Robust} variance
  meta-regression}.
\newblock {\it \bibinfo{journal}{R package version 2.0}\/}, .
\newblock
  \bibinfo{note}{\url{https://cran.r-project.org/web/packages/robumeta/index.html}}.
\bibitem[{Follmann \& Proschan(1999)}]{follmann1999valid}
\bibinfo{author}{Follmann, D.~A.}, \& \bibinfo{author}{Proschan, M.~A.}
  (\bibinfo{year}{1999}).
\newblock \bibinfo{title}{Valid inference in random effects meta-analysis}.
\newblock {\it \bibinfo{journal}{Biometrics}\/},  {\it \bibinfo{volume}{55}\/},
  \bibinfo{pages}{732--737}.
\bibitem[{Galbraith(1988)}]{galbraith1988note}
\bibinfo{author}{Galbraith, R.} (\bibinfo{year}{1988}).
\newblock \bibinfo{title}{A note on graphical presentation of estimated odds
  ratios from several clinical trials}.
\newblock {\it \bibinfo{journal}{Statistics in Medicine}\/},  {\it
  \bibinfo{volume}{7}\/}, \bibinfo{pages}{889--894}.
\bibitem[{Glass(1976)}]{glass1976primary}
\bibinfo{author}{Glass, G.~V.} (\bibinfo{year}{1976}).
\newblock \bibinfo{title}{Primary, secondary, and meta-analysis of research}.
\newblock {\it \bibinfo{journal}{Educational Researcher}\/},  {\it
  \bibinfo{volume}{5}\/}, \bibinfo{pages}{3--8}.
\bibitem[{Gorg \& Strobl(2001)}]{gorg2001multinational}
\bibinfo{author}{Gorg, H.}, \& \bibinfo{author}{Strobl, E.}
  (\bibinfo{year}{2001}).
\newblock \bibinfo{title}{Multinational companies and productivity spillovers:
  {A} meta-analysis}.
\newblock {\it \bibinfo{journal}{The Economic Journal}\/},  {\it
  \bibinfo{volume}{111}\/}, \bibinfo{pages}{723--739}.
\bibitem[{G{\o}tzsche(1987)}]{gotzsche1987reference}
\bibinfo{author}{G{\o}tzsche, P.~C.} (\bibinfo{year}{1987}).
\newblock \bibinfo{title}{Reference bias in reports of drug trials}.
\newblock {\it \bibinfo{journal}{BMJ: British Medical Journal}\/},  {\it
  \bibinfo{volume}{295}\/}, \bibinfo{pages}{654--656}.
\bibitem[{Greene(2012)}]{greene2012econometric}
\bibinfo{author}{Greene, W.~H.} (\bibinfo{year}{2012}).
\newblock {\it \bibinfo{title}{Econometric analysis}\/}.
\newblock (\bibinfo{edition}{7th} ed.).
\newblock \bibinfo{publisher}{Pearson Education}.
\bibitem[{Guolo \& Varin(2017)}]{guolo2017random}
\bibinfo{author}{Guolo, A.}, \& \bibinfo{author}{Varin, C.}
  (\bibinfo{year}{2017}).
\newblock \bibinfo{title}{Random-effects meta-analysis: {The} number of studies
  matters}.
\newblock {\it \bibinfo{journal}{Statistical Methods in Medical Research}\/},
  {\it \bibinfo{volume}{26}\/}, \bibinfo{pages}{1500--1518}.
\bibitem[{Hahn et~al.(2002)Hahn, Williamson \& Hutton}]{hahn2002investigation}
\bibinfo{author}{Hahn, S.}, \bibinfo{author}{Williamson, P.}, \&
  \bibinfo{author}{Hutton, J.} (\bibinfo{year}{2002}).
\newblock \bibinfo{title}{Investigation of within-study selective reporting in
  clinical research: {Follow-up} of applications submitted to a local research
  ethics committee}.
\newblock {\it \bibinfo{journal}{Journal of Evaluation in Clinical
  Practice}\/},  {\it \bibinfo{volume}{8}\/}, \bibinfo{pages}{353--359}.
\bibitem[{Hanji(2017)}]{hanji2017meta}
\bibinfo{author}{Hanji, M.~B.} (\bibinfo{year}{2017}).
\newblock {\it \bibinfo{title}{Meta-analysis in psychiatry research:
  {Fundamental} and advanced methods}\/}.
\newblock \bibinfo{publisher}{Apple Academic Press}.
\bibitem[{Harbord \& Higgins(2008)}]{harbord2008meta}
\bibinfo{author}{Harbord, R.~M.}, \& \bibinfo{author}{Higgins, J.}
  (\bibinfo{year}{2008}).
\newblock \bibinfo{title}{Meta-regression in stata}.
\newblock {\it \bibinfo{journal}{Meta}\/},  {\it \bibinfo{volume}{8}\/},
  \bibinfo{pages}{493--519}.
\bibitem[{Hardy \& Thompson(1996)}]{hardy1996likelihood}
\bibinfo{author}{Hardy, R.~J.}, \& \bibinfo{author}{Thompson, S.~G.}
  (\bibinfo{year}{1996}).
\newblock \bibinfo{title}{A likelihood approach to meta-analysis with random
  effects}.
\newblock {\it \bibinfo{journal}{Statistics in Medicine}\/},  {\it
  \bibinfo{volume}{15}\/}, \bibinfo{pages}{619--629}.
\bibitem[{Hardy \& Thompson(1998)}]{hardy1998detecting}
\bibinfo{author}{Hardy, R.~J.}, \& \bibinfo{author}{Thompson, S.~G.}
  (\bibinfo{year}{1998}).
\newblock \bibinfo{title}{Detecting and describing heterogeneity in
  meta-analysis}.
\newblock {\it \bibinfo{journal}{Statistics in Medicine}\/},  {\it
  \bibinfo{volume}{17}\/}, \bibinfo{pages}{841--856}.
\bibitem[{Hartung \& Knapp(2001{\natexlab{a}})}]{hartung2001tests}
\bibinfo{author}{Hartung, J.}, \& \bibinfo{author}{Knapp, G.}
  (\bibinfo{year}{2001}{\natexlab{a}}).
\newblock \bibinfo{title}{On tests of the overall treatment effect in
  meta-analysis with normally distributed responses}.
\newblock {\it \bibinfo{journal}{Statistics in Medicine}\/},  {\it
  \bibinfo{volume}{20}\/}, \bibinfo{pages}{1771--1782}.
\bibitem[{Hartung \& Knapp(2001{\natexlab{b}})}]{hartung2001refined}
\bibinfo{author}{Hartung, J.}, \& \bibinfo{author}{Knapp, G.}
  (\bibinfo{year}{2001}{\natexlab{b}}).
\newblock \bibinfo{title}{A refined method for the meta-analysis of controlled
  clinical trials with binary outcome}.
\newblock {\it \bibinfo{journal}{Statistics in Medicine}\/},  {\it
  \bibinfo{volume}{20}\/}, \bibinfo{pages}{3875--3889}.
\bibitem[{Hedberg(2011)}]{hedberg2011robumeta}
\bibinfo{author}{Hedberg, E.~C.} (\bibinfo{year}{2011}).
\newblock \bibinfo{title}{{ROBUMETA: {Stata} module to perform robust variance
  estimation in meta-regression with dependent effect size estimates}}.
\newblock \URLprefix \url{https://ideas.repec.org/c/boc/bocode/s457219.html}
  \bibinfo{note}{statistical Software Components S457219, Boston College
  Department of Economics, revised 23 Apr 2014}.
\bibitem[{Hedges(1984)}]{hedges1984estimation}
\bibinfo{author}{Hedges, L.~V.} (\bibinfo{year}{1984}).
\newblock \bibinfo{title}{Estimation of effect size under nonrandom sampling:
  {The} effects of censoring studies yielding statistically insignificant mean
  differences}.
\newblock {\it \bibinfo{journal}{Journal of Educational Statistics}\/},  {\it
  \bibinfo{volume}{9}\/}, \bibinfo{pages}{61--85}.
\bibitem[{Hedges(1992)}]{hedges1992modeling}
\bibinfo{author}{Hedges, L.~V.} (\bibinfo{year}{1992}).
\newblock \bibinfo{title}{Modeling publication selection effects in
  meta-analysis}.
\newblock {\it \bibinfo{journal}{Statistical Science}\/},  {\it
  \bibinfo{volume}{7}\/}, \bibinfo{pages}{246--255}.
\bibitem[{Hedges \& Olkin(1985)}]{hedges1985statistical}
\bibinfo{author}{Hedges, L.~V.}, \& \bibinfo{author}{Olkin, I.}
  (\bibinfo{year}{1985}).
\newblock {\it \bibinfo{title}{Statistical methods for meta-analysis}\/}.
\newblock \bibinfo{publisher}{Academic Press}.
\bibitem[{Hedges et~al.(2010)Hedges, Tipton \& Johnson}]{hedges2010robust}
\bibinfo{author}{Hedges, L.~V.}, \bibinfo{author}{Tipton, E.}, \&
  \bibinfo{author}{Johnson, M.~C.} (\bibinfo{year}{2010}).
\newblock \bibinfo{title}{Robust variance estimation in meta-regression with
  dependent effect size estimates}.
\newblock {\it \bibinfo{journal}{Research Synthesis Methods}\/},  {\it
  \bibinfo{volume}{1}\/}, \bibinfo{pages}{39--65}.
\bibitem[{Hedges \& Vevea(1998)}]{hedges1998fixed}
\bibinfo{author}{Hedges, L.~V.}, \& \bibinfo{author}{Vevea, J.~L.}
  (\bibinfo{year}{1998}).
\newblock \bibinfo{title}{Fixed-and random-effects models in meta-analysis}.
\newblock {\it \bibinfo{journal}{Psychological Methods}\/},  {\it
  \bibinfo{volume}{3}\/}, \bibinfo{pages}{486--504}.
\bibitem[{Higgins \& Thompson(2002)}]{higgins2002quantifying}
\bibinfo{author}{Higgins, J.}, \& \bibinfo{author}{Thompson, S.~G.}
  (\bibinfo{year}{2002}).
\newblock \bibinfo{title}{Quantifying heterogeneity in a meta-analysis}.
\newblock {\it \bibinfo{journal}{Statistics in Medicine}\/},  {\it
  \bibinfo{volume}{21}\/}, \bibinfo{pages}{1539--1558}.
\bibitem[{Higgins \& Thompson(2004)}]{higgins2004controlling}
\bibinfo{author}{Higgins, J.}, \& \bibinfo{author}{Thompson, S.~G.}
  (\bibinfo{year}{2004}).
\newblock \bibinfo{title}{Controlling the risk of spurious findings from
  meta-regression}.
\newblock {\it \bibinfo{journal}{Statistics in Medicine}\/},  {\it
  \bibinfo{volume}{23}\/}, \bibinfo{pages}{1663--1682}.
\bibitem[{Higgins et~al.(2009)Higgins, Thompson \&
  Spiegelhalter}]{higgins2009re}
\bibinfo{author}{Higgins, J.}, \bibinfo{author}{Thompson, S.~G.}, \&
  \bibinfo{author}{Spiegelhalter, D.~J.} (\bibinfo{year}{2009}).
\newblock \bibinfo{title}{A re-evaluation of random-effects meta-analysis}.
\newblock {\it \bibinfo{journal}{Journal of the Royal Statistical Society:
  Series A (Statistics in Society)}\/},  {\it \bibinfo{volume}{172}\/},
  \bibinfo{pages}{137--159}.
\bibitem[{Higgins(2008)}]{higgins2008commentary}
\bibinfo{author}{Higgins, J.~P.} (\bibinfo{year}{2008}).
\newblock \bibinfo{title}{Commentary: {Heterogeneity} in meta-analysis should
  be expected and appropriately quantified}.
\newblock {\it \bibinfo{journal}{International Journal of Epidemiology}\/},
  {\it \bibinfo{volume}{37}\/}, \bibinfo{pages}{1158--1160}.
\bibitem[{Higgins et~al.(2003)Higgins, Thompson, Deeks \&
  Altman}]{higgins2003measuring}
\bibinfo{author}{Higgins, J.~P.}, \bibinfo{author}{Thompson, S.~G.},
  \bibinfo{author}{Deeks, J.~J.}, \& \bibinfo{author}{Altman, D.~G.}
  (\bibinfo{year}{2003}).
\newblock \bibinfo{title}{Measuring inconsistency in meta-analyses}.
\newblock {\it \bibinfo{journal}{BMJ: British Medical Journal}\/},  {\it
  \bibinfo{volume}{327}\/}, \bibinfo{pages}{557--560}.
\bibitem[{Hox et~al.(2010)Hox, Moerbeek \& Van~de Schoot}]{hox2010multilevel}
\bibinfo{author}{Hox, J.~J.}, \bibinfo{author}{Moerbeek, M.}, \&
  \bibinfo{author}{Van~de Schoot, R.} (\bibinfo{year}{2010}).
\newblock {\it \bibinfo{title}{Multilevel analysis: {Techniques} and
  applications}\/}.
\newblock Quantitative Methodology Series (\bibinfo{edition}{2nd} ed.).
\newblock \bibinfo{address}{New York}: \bibinfo{publisher}{Routledge}.
\bibitem[{Huedo-Medina et~al.(2006)Huedo-Medina, S{\'a}nchez-Meca,
  Mar{\'\i}n-Mart{\'\i}nez \& Botella}]{huedo2006assessing}
\bibinfo{author}{Huedo-Medina, T.~B.}, \bibinfo{author}{S{\'a}nchez-Meca, J.},
  \bibinfo{author}{Mar{\'\i}n-Mart{\'\i}nez, F.}, \& \bibinfo{author}{Botella,
  J.} (\bibinfo{year}{2006}).
\newblock \bibinfo{title}{Assessing heterogeneity in meta-analysis: {Q}
  statistic or {$I^2$} index?}
\newblock {\it \bibinfo{journal}{Psychological Methods}\/},  {\it
  \bibinfo{volume}{11}\/}, \bibinfo{pages}{193--206}.
\bibitem[{Huizenga et~al.(2011)Huizenga, Visser \& Dolan}]{huizenga2011testing}
\bibinfo{author}{Huizenga, H.~M.}, \bibinfo{author}{Visser, I.}, \&
  \bibinfo{author}{Dolan, C.~V.} (\bibinfo{year}{2011}).
\newblock \bibinfo{title}{Testing overall and moderator effects in random
  effects meta-regression}.
\newblock {\it \bibinfo{journal}{British Journal of Mathematical and
  Statistical Psychology}\/},  {\it \bibinfo{volume}{64}\/},
  \bibinfo{pages}{1--19}.
\bibitem[{Hunter et~al.(1996)Hunter, Schmidt \& Coggin}]{hunter1996meta}
\bibinfo{author}{Hunter, J.}, \bibinfo{author}{Schmidt, F.}, \&
  \bibinfo{author}{Coggin, T.} (\bibinfo{year}{1996}).
\newblock \bibinfo{title}{Meta-analysis of correlations: {Bias} in the
  correlation coefficient and the {Fisher} z transformation}.
\newblock \bibinfo{note}{University of Iowa}.
\bibitem[{Hunter \& Schmidt(2000)}]{hunter2000fixed}
\bibinfo{author}{Hunter, J.~E.}, \& \bibinfo{author}{Schmidt, F.~L.}
  (\bibinfo{year}{2000}).
\newblock \bibinfo{title}{Fixed effects vs. random effects meta-analysis
  models: {Implications} for cumulative research knowledge}.
\newblock {\it \bibinfo{journal}{International Journal of Selection and
  Assessment}\/},  {\it \bibinfo{volume}{8}\/}, \bibinfo{pages}{275--292}.
\bibitem[{Hunter \& Schmidt(2004)}]{hunter2004methods}
\bibinfo{author}{Hunter, J.~E.}, \& \bibinfo{author}{Schmidt, F.~L.}
  (\bibinfo{year}{2004}).
\newblock {\it \bibinfo{title}{Methods of meta-analysis: {Correcting} error and
  bias in research findings}\/}.
\newblock \bibinfo{address}{New York}: \bibinfo{publisher}{SAGE Publications,
  Inc}.
\bibitem[{Hutton \& Williamson(2000)}]{hutton2000bias}
\bibinfo{author}{Hutton, J.}, \& \bibinfo{author}{Williamson, P.~R.}
  (\bibinfo{year}{2000}).
\newblock \bibinfo{title}{Bias in meta-analysis due to outcome variable
  selection within studies}.
\newblock {\it \bibinfo{journal}{Journal of the Royal Statistical Society:
  Series C (Applied Statistics)}\/},  {\it \bibinfo{volume}{49}\/},
  \bibinfo{pages}{359--370}.
\bibitem[{Ioannidis(1998)}]{ioannidis1998effect}
\bibinfo{author}{Ioannidis, J.~P.} (\bibinfo{year}{1998}).
\newblock \bibinfo{title}{Effect of the statistical significance of results on
  the time to completion and publication of randomized efficacy trials}.
\newblock {\it \bibinfo{journal}{Jama}\/},  {\it \bibinfo{volume}{279}\/},
  \bibinfo{pages}{281--286}.
\bibitem[{Ioannidis et~al.(2007)Ioannidis, Patsopoulos \&
  Evangelou}]{ioannidis2007uncertainty}
\bibinfo{author}{Ioannidis, J.~P.}, \bibinfo{author}{Patsopoulos, N.~A.}, \&
  \bibinfo{author}{Evangelou, E.} (\bibinfo{year}{2007}).
\newblock \bibinfo{title}{Uncertainty in heterogeneity estimates in
  meta-analyses}.
\newblock {\it \bibinfo{journal}{BMJ: British Medical Journal}\/},  {\it
  \bibinfo{volume}{335}\/}, \bibinfo{pages}{914--916}.
\bibitem[{Iyengar \& Greenhouse(1988)}]{iyengar1988selection}
\bibinfo{author}{Iyengar, S.}, \& \bibinfo{author}{Greenhouse, J.~B.}
  (\bibinfo{year}{1988}).
\newblock \bibinfo{title}{Selection models and the file drawer problem}.
\newblock {\it \bibinfo{journal}{Statistical Science}\/},  {\it
  \bibinfo{volume}{3}\/}, \bibinfo{pages}{109--117}.
\bibitem[{Jackson(2013)}]{jackson2013confidence}
\bibinfo{author}{Jackson, D.} (\bibinfo{year}{2013}).
\newblock \bibinfo{title}{Confidence intervals for the between-study variance
  in random effects meta-analysis using generalised {Cochran} heterogeneity
  statistics}.
\newblock {\it \bibinfo{journal}{Research Synthesis Methods}\/},  {\it
  \bibinfo{volume}{4}\/}, \bibinfo{pages}{220--229}.
\bibitem[{Jackson \& Bowden(2016)}]{jackson2016confidence}
\bibinfo{author}{Jackson, D.}, \& \bibinfo{author}{Bowden, J.}
  (\bibinfo{year}{2016}).
\newblock \bibinfo{title}{Confidence intervals for the between-study variance
  in random-effects meta-analysis using generalised heterogeneity statistics:
  {Should} we use unequal tails?}
\newblock {\it \bibinfo{journal}{BMC Medical Research Methodology}\/},  {\it
  \bibinfo{volume}{16}\/}, \bibinfo{pages}{118}.
\bibitem[{Jackson et~al.(2015)Jackson, Bowden \&
  Baker}]{jackson2015approximate}
\bibinfo{author}{Jackson, D.}, \bibinfo{author}{Bowden, J.}, \&
  \bibinfo{author}{Baker, R.} (\bibinfo{year}{2015}).
\newblock \bibinfo{title}{Approximate confidence intervals for moment-based
  estimators of the between-study variance in random effects meta-analysis}.
\newblock {\it \bibinfo{journal}{Research Synthesis Methods}\/},  {\it
  \bibinfo{volume}{6}\/}, \bibinfo{pages}{372--382}.
\bibitem[{Jackson et~al.(2011)Jackson, Riley \&
  White}]{jackson2011multivariate}
\bibinfo{author}{Jackson, D.}, \bibinfo{author}{Riley, R.}, \&
  \bibinfo{author}{White, I.~R.} (\bibinfo{year}{2011}).
\newblock \bibinfo{title}{Multivariate meta-analysis: {Potential} and promise}.
\newblock {\it \bibinfo{journal}{Statistics in Medicine}\/},  {\it
  \bibinfo{volume}{30}\/}, \bibinfo{pages}{2481--2498}.
\bibitem[{Jackson et~al.(2014)Jackson, Turner, Rhodes \&
  Viechtbauer}]{jackson2014methods}
\bibinfo{author}{Jackson, D.}, \bibinfo{author}{Turner, R.},
  \bibinfo{author}{Rhodes, K.}, \& \bibinfo{author}{Viechtbauer, W.}
  (\bibinfo{year}{2014}).
\newblock \bibinfo{title}{Methods for calculating confidence and credible
  intervals for the residual between-study variance in random effects
  meta-regression models}.
\newblock {\it \bibinfo{journal}{BMC Medical Research Methodology}\/},  {\it
  \bibinfo{volume}{14}\/}, \bibinfo{pages}{103}.
\bibitem[{Jennions et~al.(2013)Jennions, Lortie, Rosenberg \&
  Rothstein}]{jennions2013publication}
\bibinfo{author}{Jennions, M.~D.}, \bibinfo{author}{Lortie, C.~J.},
  \bibinfo{author}{Rosenberg, M.~S.}, \& \bibinfo{author}{Rothstein, H.~R.}
  (\bibinfo{year}{2013}).
\newblock \bibinfo{title}{Publication and related biases}.
\newblock In \bibinfo{editor}{J.~Koricheva}, \bibinfo{editor}{J.~Gurevitch}, \&
  \bibinfo{editor}{K.~Mengersen} (Eds.), {\it \bibinfo{booktitle}{Handbook of
  Meta-analysis in Ecology and Evolution}\/} chapter~\bibinfo{chapter}{14}.
  (pp. \bibinfo{pages}{207--236}).
\newblock \bibinfo{address}{Princeton and Oxford}:
  \bibinfo{publisher}{Princeton University Press}.
\bibitem[{Jennions \& M{\o}ller(2002)}]{jennions2002relationships}
\bibinfo{author}{Jennions, M.~D.}, \& \bibinfo{author}{M{\o}ller, A.~P.}
  (\bibinfo{year}{2002}).
\newblock \bibinfo{title}{Relationships fade with time: {A} meta-analysis of
  temporal trends in publication in ecology and evolution}.
\newblock {\it \bibinfo{journal}{Proceedings of the Royal Society of London B:
  Biological Sciences}\/},  {\it \bibinfo{volume}{269}\/},
  \bibinfo{pages}{43--48}.
\bibitem[{J{\"u}ni et~al.(2002)J{\"u}ni, Holenstein, Sterne, Bartlett \&
  Egger}]{juni2002direction}
\bibinfo{author}{J{\"u}ni, P.}, \bibinfo{author}{Holenstein, F.},
  \bibinfo{author}{Sterne, J.}, \bibinfo{author}{Bartlett, C.}, \&
  \bibinfo{author}{Egger, M.} (\bibinfo{year}{2002}).
\newblock \bibinfo{title}{Direction and impact of language bias in
  meta-analyses of controlled trials: {Empirical} study}.
\newblock {\it \bibinfo{journal}{International Journal of Epidemiology}\/},
  {\it \bibinfo{volume}{31}\/}, \bibinfo{pages}{115--123}.
\bibitem[{Kalaian \& Raudenbush(1996)}]{kalaian1996multivariate}
\bibinfo{author}{Kalaian, H.~A.}, \& \bibinfo{author}{Raudenbush, S.~W.}
  (\bibinfo{year}{1996}).
\newblock \bibinfo{title}{A multivariate mixed linear model for meta-analysis}.
\newblock {\it \bibinfo{journal}{Psychological Methods}\/},  {\it
  \bibinfo{volume}{1}\/}, \bibinfo{pages}{227--235}.
\bibitem[{Kenward \& Roger(1997)}]{kenward1997small}
\bibinfo{author}{Kenward, M.~G.}, \& \bibinfo{author}{Roger, J.~H.}
  (\bibinfo{year}{1997}).
\newblock \bibinfo{title}{Small sample inference for fixed effects from
  restricted maximum likelihood}.
\newblock {\it \bibinfo{journal}{Biometrics}\/},  {\it \bibinfo{volume}{53}\/},
  \bibinfo{pages}{983--997}.
\bibitem[{Kjaergard \& Gluud(2002)}]{kjaergard2002citation}
\bibinfo{author}{Kjaergard, L.~L.}, \& \bibinfo{author}{Gluud, C.}
  (\bibinfo{year}{2002}).
\newblock \bibinfo{title}{Citation bias of hepato-biliary randomized clinical
  trials}.
\newblock {\it \bibinfo{journal}{Journal of Clinical Epidemiology}\/},  {\it
  \bibinfo{volume}{55}\/}, \bibinfo{pages}{407--410}.
\bibitem[{Knapp et~al.(2006)Knapp, Biggerstaff \& Hartung}]{knapp2006assessing}
\bibinfo{author}{Knapp, G.}, \bibinfo{author}{Biggerstaff, B.~J.}, \&
  \bibinfo{author}{Hartung, J.} (\bibinfo{year}{2006}).
\newblock \bibinfo{title}{Assessing the amount of heterogeneity in
  random-effects meta-analysis}.
\newblock {\it \bibinfo{journal}{Biometrical Journal: Journal of Mathematical
  Methods in Biosciences}\/},  {\it \bibinfo{volume}{48}\/},
  \bibinfo{pages}{271--285}.
\bibitem[{Knapp \& Hartung(2003)}]{knapp2003improved}
\bibinfo{author}{Knapp, G.}, \& \bibinfo{author}{Hartung, J.}
  (\bibinfo{year}{2003}).
\newblock \bibinfo{title}{Improved tests for a random effects meta-regression
  with a single covariate}.
\newblock {\it \bibinfo{journal}{Statistics in Medicine}\/},  {\it
  \bibinfo{volume}{22}\/}, \bibinfo{pages}{2693--2710}.
\bibitem[{Konstantopoulos(2011)}]{konstantopoulos2011fixed}
\bibinfo{author}{Konstantopoulos, S.} (\bibinfo{year}{2011}).
\newblock \bibinfo{title}{Fixed effects and variance components estimation in
  three-level meta-analysis}.
\newblock {\it \bibinfo{journal}{Research Synthesis Methods}\/},  {\it
  \bibinfo{volume}{2}\/}, \bibinfo{pages}{61--76}.
\bibitem[{Langan et~al.(2017)Langan, Higgins \&
  Simmonds}]{langan2017comparative}
\bibinfo{author}{Langan, D.}, \bibinfo{author}{Higgins, J.}, \&
  \bibinfo{author}{Simmonds, M.} (\bibinfo{year}{2017}).
\newblock \bibinfo{title}{Comparative performance of heterogeneity variance
  estimators in meta-analysis: a review of simulation studies}.
\newblock {\it \bibinfo{journal}{Research Synthesis Methods}\/},  {\it
  \bibinfo{volume}{8}\/}, \bibinfo{pages}{181--198}.
\bibitem[{Lewis \& Clarke(2001)}]{lewis2001forest}
\bibinfo{author}{Lewis, S.}, \& \bibinfo{author}{Clarke, M.}
  (\bibinfo{year}{2001}).
\newblock \bibinfo{title}{Forest plots: {Trying} to see the wood and the
  trees}.
\newblock {\it \bibinfo{journal}{BMJ: British Medical Journal}\/},  {\it
  \bibinfo{volume}{322}\/}, \bibinfo{pages}{1479--1480}.
\bibitem[{Light \& Pillemer(1984)}]{light1984summing}
\bibinfo{author}{Light, R.~J.}, \& \bibinfo{author}{Pillemer, D.~B.}
  (\bibinfo{year}{1984}).
\newblock {\it \bibinfo{title}{Summing up: {The} science of reviewing
  research}\/}.
\newblock \bibinfo{address}{Cambridge, MA}: \bibinfo{publisher}{Harvard
  University Press}.
\bibitem[{Light et~al.(1994)Light, Singer \& Willett}]{light1994visual}
\bibinfo{author}{Light, R.~J.}, \bibinfo{author}{Singer, J.~D.}, \&
  \bibinfo{author}{Willett, J.~B.} (\bibinfo{year}{1994}).
\newblock \bibinfo{title}{The visual presentation and interpretation of
  meta-analyses}.
\newblock In \bibinfo{editor}{H.~Kooper}, \& \bibinfo{editor}{L.~V. Hedges}
  (Eds.), {\it \bibinfo{booktitle}{The handbook of research synthesis}\/} (pp.
  \bibinfo{pages}{439--453}).
\newblock \bibinfo{address}{New York}: \bibinfo{publisher}{Russell Sage
  Foundation}. (\bibinfo{edition}{1st} ed.).
\bibitem[{Lipsey \& Wilson(2001)}]{lipsey2001practical}
\bibinfo{author}{Lipsey, M.~W.}, \& \bibinfo{author}{Wilson, D.~B.}
  (\bibinfo{year}{2001}).
\newblock {\it \bibinfo{title}{Practical Meta-Analysis}\/}.
\newblock \bibinfo{publisher}{SAGE Publications, Inc.}
\bibitem[{L{\'o}pez-L{\'o}pez et~al.(2014)L{\'o}pez-L{\'o}pez,
  Mar{\'\i}n-Mart{\'\i}nez, S{\'a}nchez-Meca, Noortgate \&
  Viechtbauer}]{lopez2014estimation}
\bibinfo{author}{L{\'o}pez-L{\'o}pez, J.~A.},
  \bibinfo{author}{Mar{\'\i}n-Mart{\'\i}nez, F.},
  \bibinfo{author}{S{\'a}nchez-Meca, J.}, \bibinfo{author}{Noortgate, W.}, \&
  \bibinfo{author}{Viechtbauer, W.} (\bibinfo{year}{2014}).
\newblock \bibinfo{title}{Estimation of the predictive power of the model in
  mixed-effects meta-regression: {A} simulation study}.
\newblock {\it \bibinfo{journal}{British Journal of Mathematical and
  Statistical Psychology}\/},  {\it \bibinfo{volume}{67}\/},
  \bibinfo{pages}{30--48}.
\bibitem[{L{\'o}pez-L{\'o}pez et~al.(2017)L{\'o}pez-L{\'o}pez, Van~den
  Noortgate, Tanner-Smith, Wilson \& Lipsey}]{lopez2017assessing}
\bibinfo{author}{L{\'o}pez-L{\'o}pez, J.~A.}, \bibinfo{author}{Van~den
  Noortgate, W.}, \bibinfo{author}{Tanner-Smith, E.~E.},
  \bibinfo{author}{Wilson, S.~J.}, \& \bibinfo{author}{Lipsey, M.~W.}
  (\bibinfo{year}{2017}).
\newblock \bibinfo{title}{Assessing meta-regression methods for examining
  moderator relationships with dependent effect sizes: {AM onte C arlo
  simulation}}.
\newblock {\it \bibinfo{journal}{Research Synthesis Methods}\/},  {\it
  \bibinfo{volume}{8}\/}, \bibinfo{pages}{435--450}.
\bibitem[{Maier et~al.(2023)Maier, Bartoš \& Wagenmakers}]{maier2023robust}
\bibinfo{author}{Maier, M.}, \bibinfo{author}{Bartoš, F.}, \&
  \bibinfo{author}{Wagenmakers, E.-J.} (\bibinfo{year}{2023}).
\newblock \bibinfo{title}{Robust bayesian meta-analysis: {Addressing}
  publication bias with model-averaging}.
\newblock {\it \bibinfo{journal}{Psychological Methods}\/},  {\it
  \bibinfo{volume}{28}\/}, \bibinfo{pages}{107--122}.
\bibitem[{Maier et~al.(2022)Maier, VanderWeele \& Mathur}]{maier2022using}
\bibinfo{author}{Maier, M.}, \bibinfo{author}{VanderWeele, T.~J.}, \&
  \bibinfo{author}{Mathur, M.~B.} (\bibinfo{year}{2022}).
\newblock \bibinfo{title}{Using selection models to assess sensitivity to
  publication bias: {A} tutorial and call for more routine use}.
\newblock {\it \bibinfo{journal}{Campbell Systematic Reviews}\/},  {\it
  \bibinfo{volume}{18}\/}, \bibinfo{pages}{e1256}.
\bibitem[{Mar{\'\i}n-Mart{\'\i}nez \&
  S{\'a}nchez-Meca(1999)}]{marin1999averaging}
\bibinfo{author}{Mar{\'\i}n-Mart{\'\i}nez, F.}, \&
  \bibinfo{author}{S{\'a}nchez-Meca, J.} (\bibinfo{year}{1999}).
\newblock \bibinfo{title}{Averaging dependent effect sizes in meta-analysis: A
  cautionary note about procedures}.
\newblock {\it \bibinfo{journal}{The Spanish Journal of Psychology}\/},  {\it
  \bibinfo{volume}{2}\/}, \bibinfo{pages}{32--38}.
\bibitem[{Marks-Anglin \& Chen(2020)}]{ioannidis2020historical}
\bibinfo{author}{Marks-Anglin, A.}, \& \bibinfo{author}{Chen, Y.}
  (\bibinfo{year}{2020}).
\newblock \bibinfo{title}{A historical review of publication bias}.
\newblock {\it \bibinfo{journal}{Research Synthesis Methods}\/},  {\it
  \bibinfo{volume}{11}\/}, \bibinfo{pages}{725--742}.
\bibitem[{Mart{\'\i}nez \& Mart{\'\i}nez(2021)}]{martinez2021effects}
\bibinfo{author}{Mart{\'\i}nez, M.~J.}, \& \bibinfo{author}{Mart{\'\i}nez,
  M.~J.} (\bibinfo{year}{2021}).
\newblock \bibinfo{title}{Are the effects of minimum wage on the labour market
  the same across countries? {A} meta-analysis spanning a century}.
\newblock {\it \bibinfo{journal}{Economic Systems}\/},  {\it
  \bibinfo{volume}{45}\/}, \bibinfo{pages}{100849}.
\bibitem[{Martyn-St~James \& Carroll(2010)}]{martyn2010effects}
\bibinfo{author}{Martyn-St~James, M.}, \& \bibinfo{author}{Carroll, S.}
  (\bibinfo{year}{2010}).
\newblock \bibinfo{title}{Effects of different impact exercise modalities on
  bone mineral density in premenopausal women: {A} meta-analysis}.
\newblock {\it \bibinfo{journal}{Journal of Bone and Mineral Metabolism}\/},
  {\it \bibinfo{volume}{28}\/}, \bibinfo{pages}{251--267}.
\bibitem[{Mavridis \& Salanti(2013)}]{mavridis2013practical}
\bibinfo{author}{Mavridis, D.}, \& \bibinfo{author}{Salanti, G.}
  (\bibinfo{year}{2013}).
\newblock \bibinfo{title}{A practical introduction to multivariate
  meta-analysis}.
\newblock {\it \bibinfo{journal}{Statistical Methods in Medical Research}\/},
  {\it \bibinfo{volume}{22}\/}, \bibinfo{pages}{133--158}.
\bibitem[{McShane et~al.(2016)McShane, B{\"o}ckenholt \&
  Hansen}]{mcshane2016adjusting}
\bibinfo{author}{McShane, B.~B.}, \bibinfo{author}{B{\"o}ckenholt, U.}, \&
  \bibinfo{author}{Hansen, K.~T.} (\bibinfo{year}{2016}).
\newblock \bibinfo{title}{Adjusting for publication bias in meta-analysis: {An}
  evaluation of selection methods and some cautionary notes}.
\newblock {\it \bibinfo{journal}{Perspectives on Psychological Science}\/},
  {\it \bibinfo{volume}{11}\/}, \bibinfo{pages}{730--749}.
\bibitem[{Moreno et~al.(2009)Moreno, Sutton, Ades, Stanley, Abrams, Peters \&
  Cooper}]{moreno2009assessment}
\bibinfo{author}{Moreno, S.~G.}, \bibinfo{author}{Sutton, A.~J.},
  \bibinfo{author}{Ades, A.}, \bibinfo{author}{Stanley, T.~D.},
  \bibinfo{author}{Abrams, K.~R.}, \bibinfo{author}{Peters, J.~L.}, \&
  \bibinfo{author}{Cooper, N.~J.} (\bibinfo{year}{2009}).
\newblock \bibinfo{title}{Assessment of regression-based methods to adjust for
  publication bias through a comprehensive simulation study}.
\newblock {\it \bibinfo{journal}{BMC Medical Research Methodology}\/},  {\it
  \bibinfo{volume}{9}\/}, \bibinfo{pages}{2}.
\bibitem[{Nagashima et~al.(2019)Nagashima, Noma \&
  Furukawa}]{nagashima2019prediction}
\bibinfo{author}{Nagashima, K.}, \bibinfo{author}{Noma, H.}, \&
  \bibinfo{author}{Furukawa, T.~A.} (\bibinfo{year}{2019}).
\newblock \bibinfo{title}{Prediction intervals for random-effects
  meta-analysis: {A} confidence distribution approach}.
\newblock {\it \bibinfo{journal}{Statistical Methods in Medical Research}\/},
  {\it \bibinfo{volume}{28}\/}, \bibinfo{pages}{1689--1702}.
\bibitem[{Neumark \& Wascher(2006)}]{neumark2006minimum}
\bibinfo{author}{Neumark, D.}, \& \bibinfo{author}{Wascher, W.}
  (\bibinfo{year}{2006}).
\newblock {\it \bibinfo{title}{Minimum wages and employment: {A} review of
  evidence from the new minimum wage research}\/}.
\newblock \bibinfo{type}{Working Paper} \bibinfo{number}{060708} University of
  California-Irvine, Department of Economics.
\newblock \bibinfo{note}{Revised January 2007}.
\bibitem[{Van~den Noortgate et~al.(2013)Van~den Noortgate, L{\'o}pez-L{\'o}pez,
  Mar{\'\i}n-Mart{\'\i}nez \& S{\'a}nchez-Meca}]{van2013three}
\bibinfo{author}{Van~den Noortgate, W.}, \bibinfo{author}{L{\'o}pez-L{\'o}pez,
  J.~A.}, \bibinfo{author}{Mar{\'\i}n-Mart{\'\i}nez, F.}, \&
  \bibinfo{author}{S{\'a}nchez-Meca, J.} (\bibinfo{year}{2013}).
\newblock \bibinfo{title}{Three-level meta-analysis of dependent effect sizes}.
\newblock {\it \bibinfo{journal}{Behavior Research Methods}\/},  {\it
  \bibinfo{volume}{45}\/}, \bibinfo{pages}{576--594}.
\bibitem[{Van~den Noortgate et~al.(2015)Van~den Noortgate, L{\'o}pez-L{\'o}pez,
  Mar{\'\i}n-Mart{\'\i}nez \& S{\'a}nchez-Meca}]{van2015meta}
\bibinfo{author}{Van~den Noortgate, W.}, \bibinfo{author}{L{\'o}pez-L{\'o}pez,
  J.~A.}, \bibinfo{author}{Mar{\'\i}n-Mart{\'\i}nez, F.}, \&
  \bibinfo{author}{S{\'a}nchez-Meca, J.} (\bibinfo{year}{2015}).
\newblock \bibinfo{title}{Meta-analysis of multiple outcomes: {A} multilevel
  approach}.
\newblock {\it \bibinfo{journal}{Behavior Research Methods}\/},  {\it
  \bibinfo{volume}{47}\/}, \bibinfo{pages}{1274--1294}.
\bibitem[{Olkin \& Gleser(2009)}]{olkin2009stochastically}
\bibinfo{author}{Olkin, I.}, \& \bibinfo{author}{Gleser, L.}
  (\bibinfo{year}{2009}).
\newblock \bibinfo{title}{Stochastically dependent effect sizes}.
\newblock In \bibinfo{editor}{C.~Harris}, \bibinfo{editor}{L.~Hedges}, \&
  \bibinfo{editor}{J.~Valentine} (Eds.), {\it \bibinfo{booktitle}{Handbook of
  research synthesis and meta-analysis}\/} chapter~\bibinfo{chapter}{19}. (pp.
  \bibinfo{pages}{357--376}).
\newblock \bibinfo{address}{New York}: \bibinfo{publisher}{Russell Sage
  Foundation: New York, NY}. (\bibinfo{edition}{2nd} ed.).
\bibitem[{Onishi \& Furukawa(2014)}]{onishi2014publication}
\bibinfo{author}{Onishi, A.}, \& \bibinfo{author}{Furukawa, T.~A.}
  (\bibinfo{year}{2014}).
\newblock \bibinfo{title}{Publication bias is underreported in systematic
  reviews published in high-impact-factor journals: {Metaepidemiologic} study}.
\newblock {\it \bibinfo{journal}{Journal of Clinical Epidemiology}\/},  {\it
  \bibinfo{volume}{67}\/}, \bibinfo{pages}{1320--1326}.
\bibitem[{O'rourke(2007)}]{o2007historical}
\bibinfo{author}{O'rourke, K.} (\bibinfo{year}{2007}).
\newblock \bibinfo{title}{An historical perspective on meta-analysis: {Dealing}
  quantitatively with varying study results}.
\newblock {\it \bibinfo{journal}{Journal of the Royal Society of Medicine}\/},
  {\it \bibinfo{volume}{100}\/}, \bibinfo{pages}{579--582}.
\bibitem[{Page et~al.(2022)Page, Sterne, Higgins \&
  Egger}]{page2022investigating}
\bibinfo{author}{Page, M.~J.}, \bibinfo{author}{Sterne, J. A.~C.},
  \bibinfo{author}{Higgins, J. P.~T.}, \& \bibinfo{author}{Egger, M.}
  (\bibinfo{year}{2022}).
\newblock \bibinfo{title}{Investigating and dealing with publication bias and
  other reporting biases}.
\newblock In \bibinfo{editor}{M.~Egger}, \bibinfo{editor}{J.~P.~T. Higgins}, \&
  \bibinfo{editor}{G.~D. Smith} (Eds.), {\it \bibinfo{booktitle}{Systematic
  reviews in health research: Meta-analysis in context}\/}
  chapter~\bibinfo{chapter}{5}. (pp. \bibinfo{pages}{74--90}).
\newblock \bibinfo{publisher}{BMJ Books}. (\bibinfo{edition}{3rd} ed.).
\bibitem[{Panesar et~al.(2010)Panesar, Siow \&
  Athanasiou}]{panesar2010systematic}
\bibinfo{author}{Panesar, S.~S.}, \bibinfo{author}{Siow, W.}, \&
  \bibinfo{author}{Athanasiou, T.} (\bibinfo{year}{2010}).
\newblock \bibinfo{title}{Systematic reviews and meta-analyses in surgery}.
\newblock In {\it \bibinfo{booktitle}{Key topics in surgical research and
  methodology}\/} (pp. \bibinfo{pages}{375--397}).
\newblock \bibinfo{publisher}{Springer}.
\bibitem[{Panityakul et~al.(2013)Panityakul, Bumrungsup \&
  Knapp}]{panityakul2013estimating}
\bibinfo{author}{Panityakul, T.}, \bibinfo{author}{Bumrungsup, C.}, \&
  \bibinfo{author}{Knapp, G.} (\bibinfo{year}{2013}).
\newblock \bibinfo{title}{On estimating residual heterogeneity in
  random-effects meta-regression: {A} comparative study}.
\newblock {\it \bibinfo{journal}{J Stat Theory Appl}\/},  {\it
  \bibinfo{volume}{12}\/}, \bibinfo{pages}{253--265}.
\bibitem[{Partlett \& Riley(2017)}]{partlett2017random}
\bibinfo{author}{Partlett, C.}, \& \bibinfo{author}{Riley, R.~D.}
  (\bibinfo{year}{2017}).
\newblock \bibinfo{title}{Random effects meta-analysis: {C}overage performance
  of 95\% confidence and prediction intervals following {REML} estimation}.
\newblock {\it \bibinfo{journal}{Statistics in Medicine}\/},  {\it
  \bibinfo{volume}{36}\/}, \bibinfo{pages}{301--317}.
\bibitem[{Pastor \& Lazowski(2018)}]{pastor2018multilevel}
\bibinfo{author}{Pastor, D.~A.}, \& \bibinfo{author}{Lazowski, R.~A.}
  (\bibinfo{year}{2018}).
\newblock \bibinfo{title}{On the multilevel nature of meta-analysis: {A}
  tutorial, comparison of software programs, and discussion of analytic
  choices}.
\newblock {\it \bibinfo{journal}{Multivariate Behavioral Research}\/},  {\it
  \bibinfo{volume}{53}\/}, \bibinfo{pages}{74--89}.
\bibitem[{Pearson(1904)}]{Pearson1904report}
\bibinfo{author}{Pearson, K.} (\bibinfo{year}{1904}).
\newblock \bibinfo{title}{Report on certain enteric fever inoculation
  statistics}.
\newblock {\it \bibinfo{journal}{BMJ: British Medical Journal}\/},  {\it
  \bibinfo{volume}{2}\/}, \bibinfo{pages}{1243--1246}.
\bibitem[{Peters et~al.(2008)Peters, Sutton, Jones, Abrams \&
  Rushton}]{peters2008contour}
\bibinfo{author}{Peters, J.~L.}, \bibinfo{author}{Sutton, A.~J.},
  \bibinfo{author}{Jones, D.~R.}, \bibinfo{author}{Abrams, K.~R.}, \&
  \bibinfo{author}{Rushton, L.} (\bibinfo{year}{2008}).
\newblock \bibinfo{title}{Contour-enhanced meta-analysis funnel plots help
  distinguish publication bias from other causes of asymmetry}.
\newblock {\it \bibinfo{journal}{Journal of Clinical Epidemiology}\/},  {\it
  \bibinfo{volume}{61}\/}, \bibinfo{pages}{991--996}.
\bibitem[{Peto et~al.(1988)Peto, Gray, Collins, Wheatley, Hennekens, Jamrozik,
  Warlow, Hafner, Thompson, Norton, Gilliland \& Doll}]{peto1988randomised}
\bibinfo{author}{Peto, R.}, \bibinfo{author}{Gray, R.},
  \bibinfo{author}{Collins, R.}, \bibinfo{author}{Wheatley, K.},
  \bibinfo{author}{Hennekens, C.}, \bibinfo{author}{Jamrozik, K.},
  \bibinfo{author}{Warlow, C.}, \bibinfo{author}{Hafner, B.},
  \bibinfo{author}{Thompson, E.}, \bibinfo{author}{Norton, S.},
  \bibinfo{author}{Gilliland, J.}, \& \bibinfo{author}{Doll, R.}
  (\bibinfo{year}{1988}).
\newblock \bibinfo{title}{Randomised trial of prophylactic daily aspirin in
  british male doctors}.
\newblock {\it \bibinfo{journal}{BMJ: British Medical Journal}\/},  {\it
  \bibinfo{volume}{296}\/}, \bibinfo{pages}{313--316}.
\bibitem[{Petropoulou \& Mavridis(2017)}]{petropoulou2017comparison}
\bibinfo{author}{Petropoulou, M.}, \& \bibinfo{author}{Mavridis, D.}
  (\bibinfo{year}{2017}).
\newblock \bibinfo{title}{A comparison of 20 heterogeneity variance estimators
  in statistical synthesis of results from studies: {A} simulation study}.
\newblock {\it \bibinfo{journal}{Statistics in Medicine}\/},  {\it
  \bibinfo{volume}{36}\/}, \bibinfo{pages}{4266--4280}.
\bibitem[{Pigott et~al.(2013)Pigott, Valentine, Polanin, Williams \&
  Canada}]{pigott2013outcome}
\bibinfo{author}{Pigott, T.~D.}, \bibinfo{author}{Valentine, J.~C.},
  \bibinfo{author}{Polanin, J.~R.}, \bibinfo{author}{Williams, R.~T.}, \&
  \bibinfo{author}{Canada, D.~D.} (\bibinfo{year}{2013}).
\newblock \bibinfo{title}{Outcome-reporting bias in education research}.
\newblock {\it \bibinfo{journal}{Educational Researcher}\/},  {\it
  \bibinfo{volume}{42}\/}, \bibinfo{pages}{424--432}.
\bibitem[{Pinheiro \& Bates(2000)}]{pinheiro2000mixed}
\bibinfo{author}{Pinheiro, J.~C.}, \& \bibinfo{author}{Bates, D.~M.}
  (\bibinfo{year}{2000}).
\newblock {\it \bibinfo{title}{Mixed-effects models in {S} and {S-PLUS}}\/}.
\newblock (\bibinfo{edition}{1st} ed.).
\newblock \bibinfo{address}{New York}: \bibinfo{publisher}{Springer-Verlag}.
\bibitem[{Polanin et~al.(2017)Polanin, Hennessy \&
  Tanner-Smith}]{polanin2017review}
\bibinfo{author}{Polanin, J.~R.}, \bibinfo{author}{Hennessy, E.~A.}, \&
  \bibinfo{author}{Tanner-Smith, E.~E.} (\bibinfo{year}{2017}).
\newblock \bibinfo{title}{A review of meta-analysis packages in {R}}.
\newblock {\it \bibinfo{journal}{Journal of Educational and Behavioral
  Statistics}\/},  {\it \bibinfo{volume}{42}\/}, \bibinfo{pages}{206--242}.
\bibitem[{Raghunathan \& Li(1993)}]{raghunathan1993analysis}
\bibinfo{author}{Raghunathan, T.}, \& \bibinfo{author}{Li, Y.}
  (\bibinfo{year}{1993}).
\newblock \bibinfo{title}{Analysis of binary data from a multicentre clinical
  trial}.
\newblock {\it \bibinfo{journal}{Biometrika}\/},  {\it \bibinfo{volume}{80}\/},
  \bibinfo{pages}{127--139}.
\bibitem[{Raudenbush(2009)}]{raudenbush2009analyzing}
\bibinfo{author}{Raudenbush, S.~W.} (\bibinfo{year}{2009}).
\newblock \bibinfo{title}{Relationship banking, deposit insurance and bank
  portfolio choice}.
\newblock In \bibinfo{editor}{C.~Harris}, \bibinfo{editor}{L.~Hedges}, \&
  \bibinfo{editor}{J.~Valentine} (Eds.), {\it \bibinfo{booktitle}{Handbook of
  research synthesis and meta-analysis}\/} chapter~\bibinfo{chapter}{16}. (pp.
  \bibinfo{pages}{295--316}).
\newblock \bibinfo{address}{New York}: \bibinfo{publisher}{Russell Sage
  Foundation: New York, NY}. (\bibinfo{edition}{2nd} ed.).
\bibitem[{Raudenbush et~al.(1988)Raudenbush, Becker \&
  Kalaian}]{raudenbush1988modeling}
\bibinfo{author}{Raudenbush, S.~W.}, \bibinfo{author}{Becker, B.~J.}, \&
  \bibinfo{author}{Kalaian, H.} (\bibinfo{year}{1988}).
\newblock \bibinfo{title}{Modeling multivariate effect sizes}.
\newblock {\it \bibinfo{journal}{Psychological Bulletin}\/},  {\it
  \bibinfo{volume}{103}\/}, \bibinfo{pages}{111--120}.
\bibitem[{Raudenbush \& Bryk(2002)}]{raudenbush2002hierarchical}
\bibinfo{author}{Raudenbush, S.~W.}, \& \bibinfo{author}{Bryk, A.~S.}
  (\bibinfo{year}{2002}).
\newblock {\it \bibinfo{title}{Hierarchical linear models: {Applications} and
  data analysis methods}\/} volume~\bibinfo{volume}{1}.
\newblock (\bibinfo{edition}{2nd} ed.).
\newblock \bibinfo{publisher}{SAGE Publications, Inc}.
\bibitem[{Ravnskov(1992)}]{ravnskov1992frequency}
\bibinfo{author}{Ravnskov, U.} (\bibinfo{year}{1992}).
\newblock \bibinfo{title}{Frequency of citation and outcome of cholesterol
  lowering trials}.
\newblock {\it \bibinfo{journal}{BMJ: British Medical Journal}\/},  {\it
  \bibinfo{volume}{305}\/}, \bibinfo{pages}{717}.
\bibitem[{Rosenthal(1979)}]{rosenthal1979file}
\bibinfo{author}{Rosenthal, R.} (\bibinfo{year}{1979}).
\newblock \bibinfo{title}{The file drawer problem and tolerance for null
  results}.
\newblock {\it \bibinfo{journal}{Psychological Bulletin}\/},  {\it
  \bibinfo{volume}{86}\/}, \bibinfo{pages}{638--641}.
\bibitem[{Rosenthal(1991)}]{rosenthal1991meta}
\bibinfo{author}{Rosenthal, R.} (\bibinfo{year}{1991}).
\newblock {\it \bibinfo{title}{Meta-analytic procedures for social
  research}\/}.
\newblock (\bibinfo{edition}{2nd} ed.).
\newblock \bibinfo{address}{Beverly Hills, California}:
  \bibinfo{publisher}{Sage Publications}.
\bibitem[{Rosenthal \& Rubin(1986)}]{rosenthal1986meta}
\bibinfo{author}{Rosenthal, R.}, \& \bibinfo{author}{Rubin, D.~B.}
  (\bibinfo{year}{1986}).
\newblock \bibinfo{title}{Meta-analytic procedures for combining studies with
  multiple effect sizes}.
\newblock {\it \bibinfo{journal}{Psychological Bulletin}\/},  {\it
  \bibinfo{volume}{99}\/}, \bibinfo{pages}{400--406}.
\bibitem[{Rothstein(2008)}]{rothstein2008publication}
\bibinfo{author}{Rothstein, H.~R.} (\bibinfo{year}{2008}).
\newblock \bibinfo{title}{Publication bias as a threat to the validity of
  meta-analytic results}.
\newblock {\it \bibinfo{journal}{Journal of Experimental Criminology}\/},  {\it
  \bibinfo{volume}{4}\/}, \bibinfo{pages}{61--81}.
\bibitem[{Rothstein et~al.(2005{\natexlab{a}})Rothstein, Sutton \&
  Borenstein}]{rothstein2005publicationch1}
\bibinfo{author}{Rothstein, H.~R.}, \bibinfo{author}{Sutton, A.~J.}, \&
  \bibinfo{author}{Borenstein, M.} (\bibinfo{year}{2005}{\natexlab{a}}).
\newblock \bibinfo{title}{Publication bias in meta-analysis}.
\newblock In \bibinfo{editor}{H.~R. Rothstein}, \bibinfo{editor}{A.~J. Sutton},
  \& \bibinfo{editor}{M.~Borenstein} (Eds.), {\it
  \bibinfo{booktitle}{Publication bias in meta-analysis: {Prevention},
  assessment and adjustments}\/} chapter~\bibinfo{chapter}{1}. (pp.
  \bibinfo{pages}{1--7}).
\newblock \bibinfo{publisher}{John Wiley \& Sons}.
\bibitem[{Rothstein et~al.(2005{\natexlab{b}})Rothstein, Sutton \&
  Borenstein}]{rothstein2005publication}
\bibinfo{author}{Rothstein, H.~R.}, \bibinfo{author}{Sutton, A.~J.}, \&
  \bibinfo{author}{Borenstein, M.} (\bibinfo{year}{2005}{\natexlab{b}}).
\newblock {\it \bibinfo{title}{Publication bias in meta-analysis: {Prevention},
  assessment and adjustments}\/}.
\newblock \bibinfo{address}{Chichester}: \bibinfo{publisher}{John Wiley \&
  Sons}.
\bibitem[{R{\"o}ver et~al.(2015)R{\"o}ver, Knapp \& Friede}]{rover2015hartung}
\bibinfo{author}{R{\"o}ver, C.}, \bibinfo{author}{Knapp, G.}, \&
  \bibinfo{author}{Friede, T.} (\bibinfo{year}{2015}).
\newblock \bibinfo{title}{{Hartung-Knapp-Sidik-Jonkman} approach and its
  modification for random-effects meta-analysis with few studies}.
\newblock {\it \bibinfo{journal}{BMC Medical Research Methodology}\/},  {\it
  \bibinfo{volume}{15}\/}, \bibinfo{pages}{99}.
\bibitem[{Sackett et~al.(1986)Sackett, Harris \& Orr}]{sackett1986seeking}
\bibinfo{author}{Sackett, P.~R.}, \bibinfo{author}{Harris, M.~M.}, \&
  \bibinfo{author}{Orr, J.~M.} (\bibinfo{year}{1986}).
\newblock \bibinfo{title}{On seeking moderator variables in the meta-analysis
  of correlational data: {A Monte Carlo} investigation of statistical power and
  resistance to {Type I} error}.
\newblock {\it \bibinfo{journal}{Journal of Applied Psychology}\/},  {\it
  \bibinfo{volume}{71}\/}, \bibinfo{pages}{302--310}.
\bibitem[{S{\'a}nchez-Meca \&
  Mar{\'\i}n-Mart{\'\i}nez(1997)}]{sanchez1997homogeneity}
\bibinfo{author}{S{\'a}nchez-Meca, J.}, \&
  \bibinfo{author}{Mar{\'\i}n-Mart{\'\i}nez, F.} (\bibinfo{year}{1997}).
\newblock \bibinfo{title}{Homogeneity tests in meta-analysis: {A Monte Carlo}
  comparison of statistical power and {Type I} error}.
\newblock {\it \bibinfo{journal}{Quality and Quantity}\/},  {\it
  \bibinfo{volume}{31}\/}, \bibinfo{pages}{385--399}.
\bibitem[{S{\'a}nchez-Meca \&
  Mar{\'\i}n-Mart{\'\i}nez(2008)}]{sanchez2008confidence}
\bibinfo{author}{S{\'a}nchez-Meca, J.}, \&
  \bibinfo{author}{Mar{\'\i}n-Mart{\'\i}nez, F.} (\bibinfo{year}{2008}).
\newblock \bibinfo{title}{Confidence intervals for the overall effect size in
  random-effects meta-analysis}.
\newblock {\it \bibinfo{journal}{Psychological Methods}\/},  {\it
  \bibinfo{volume}{13}\/}, \bibinfo{pages}{31--48}.
\bibitem[{Scammacca et~al.(2014)Scammacca, Roberts \&
  Stuebing}]{scammacca2014meta}
\bibinfo{author}{Scammacca, N.}, \bibinfo{author}{Roberts, G.}, \&
  \bibinfo{author}{Stuebing, K.~K.} (\bibinfo{year}{2014}).
\newblock \bibinfo{title}{Meta-analysis with complex research designs:
  {Dealing} with dependence from multiple measures and multiple group
  comparisons}.
\newblock {\it \bibinfo{journal}{Review of Educational Research}\/},  {\it
  \bibinfo{volume}{84}\/}, \bibinfo{pages}{328--364}.
\bibitem[{Scherer et~al.(2007)Scherer, Langenberg \& Von~Elm}]{scherer2007full}
\bibinfo{author}{Scherer, R.~W.}, \bibinfo{author}{Langenberg, P.}, \&
  \bibinfo{author}{Von~Elm, E.} (\bibinfo{year}{2007}).
\newblock \bibinfo{title}{Full publication of results initially presented in
  abstracts}.
\newblock {\it \bibinfo{journal}{Cochrane Database of Systematic Reviews}\/},
  {\it \bibinfo{volume}{18}\/}.
\bibitem[{Schmidt et~al.(2009)Schmidt, Oh \& Hayes}]{schmidt2009fixed}
\bibinfo{author}{Schmidt, F.~L.}, \bibinfo{author}{Oh, I.-S.}, \&
  \bibinfo{author}{Hayes, T.~L.} (\bibinfo{year}{2009}).
\newblock \bibinfo{title}{Fixed-versus random-effects models in meta-analysis:
  {Model} properties and an empirical comparison of differences in results}.
\newblock {\it \bibinfo{journal}{British Journal of Mathematical and
  Statistical Psychology}\/},  {\it \bibinfo{volume}{62}\/},
  \bibinfo{pages}{97--128}.
\bibitem[{Sidik \& Jonkman(2002)}]{sidik2002simple}
\bibinfo{author}{Sidik, K.}, \& \bibinfo{author}{Jonkman, J.~N.}
  (\bibinfo{year}{2002}).
\newblock \bibinfo{title}{A simple confidence interval for meta-analysis}.
\newblock {\it \bibinfo{journal}{Statistics in Medicine}\/},  {\it
  \bibinfo{volume}{21}\/}, \bibinfo{pages}{3153--3159}.
\bibitem[{Sidik \& Jonkman(2003)}]{sidik2003constructing}
\bibinfo{author}{Sidik, K.}, \& \bibinfo{author}{Jonkman, J.~N.}
  (\bibinfo{year}{2003}).
\newblock \bibinfo{title}{On constructing confidence intervals for a
  standardized mean difference in meta-analysis}.
\newblock {\it \bibinfo{journal}{Communications in Statistics-Simulation and
  Computation}\/},  {\it \bibinfo{volume}{32}\/}, \bibinfo{pages}{1191--1203}.
\bibitem[{Sidik \& Jonkman(2006)}]{sidik2006robust}
\bibinfo{author}{Sidik, K.}, \& \bibinfo{author}{Jonkman, J.~N.}
  (\bibinfo{year}{2006}).
\newblock \bibinfo{title}{Robust variance estimation for random effects
  meta-analysis}.
\newblock {\it \bibinfo{journal}{Computational Statistics \& Data Analysis}\/},
   {\it \bibinfo{volume}{50}\/}, \bibinfo{pages}{3681--3701}.
\bibitem[{Sidik \& Jonkman(2007)}]{sidik2007comparison}
\bibinfo{author}{Sidik, K.}, \& \bibinfo{author}{Jonkman, J.~N.}
  (\bibinfo{year}{2007}).
\newblock \bibinfo{title}{A comparison of heterogeneity variance estimators in
  combining results of studies}.
\newblock {\it \bibinfo{journal}{Statistics in Medicine}\/},  {\it
  \bibinfo{volume}{26}\/}, \bibinfo{pages}{1964--1981}.
\bibitem[{Simonsohn et~al.(2014{\natexlab{a}})Simonsohn, Nelson \&
  Simmons}]{simonsohn2014pcurve}
\bibinfo{author}{Simonsohn, U.}, \bibinfo{author}{Nelson, L.~D.}, \&
  \bibinfo{author}{Simmons, J.~P.} (\bibinfo{year}{2014}{\natexlab{a}}).
\newblock \bibinfo{title}{P-curve: {A} key to the file-drawer}.
\newblock {\it \bibinfo{journal}{Journal of Experimental Psychology:
  General}\/},  {\it \bibinfo{volume}{143}\/}, \bibinfo{pages}{534--547}.
\bibitem[{Simonsohn et~al.(2014{\natexlab{b}})Simonsohn, Nelson \&
  Simmons}]{simonsohn2014effect}
\bibinfo{author}{Simonsohn, U.}, \bibinfo{author}{Nelson, L.~D.}, \&
  \bibinfo{author}{Simmons, J.~P.} (\bibinfo{year}{2014}{\natexlab{b}}).
\newblock \bibinfo{title}{P-curve and effect size: {Correcting} for publication
  bias using only significant results}.
\newblock {\it \bibinfo{journal}{Perspectives on Psychological Science}\/},
  {\it \bibinfo{volume}{9}\/}, \bibinfo{pages}{666--681}.
\bibitem[{Smith \& Glass(1977)}]{smith1977meta}
\bibinfo{author}{Smith, M.~L.}, \& \bibinfo{author}{Glass, G.~V.}
  (\bibinfo{year}{1977}).
\newblock \bibinfo{title}{Meta-analysis of psychotherapy outcome studies}.
\newblock {\it \bibinfo{journal}{American Psychologist}\/},  {\it
  \bibinfo{volume}{32}\/}, \bibinfo{pages}{752--760}.
\bibitem[{Smyth et~al.(2011)Smyth, Kirkham, Jacoby, Altman, Gamble \&
  Williamson}]{smyth2011frequency}
\bibinfo{author}{Smyth, R.}, \bibinfo{author}{Kirkham, J.},
  \bibinfo{author}{Jacoby, A.}, \bibinfo{author}{Altman, D.},
  \bibinfo{author}{Gamble, C.}, \& \bibinfo{author}{Williamson, P.}
  (\bibinfo{year}{2011}).
\newblock \bibinfo{title}{Frequency and reasons for outcome reporting bias in
  clinical trials: {Interviews} with trialists}.
\newblock {\it \bibinfo{journal}{Bmj}\/},  {\it \bibinfo{volume}{342}\/},
  \bibinfo{pages}{c7153}.
\bibitem[{Song et~al.(2000)Song, Eastwood, Gilbody, Duley \&
  Sutton}]{song2000publication}
\bibinfo{author}{Song, F.}, \bibinfo{author}{Eastwood, A.},
  \bibinfo{author}{Gilbody, S.}, \bibinfo{author}{Duley, L.}, \&
  \bibinfo{author}{Sutton, A.} (\bibinfo{year}{2000}).
\newblock \bibinfo{title}{Publication and related biases: {A} review}.
\newblock {\it \bibinfo{journal}{Health Technology Assessment}\/},  {\it
  \bibinfo{volume}{4}\/}.
\bibitem[{Song et~al.(2002)Song, Khan, Dinnes \& Sutton}]{song2002asymmetric}
\bibinfo{author}{Song, F.}, \bibinfo{author}{Khan, K.~S.},
  \bibinfo{author}{Dinnes, J.}, \& \bibinfo{author}{Sutton, A.~J.}
  (\bibinfo{year}{2002}).
\newblock \bibinfo{title}{Asymmetric funnel plots and publication bias in
  meta-analyses of diagnostic accuracy}.
\newblock {\it \bibinfo{journal}{International Journal of Epidemiology}\/},
  {\it \bibinfo{volume}{31}\/}, \bibinfo{pages}{88--95}.
\bibitem[{Spector \& Levine(1987)}]{spector1987meta}
\bibinfo{author}{Spector, P.~E.}, \& \bibinfo{author}{Levine, E.~L.}
  (\bibinfo{year}{1987}).
\newblock \bibinfo{title}{Meta-analysis for integrating study outcomes: {A
  Monte Carlo} study of its susceptibility to {Type I and Type II errors}}.
\newblock {\it \bibinfo{journal}{Journal of Applied Psychology}\/},  {\it
  \bibinfo{volume}{72}\/}, \bibinfo{pages}{3--9}.
\bibitem[{Stanley \& Doucouliagos(2007)}]{stanley2007identifying}
\bibinfo{author}{Stanley, T.}, \& \bibinfo{author}{Doucouliagos, C.}
  (\bibinfo{year}{2007}).
\newblock \bibinfo{title}{Identifying and correcting publication selection bias
  in the efficiency-wage literature: {Heckman} meta-regression}.
\newblock \bibinfo{note}{School Working Paper No. 2007/11}.
\bibitem[{Stanley(2001)}]{stanley2001wheat}
\bibinfo{author}{Stanley, T.~D.} (\bibinfo{year}{2001}).
\newblock \bibinfo{title}{Wheat from chaff: {Meta-analysis} as quantitative
  literature review}.
\newblock {\it \bibinfo{journal}{Journal of Economic Perspectives}\/},  {\it
  \bibinfo{volume}{15}\/}, \bibinfo{pages}{131--150}.
\bibitem[{Stanley(2004)}]{stanley2004does}
\bibinfo{author}{Stanley, T.~D.} (\bibinfo{year}{2004}).
\newblock \bibinfo{title}{Does unemployment hysteresis falsify the natural rate
  hypothesis? {A} meta-regression analysis}.
\newblock {\it \bibinfo{journal}{Journal of Economic Surveys}\/},  {\it
  \bibinfo{volume}{18}\/}, \bibinfo{pages}{589--612}.
\bibitem[{Stanley(2005)}]{stanley2005beyond}
\bibinfo{author}{Stanley, T.~D.} (\bibinfo{year}{2005}).
\newblock \bibinfo{title}{Beyond publication bias}.
\newblock {\it \bibinfo{journal}{Journal of Economic Surveys}\/},  {\it
  \bibinfo{volume}{19}\/}, \bibinfo{pages}{309--345}.
\bibitem[{Stanley(2008)}]{stanley2008meta2}
\bibinfo{author}{Stanley, T.~D.} (\bibinfo{year}{2008}).
\newblock \bibinfo{title}{Meta-regression methods for detecting and estimating
  empirical effects in the presence of publication selection}.
\newblock {\it \bibinfo{journal}{Oxford Bulletin of Economics and
  Statistics}\/},  {\it \bibinfo{volume}{70}\/}, \bibinfo{pages}{103--127}.
\bibitem[{Stanley et~al.(2008)Stanley, Doucouliagos \&
  Jarrell}]{stanley2008meta1}
\bibinfo{author}{Stanley, T.~D.}, \bibinfo{author}{Doucouliagos, C.}, \&
  \bibinfo{author}{Jarrell, S.~B.} (\bibinfo{year}{2008}).
\newblock \bibinfo{title}{Meta-regression analysis as the socio-economics of
  economics research}.
\newblock {\it \bibinfo{journal}{The Journal of Socio-Economics}\/},  {\it
  \bibinfo{volume}{37}\/}, \bibinfo{pages}{276--292}.
\bibitem[{Stanley \& Doucouliagos(2010)}]{stanley2010picture}
\bibinfo{author}{Stanley, T.~D.}, \& \bibinfo{author}{Doucouliagos, H.}
  (\bibinfo{year}{2010}).
\newblock \bibinfo{title}{Picture this: {A} simple graph that reveals much ado
  about research}.
\newblock {\it \bibinfo{journal}{Journal of Economic Surveys}\/},  {\it
  \bibinfo{volume}{24}\/}, \bibinfo{pages}{170--191}.
\bibitem[{Stanley \& Doucouliagos(2012)}]{stanley2012meta}
\bibinfo{author}{Stanley, T.~D.}, \& \bibinfo{author}{Doucouliagos, H.}
  (\bibinfo{year}{2012}).
\newblock {\it \bibinfo{title}{Meta-regression analysis in economics and
  business}\/}.
\newblock \bibinfo{address}{Oxon, England}: \bibinfo{publisher}{Routledge}.
\bibitem[{Stanley \& Doucouliagos(2013)}]{stanley2013better}
\bibinfo{author}{Stanley, T.~D.}, \& \bibinfo{author}{Doucouliagos, H.}
  (\bibinfo{year}{2013}).
\newblock {\it \bibinfo{title}{Better than random: {Weighted} least squares
  meta-regression analysis}\/}.
\newblock \bibinfo{type}{Working Paper} Deakin University.
\bibitem[{Stanley \& Doucouliagos(2014)}]{stanley2014meta}
\bibinfo{author}{Stanley, T.~D.}, \& \bibinfo{author}{Doucouliagos, H.}
  (\bibinfo{year}{2014}).
\newblock \bibinfo{title}{Meta-regression approximations to reduce publication
  selection bias}.
\newblock {\it \bibinfo{journal}{Research Synthesis Methods}\/},  {\it
  \bibinfo{volume}{5}\/}, \bibinfo{pages}{60--78}.
\bibitem[{Stanley \& Doucouliagos(2015)}]{stanley2015neither}
\bibinfo{author}{Stanley, T.~D.}, \& \bibinfo{author}{Doucouliagos, H.}
  (\bibinfo{year}{2015}).
\newblock \bibinfo{title}{Neither fixed nor random: {W}eighted least squares
  meta-analysis}.
\newblock {\it \bibinfo{journal}{Statistics in Medicine}\/},  {\it
  \bibinfo{volume}{34}\/}, \bibinfo{pages}{2116--2127}.
\bibitem[{Stanley \& Doucouliagos(2016)}]{stanley2016neither}
\bibinfo{author}{Stanley, T.~D.}, \& \bibinfo{author}{Doucouliagos, H.}
  (\bibinfo{year}{2016}).
\newblock \bibinfo{title}{Neither fixed nor random: {Weighted} least squares
  meta-regression}.
\newblock {\it \bibinfo{journal}{Research Synthesis Methods}\/},  {\it
  \bibinfo{volume}{8}\/}, \bibinfo{pages}{19--42}.
\bibitem[{Stanley et~al.(2017)Stanley, Doucouliagos \&
  Ioannidis}]{stanley2017finding}
\bibinfo{author}{Stanley, T.~D.}, \bibinfo{author}{Doucouliagos, H.}, \&
  \bibinfo{author}{Ioannidis, J. P.~A.} (\bibinfo{year}{2017}).
\newblock \bibinfo{title}{Finding the power to reduce publication bias}.
\newblock {\it \bibinfo{journal}{Statistics in Medicine}\/},  {\it
  \bibinfo{volume}{36}\/}, \bibinfo{pages}{1580--1598}.
\bibitem[{Stanley et~al.(2021)Stanley, Doucouliagos, Ioannidis \&
  Carter}]{stanley2021detecting}
\bibinfo{author}{Stanley, T.~D.}, \bibinfo{author}{Doucouliagos, H.},
  \bibinfo{author}{Ioannidis, J. P.~A.}, \& \bibinfo{author}{Carter, E.~C.}
  (\bibinfo{year}{2021}).
\newblock \bibinfo{title}{Detecting publication selection bias through excess
  statistical significance}.
\newblock {\it \bibinfo{journal}{Research Synthesis Methods}\/},  {\it
  \bibinfo{volume}{12}\/}, \bibinfo{pages}{776--795}.
\bibitem[{Stanley et~al.(2023)Stanley, Ioannidis, Maier, Doucouliagos, Otte \&
  Bartoš}]{stanley2023unrestricted}
\bibinfo{author}{Stanley, T.~D.}, \bibinfo{author}{Ioannidis, J. P.~A.},
  \bibinfo{author}{Maier, M.}, \bibinfo{author}{Doucouliagos, H.},
  \bibinfo{author}{Otte, W.~M.}, \& \bibinfo{author}{Bartoš, F.}
  (\bibinfo{year}{2023}).
\newblock \bibinfo{title}{Unrestricted weighted least squares represent medical
  research better than random effects in 67,308 cochrane meta-analyses}.
\newblock {\it \bibinfo{journal}{Journal of Clinical Epidemiology}\/},  {\it
  \bibinfo{volume}{157}\/}, \bibinfo{pages}{53--58}.
\bibitem[{Stanley et~al.(2010)Stanley, Jarrell \&
  Doucouliagos}]{stanley2010discard}
\bibinfo{author}{Stanley, T.~D.}, \bibinfo{author}{Jarrell, S.~B.}, \&
  \bibinfo{author}{Doucouliagos, H.} (\bibinfo{year}{2010}).
\newblock \bibinfo{title}{Could it be better to discard 90\% of the data? {A}
  statistical paradox}.
\newblock {\it \bibinfo{journal}{The American Statistician}\/},  {\it
  \bibinfo{volume}{64}\/}, \bibinfo{pages}{70--77}.
\bibitem[{Stern \& Simes(1997)}]{stern1997publication}
\bibinfo{author}{Stern, J.~M.}, \& \bibinfo{author}{Simes, R.~J.}
  (\bibinfo{year}{1997}).
\newblock \bibinfo{title}{Publication bias: {Evidence} of delayed publication
  in a cohort study of clinical research projects}.
\newblock {\it \bibinfo{journal}{Bmj}\/},  {\it \bibinfo{volume}{315}\/},
  \bibinfo{pages}{640--645}.
\bibitem[{Sterne et~al.(2005)Sterne, Becker \& Egger}]{sterne2005funnel}
\bibinfo{author}{Sterne, J.~A.}, \bibinfo{author}{Becker, B.~J.}, \&
  \bibinfo{author}{Egger, M.} (\bibinfo{year}{2005}).
\newblock \bibinfo{title}{The funnel plot}.
\newblock In \bibinfo{editor}{H.~R. Rothstein}, \bibinfo{editor}{A.~J. Sutton},
  \& \bibinfo{editor}{M.~Borenstein} (Eds.), {\it
  \bibinfo{booktitle}{Publication bias in meta-analysis: {Prevention},
  assessment and adjustments}\/} chapter~\bibinfo{chapter}{5}. (pp.
  \bibinfo{pages}{75--98}).
\newblock \bibinfo{publisher}{John Wiley \& Sons}.
\bibitem[{Sterne \& Egger(2001)}]{sterne2001funnel}
\bibinfo{author}{Sterne, J.~A.}, \& \bibinfo{author}{Egger, M.}
  (\bibinfo{year}{2001}).
\newblock \bibinfo{title}{Funnel plots for detecting bias in meta-analysis:
  {Guidelines} on choice of axis}.
\newblock {\it \bibinfo{journal}{Journal of Clinical Epidemiology}\/},  {\it
  \bibinfo{volume}{54}\/}, \bibinfo{pages}{1046--1055}.
\bibitem[{Sterne \& Egger(2005)}]{sterne2005regression}
\bibinfo{author}{Sterne, J.~A.}, \& \bibinfo{author}{Egger, M.}
  (\bibinfo{year}{2005}).
\newblock \bibinfo{title}{Regression methods to detect publication and other
  bias in meta-analysis}.
\newblock In \bibinfo{editor}{H.~R. Rothstein}, \bibinfo{editor}{A.~J. Sutton},
  \& \bibinfo{editor}{M.~Borenstein} (Eds.), {\it
  \bibinfo{booktitle}{Publication bias in meta-analysis: {Prevention},
  assessment and adjustments}\/} chapter~\bibinfo{chapter}{5}. (pp.
  \bibinfo{pages}{99--110}).
\newblock \bibinfo{publisher}{John Wiley \& Sons}.
\bibitem[{Sterne et~al.(2008)Sterne, Egger \& Moher}]{sterne2008addressing}
\bibinfo{author}{Sterne, J.~A.}, \bibinfo{author}{Egger, M.}, \&
  \bibinfo{author}{Moher, D.} (\bibinfo{year}{2008}).
\newblock \bibinfo{title}{Addressing reporting biases}.
\newblock In \bibinfo{editor}{J.~P. Higgins}, \& \bibinfo{editor}{S.~Green}
  (Eds.), {\it \bibinfo{booktitle}{Cochrane handbook for systematic reviews of
  interventions: {Cochrane} book series}\/} chapter~\bibinfo{chapter}{10}. (pp.
  \bibinfo{pages}{297--333}).
\newblock \bibinfo{publisher}{Wiley Online Library}.
\bibitem[{Sterne et~al.(2001)Sterne, Egger \& Smith}]{sterne2001investigating}
\bibinfo{author}{Sterne, J.~A.}, \bibinfo{author}{Egger, M.}, \&
  \bibinfo{author}{Smith, G.~D.} (\bibinfo{year}{2001}).
\newblock \bibinfo{title}{Investigating and dealing with publication and other
  biases}.
\newblock In \bibinfo{editor}{M.~Egger}, \bibinfo{editor}{G.~Davey-Smith}, \&
  \bibinfo{editor}{D.~Altman} (Eds.), {\it \bibinfo{booktitle}{Systematic
  reviews in health care: {Meta-analysis} in context}\/}
  chapter~\bibinfo{chapter}{11}. (pp. \bibinfo{pages}{189--208}).
\newblock \bibinfo{publisher}{Wiley Online Library}.
\bibitem[{Sterne et~al.(2000)Sterne, Gavaghan \& Egger}]{sterne2000publication}
\bibinfo{author}{Sterne, J.~A.}, \bibinfo{author}{Gavaghan, D.}, \&
  \bibinfo{author}{Egger, M.} (\bibinfo{year}{2000}).
\newblock \bibinfo{title}{Publication and related bias in meta-analysis:
  {P}ower of statistical tests and prevalence in the literature}.
\newblock {\it \bibinfo{journal}{Journal of Clinical Epidemiology}\/},  {\it
  \bibinfo{volume}{53}\/}, \bibinfo{pages}{1119--1129}.
\bibitem[{Sterne et~al.(2011)Sterne, Sutton, Ioannidis, Terrin, Jones, Lau,
  Carpenter, R{\"u}cker, Harbord \& Schmid}]{sterne2011recommendations}
\bibinfo{author}{Sterne, J.~A.}, \bibinfo{author}{Sutton, A.~J.},
  \bibinfo{author}{Ioannidis, J.~P.}, \bibinfo{author}{Terrin, N.},
  \bibinfo{author}{Jones, D.~R.}, \bibinfo{author}{Lau, J.},
  \bibinfo{author}{Carpenter, J.}, \bibinfo{author}{R{\"u}cker, G.},
  \bibinfo{author}{Harbord, R.~M.}, \& \bibinfo{author}{Schmid, C.~H.}
  (\bibinfo{year}{2011}).
\newblock \bibinfo{title}{Recommendations for examining and interpreting funnel
  plot asymmetry in meta-analyses of randomised controlled trials}.
\newblock {\it \bibinfo{journal}{Bmj}\/},  {\it \bibinfo{volume}{343}\/},
  \bibinfo{pages}{d4002}.
\bibitem[{Stevens \& Taylor(2009)}]{stevens2009hierarchical}
\bibinfo{author}{Stevens, J.~R.}, \& \bibinfo{author}{Taylor, A.~M.}
  (\bibinfo{year}{2009}).
\newblock \bibinfo{title}{Hierarchical dependence in meta-analysis}.
\newblock {\it \bibinfo{journal}{Journal of Educational and Behavioral
  Statistics}\/},  {\it \bibinfo{volume}{34}\/}, \bibinfo{pages}{46--73}.
\bibitem[{Stoel et~al.(2006)Stoel, Garre, Dolan \& Van
  Den~Wittenboer}]{stoel2006likelihood}
\bibinfo{author}{Stoel, R.~D.}, \bibinfo{author}{Garre, F.~G.},
  \bibinfo{author}{Dolan, C.}, \& \bibinfo{author}{Van Den~Wittenboer, G.}
  (\bibinfo{year}{2006}).
\newblock \bibinfo{title}{On the likelihood ratio test in structural equation
  modeling when parameters are subject to boundary constraints}.
\newblock {\it \bibinfo{journal}{Psychological Methods}\/},  {\it
  \bibinfo{volume}{11}\/}, \bibinfo{pages}{439--455}.
\bibitem[{Sutton(2009)}]{sutton2009publication}
\bibinfo{author}{Sutton, A.~J.} (\bibinfo{year}{2009}).
\newblock \bibinfo{title}{Publication bias}.
\newblock In \bibinfo{editor}{C.~Harris}, \bibinfo{editor}{L.~Hedges}, \&
  \bibinfo{editor}{J.~Valentine} (Eds.), {\it \bibinfo{booktitle}{Handbook of
  research synthesis and meta-analysis}\/} chapter~\bibinfo{chapter}{23}. (pp.
  \bibinfo{pages}{435--452}).
\newblock \bibinfo{address}{New York}: \bibinfo{publisher}{Russell Sage
  Foundation New York, NY}. (\bibinfo{edition}{2nd} ed.).
\bibitem[{Sutton et~al.(2000)Sutton, Duval, Tweedie, Abrams \&
  Jones}]{sutton2000empirical}
\bibinfo{author}{Sutton, A.~J.}, \bibinfo{author}{Duval, S.},
  \bibinfo{author}{Tweedie, R.}, \bibinfo{author}{Abrams, K.~R.}, \&
  \bibinfo{author}{Jones, D.~R.} (\bibinfo{year}{2000}).
\newblock \bibinfo{title}{Empirical assessment of effect of publication bias on
  meta-analyses}.
\newblock {\it \bibinfo{journal}{Bmj}\/},  {\it \bibinfo{volume}{320}\/},
  \bibinfo{pages}{1574--1577}.
\bibitem[{Tanner-Smith \& Tipton(2014)}]{tanner2014robust}
\bibinfo{author}{Tanner-Smith, E.~E.}, \& \bibinfo{author}{Tipton, E.}
  (\bibinfo{year}{2014}).
\newblock \bibinfo{title}{Robust variance estimation with dependent effect
  sizes: {P}ractical considerations including a software tutorial in {Stata and
  SPSS}}.
\newblock {\it \bibinfo{journal}{Research Synthesis Methods}\/},  {\it
  \bibinfo{volume}{5}\/}, \bibinfo{pages}{13--30}.
\bibitem[{Tanner-Smith et~al.(2016)Tanner-Smith, Tipton \&
  Polanin}]{tanner2016handling}
\bibinfo{author}{Tanner-Smith, E.~E.}, \bibinfo{author}{Tipton, E.}, \&
  \bibinfo{author}{Polanin, J.~R.} (\bibinfo{year}{2016}).
\newblock \bibinfo{title}{Handling complex meta-analytic data structures using
  robust variance estimates: {A tutorial in R}}.
\newblock {\it \bibinfo{journal}{Journal of Developmental and Life-Course
  Criminology}\/},  {\it \bibinfo{volume}{2}\/}, \bibinfo{pages}{85--112}.
\bibitem[{Thompson(1994)}]{thompson1994sources}
\bibinfo{author}{Thompson, S.~G.} (\bibinfo{year}{1994}).
\newblock \bibinfo{title}{Why sources of heterogeneity in meta-analysis should
  be investigated}.
\newblock {\it \bibinfo{journal}{BMJ: British Medical Journal}\/},  {\it
  \bibinfo{volume}{309}\/}, \bibinfo{pages}{1351--1355}.
\bibitem[{Thompson \& Sharp(1999)}]{thompson1999explaining}
\bibinfo{author}{Thompson, S.~G.}, \& \bibinfo{author}{Sharp, S.~J.}
  (\bibinfo{year}{1999}).
\newblock \bibinfo{title}{Explaining heterogeneity in meta-analysis: {A}
  comparison of methods}.
\newblock {\it \bibinfo{journal}{Statistics in Medicine}\/},  {\it
  \bibinfo{volume}{18}\/}, \bibinfo{pages}{2693--2708}.
\bibitem[{Tipton(2013)}]{tipton2013robust}
\bibinfo{author}{Tipton, E.} (\bibinfo{year}{2013}).
\newblock \bibinfo{title}{Robust variance estimation in meta-regression with
  binary dependent effects}.
\newblock {\it \bibinfo{journal}{Research Synthesis Methods}\/},  {\it
  \bibinfo{volume}{4}\/}, \bibinfo{pages}{169--187}.
\bibitem[{Tipton(2015)}]{tipton2015small}
\bibinfo{author}{Tipton, E.} (\bibinfo{year}{2015}).
\newblock \bibinfo{title}{Small sample adjustments for robust variance
  estimation with meta-regression}.
\newblock {\it \bibinfo{journal}{Psychological Methods}\/},  {\it
  \bibinfo{volume}{20}\/}, \bibinfo{pages}{375--393}.
\bibitem[{Tipton \& Pustejovsky(2015)}]{tipton2015smallsample}
\bibinfo{author}{Tipton, E.}, \& \bibinfo{author}{Pustejovsky, J.~E.}
  (\bibinfo{year}{2015}).
\newblock \bibinfo{title}{Small-sample adjustments for tests of moderators and
  model fit using robust variance estimation in meta-regression}.
\newblock {\it \bibinfo{journal}{Journal of Educational and Behavioral
  Statistics}\/},  {\it \bibinfo{volume}{40}\/}, \bibinfo{pages}{604--634}.
\bibitem[{Tram{\`e}r et~al.(1997)Tram{\`e}r, Reynolds, Moore \&
  McQuay}]{tramer1997impact}
\bibinfo{author}{Tram{\`e}r, M.~R.}, \bibinfo{author}{Reynolds, D. J.~M.},
  \bibinfo{author}{Moore, R.~A.}, \& \bibinfo{author}{McQuay, H.~J.}
  (\bibinfo{year}{1997}).
\newblock \bibinfo{title}{Impact of covert duplicate publication on
  meta-analysis: {A} case study}.
\newblock {\it \bibinfo{journal}{Bmj}\/},  {\it \bibinfo{volume}{315}\/},
  \bibinfo{pages}{635--640}.
\bibitem[{Van~Houwelingen et~al.(2002)Van~Houwelingen, Arends \&
  Stijnen}]{van2002advanced}
\bibinfo{author}{Van~Houwelingen, H.~C.}, \bibinfo{author}{Arends, L.~R.}, \&
  \bibinfo{author}{Stijnen, T.} (\bibinfo{year}{2002}).
\newblock \bibinfo{title}{Advanced methods in meta-analysis: {Multivariate}
  approach and meta-regression}.
\newblock {\it \bibinfo{journal}{Statistics in Medicine}\/},  {\it
  \bibinfo{volume}{21}\/}, \bibinfo{pages}{589--624}.
\bibitem[{Veroniki et~al.(2016)Veroniki, Jackson, Viechtbauer, Bender, Bowden,
  Knapp, Kuss, Higgins, Langan \& Salanti}]{veroniki2016methods}
\bibinfo{author}{Veroniki, A.~A.}, \bibinfo{author}{Jackson, D.},
  \bibinfo{author}{Viechtbauer, W.}, \bibinfo{author}{Bender, R.},
  \bibinfo{author}{Bowden, J.}, \bibinfo{author}{Knapp, G.},
  \bibinfo{author}{Kuss, O.}, \bibinfo{author}{Higgins, J.},
  \bibinfo{author}{Langan, D.}, \& \bibinfo{author}{Salanti, G.}
  (\bibinfo{year}{2016}).
\newblock \bibinfo{title}{Methods to estimate the between-study variance and
  its uncertainty in meta-analysis}.
\newblock {\it \bibinfo{journal}{Research Synthesis Methods}\/},  {\it
  \bibinfo{volume}{7}\/}, \bibinfo{pages}{55--79}.
\bibitem[{Vevea et~al.(2019)Vevea, Coburn \& Sutton}]{vevea2019publication}
\bibinfo{author}{Vevea, J.~L.}, \bibinfo{author}{Coburn, K.}, \&
  \bibinfo{author}{Sutton, A.~J.} (\bibinfo{year}{2019}).
\newblock \bibinfo{title}{Publication bias}.
\newblock In \bibinfo{editor}{H.~Cooper}, \bibinfo{editor}{L.~V. Hedges}, \&
  \bibinfo{editor}{J.~C. Valentine} (Eds.), {\it \bibinfo{booktitle}{The
  Handbook of Research Synthesis and Meta-Analysis (3rd ed.)}\/} (pp.
  \bibinfo{pages}{383--430}).
\newblock \bibinfo{publisher}{Russell Sage Foundation}.
\bibitem[{Vevea \& Hedges(1995)}]{vevea1995general}
\bibinfo{author}{Vevea, J.~L.}, \& \bibinfo{author}{Hedges, L.~V.}
  (\bibinfo{year}{1995}).
\newblock \bibinfo{title}{A general linear model for estimating effect size in
  the presence of publication bias}.
\newblock {\it \bibinfo{journal}{Psychometrika}\/},  {\it
  \bibinfo{volume}{60}\/}, \bibinfo{pages}{419--435}.
\bibitem[{Viechtbauer(2007{\natexlab{a}})}]{viechtbauer2007confidence}
\bibinfo{author}{Viechtbauer, W.} (\bibinfo{year}{2007}{\natexlab{a}}).
\newblock \bibinfo{title}{Confidence intervals for the amount of heterogeneity
  in meta-analysis}.
\newblock {\it \bibinfo{journal}{Statistics in Medicine}\/},  {\it
  \bibinfo{volume}{26}\/}, \bibinfo{pages}{37--52}.
\bibitem[{Viechtbauer(2007{\natexlab{b}})}]{viechtbauer2007hypothesis}
\bibinfo{author}{Viechtbauer, W.} (\bibinfo{year}{2007}{\natexlab{b}}).
\newblock \bibinfo{title}{Hypothesis tests for population heterogeneity in
  meta-analysis}.
\newblock {\it \bibinfo{journal}{British Journal of Mathematical and
  Statistical Psychology}\/},  {\it \bibinfo{volume}{60}\/},
  \bibinfo{pages}{29--60}.
\bibitem[{Viechtbauer(2010)}]{viechtbauer2010package}
\bibinfo{author}{Viechtbauer, W.} (\bibinfo{year}{2010}).
\newblock \bibinfo{title}{Conducting meta-analyses in {R} with the {metafor}
  package}.
\newblock {\it \bibinfo{journal}{Journal of Statistical Software}\/},  {\it
  \bibinfo{volume}{36}\/}, \bibinfo{pages}{1--48}.
\newblock \bibinfo{note}{\url {http://www.jstatsoft.org/v36/i03/}}.
\bibitem[{Viechtbauer(2023)}]{viechtbauer2023metafor}
\bibinfo{author}{Viechtbauer, W.} (\bibinfo{year}{2023}).
\newblock \bibinfo{title}{The metafor package: {A} meta-analysis package for
  {R}}.
\newblock {\it \bibinfo{journal}{R package version 3.6-1}\/}, .
\newblock \bibinfo{note}{\url{https://CRAN.R-project.org/package=metafor}}.
\bibitem[{Viechtbauer et~al.(2015)Viechtbauer, L{\'o}pez-L{\'o}pez,
  S{\'a}nchez-Meca \& Mar{\'\i}n-Mart{\'\i}nez}]{viechtbauer2015comparison}
\bibinfo{author}{Viechtbauer, W.}, \bibinfo{author}{L{\'o}pez-L{\'o}pez,
  J.~A.}, \bibinfo{author}{S{\'a}nchez-Meca, J.}, \&
  \bibinfo{author}{Mar{\'\i}n-Mart{\'\i}nez, F.} (\bibinfo{year}{2015}).
\newblock \bibinfo{title}{A comparison of procedures to test for moderators in
  mixed-effects meta-regression models}.
\newblock {\it \bibinfo{journal}{Psychological Methods}\/},  {\it
  \bibinfo{volume}{20}\/}, \bibinfo{pages}{360--374}.
\bibitem[{Vooren et~al.(2019)Vooren, Haelermans, Groot \& Maassen van~den
  Brink}]{vooren2019effectiveness}
\bibinfo{author}{Vooren, M.}, \bibinfo{author}{Haelermans, C.},
  \bibinfo{author}{Groot, W.}, \& \bibinfo{author}{Maassen van~den Brink, H.}
  (\bibinfo{year}{2019}).
\newblock \bibinfo{title}{The effectiveness of active labor market policies:
  {A} meta-analysis}.
\newblock {\it \bibinfo{journal}{Journal of Economic Surveys}\/},  {\it
  \bibinfo{volume}{33}\/}, \bibinfo{pages}{125--149}.
\bibitem[{Wade et~al.(2006)Wade, Turner, Rothstein \&
  Lavenberg}]{wade2006information}
\bibinfo{author}{Wade, C.~A.}, \bibinfo{author}{Turner, H.~M.},
  \bibinfo{author}{Rothstein, H.~R.}, \& \bibinfo{author}{Lavenberg, J.~G.}
  (\bibinfo{year}{2006}).
\newblock \bibinfo{title}{Information retrieval and the role of the information
  specialist in producing high-quality systematic reviews in the social,
  behavioural and education sciences}.
\newblock {\it \bibinfo{journal}{Evidence \& Policy: A Journal of Research,
  Debate and Practice}\/},  {\it \bibinfo{volume}{2}\/},
  \bibinfo{pages}{89--108}.
\bibitem[{Williamson \& Gamble(2005)}]{williamson2005identification}
\bibinfo{author}{Williamson, P.}, \& \bibinfo{author}{Gamble, C.}
  (\bibinfo{year}{2005}).
\newblock \bibinfo{title}{Identification and impact of outcome selection bias
  in meta-analysis}.
\newblock {\it \bibinfo{journal}{Statistics in Medicine}\/},  {\it
  \bibinfo{volume}{24}\/}, \bibinfo{pages}{1547--1561}.

\end{thebibliography}
\end{document}